%% file: main.tex
\pdfoutput=1

\documentclass[sigplan,10pt]{acmart}

\setcopyright{none}
\renewcommand\footnotetextcopyrightpermission[1]{}
\usepackage{amsmath,amssymb,amsfonts}
\usepackage{algorithmic}
\usepackage{graphicx}
\usepackage{textcomp}
\usepackage{tikz}
\usepackage{booktabs}
\usepackage{xspace}
\usepackage{mathtools}
\usepackage[ruled,vlined]{algorithm2e}
\usepackage{enumitem}
\usepackage{enumerate}
\usepackage{subcaption}
\usepackage{bbding}
\usepackage{multirow}
\usepackage{breqn}
\usepackage{soul}
\usepackage{makecell}

\newcommand{\name}{\text{Concertina}\xspace}

\DeclareMathOperator*{\argmin}{arg\,min}

\newcommand{\thetitle}{Data-Centric Adaptive Pipeline Parallelism for Efficient Heterogeneous Long-Context LLM Training}

\title{Concertina: \thetitle}

\author{Shiju Wang}
\email{wangsj25@mails.tsinghua.edu.cn}
\affiliation{
  \institution{Tsinghua University}
}

\author{Yujie Wang}
\email{alfredwang@pku.edu.cn}
\affiliation{
  \institution{Peking University}
}

\author{Fangcheng Fu}
\email{ccchengff@sjtu.edu.cn}
\affiliation{
  \institution{Shanghai Jiao Tong University}
}

\author{Ao Sun}
\email{maydomine@bupt.edu.cn}
\affiliation{
  \institution{Beijing University of Posts and Telecommunications}
}

\author{Yinxiao Feng}
\email{fyx20@mails.tsinghua.edu.cn}
\affiliation{
  \institution{Tsinghua University}
}

\author{Zijian Zhu}
\email{zhuzj23@mails.tsinghua.edu.cn}
\affiliation{
  \institution{Tsinghua University}
}

\author{Bin Cui}
\email{bin.cui@pku.edu.cn}
\affiliation{
  \institution{Peking University}
}

\author{Xu Han}
\email{thu.hanxu13@gmail.com}
\affiliation{
  \institution{Tsinghua University}
}

\author{Kaisheng Ma}
\email{kaisheng@tsinghua.edu.cn}
\affiliation{
  \institution{Tsinghua University}
}

\begin{document}
\begin{abstract}
\input{Main-Abstract}
\end{abstract}

\maketitle

\input{Main-Intro}

\input{Main-Preliminary}

\input{Main-Method}
\input{Main-Implementation}

\input{Main-Experiment}

\input{Main-Conclusion}

\clearpage
\bibliographystyle{ACM-Reference-Format}
\bibliography{main}

\end{document}

%% file: Main-Abstract.tex
Long context training is crucial for extending LLM context windows.              
Existing schemes, such as sequence parallelism, incur substantial communication overhead.                   
Pipeline parallelism (PP) reduces this cost, but its effectiveness hinges on partitioning granularity.   
Batch-level PP employing sequence packing exhibits high memory consumption in long-context scenarios, whereas token-level PP splitting sequences into slices alleviates memory overhead but may introduce performance degradation.             
Moreover, the skewed sequence-length distribution in real-world datasets defeats any monolithic, static choice of PP granularity.           
In this paper, we propose \textit{Dynamic Pipeline Parallelism} (DPP), which transforms PP granularity from a static design choice into a workload-adaptive optimization space over packed, split, and hybrid chunks.         
DPP further introduces a new coupling between heterogeneous pipeline scheduling and gradient checkpointing.                      
To solve this coupling, \name co-optimizes dynamic chunk scheduling with \textit{Stage-Aware Chunk-Level Adaptive Checkpointing}.     
Comprehensive experiments demonstrate that \name achieves up to 1.69\texttimes\ speedup over FlexSP and up to 1.40\texttimes\ over MEPipe.
The source code is available at \url{https://github.com/wsjdsg/InfiniPipe-code}.

%% file: Main-Intro.tex
\section{Introduction \label{sec:intro}}
Large language models (LLMs)~\cite{DBLP:gpt3, gpt4, DBLP:llama2, llama-3, deepseekai2024deepseekv2strongeconomicalefficient, liu2024deepseek} increasingly support
longer contexts, drawing growing attention to efficient long-context training.

To reduce the substantial activation memory of long sequences, current techniques employ \textit{sequence splitting}: Sequence Parallelism (SP)~\cite{DBLP:deepspeed-sp-Ulysses, DBLP:ring-attn, DBLP:LightSeq, DBLP:striped_attn} distributes a sequence spatially but communicates at every layer, whereas Token-Level Pipeline Parallelism (token-level PP) schedules slices temporally with negligible communication.
On modern clusters with \textit{bandwidth heterogeneity}---intra-node far exceeding inter-node---SP is bottlenecked by inter-node communication (\S~\ref{sec:sp}), making PP the communication-efficient choice.

\begin{figure}
    \centering
\includegraphics[width=0.9\linewidth, keepaspectratio]{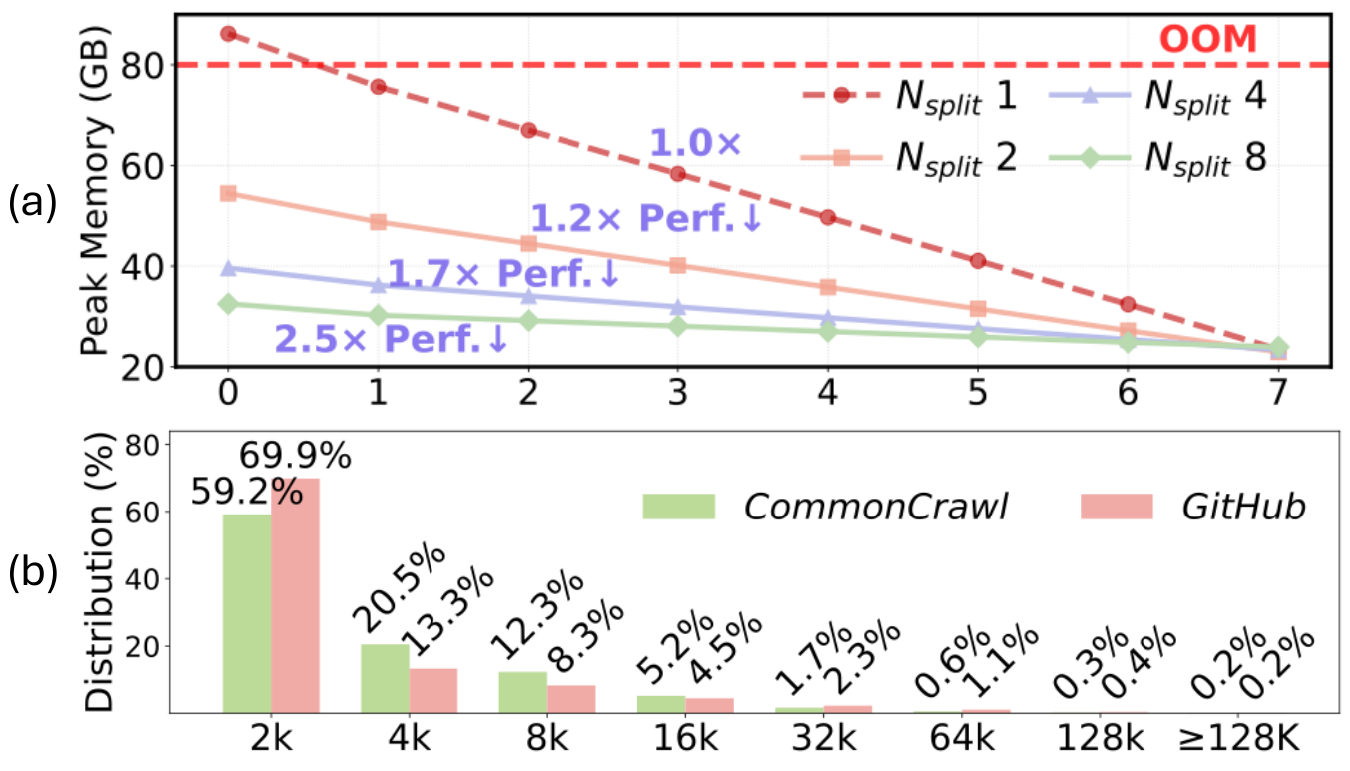}
    \caption{(a) Impact of \textit{sequence splitting} on memory footprint and computation efficiency, where $N_{split}$ refers to the number of slices to split a sequence into. Batch-Level PP ($N_{split}=1$) incurs an OOM error. 
    As Token-Level PP's $N_{split}$ increases, the finer micro-batch granularity consistently reduces the memory footprint but introduces performance degradation. (b) Length distributions of two real-world datasets, revealing a clear long-tail distribution.}
    \label{fig:moti}
\end{figure}

Unleashing PP's potential, however, hinges on its highly
\textit{workload-related} micro-batching granularity, as shown in Fig.~\ref{fig:moti}(a).
Batch-level PP (e.g., DAPPLE~\cite{DBLP:conf/ppopp/dapple}) packs short sequences into a micro-batch, but for long sequences this coarse granularity magnifies stage-wise memory
imbalance and triggers an Out-of-Memory (OOM) error. Token-level PP (e.g., MEPipe~\cite{sun2025mepipe, li2021terapipe, li2025slimpipe}) instead splits sequences, easing memory but lowering computational intensity, introducing unbalanced micro-batch workloads and performance degradation.
Navigating these trade-offs today~\cite{DBLP:conf/ppopp/dapple, sun2025mepipe} relies on manual, workload-specific tuning that does not generalize for dynamic workloads.

This tension is unavoidable because real-world workloads are \textit{skewed}: as shown in Fig.~\ref{fig:moti}(b), datasets such as \emph{GitHub} have most sequences under 16K yet less than 0.6\% exceeding 64K, and recent LLMs~\cite{yang2025qwen3, llama3, li2025minimax} deliberately train on such short/long mixtures---LLaMA3~\cite{llama3} mixes 0.1\% long-context data to optimize both regimes.
This \textit{long-tail} distribution defeats any monolithic and static PP granularity: packing alone~\cite{DBLP:conf/ppopp/dapple,pytorch_gpipe} triggers OOM for long sequences, while uniformly splitting compromises the longest one~\cite{sun2025mepipe,sun2024seq1f1b,li2025slimpipe} \textit{over-shards} the short majority and collapses performance.

The key idea of this paper is to make \textbf{PP granularity a runtime workload-adaptive optimization variable} rather than a fixed system design.
We propose \ul{\textit{Dynamic Pipeline Parallelism}} (DPP), a new PP abstraction that selects PP granularity at the sequence level and forms \textit{\textbf{heterogeneous micro-batches}}.
Instead of forcing all sequences to follow either packed micro-batches~\cite{DBLP:conf/ppopp/dapple,pytorch_gpipe} or uniformly split slices~\cite{sun2025mepipe,li2025slimpipe}, DPP allows each sequence to use a distinct packing/splitting granularity.
This exposes a new heterogeneous PP granularity space: batch-level PP and token-level PP become two boundary cases, while DPP explores intermediate granularities that adapt to the length distribution of each or even the heterogeneous workload.

This design space is not captured by prior systems.
Existing PP systems optimize within a fixed granularity: batch-level systems improve packing and scheduling of packed micro-batches~\cite{ge2024enabling,wang2025wlb,ge2025bytescale,jiang2024dynapipe}, while token-level systems split sequences into an \textit{identical} number of slices~\cite{sun2025mepipe,sun2024seq1f1b}.
They therefore miss the central optimization opportunity exposed by long-tail workloads: choosing, for each sequence or slice, whether to pack, how to split, or combine both to strike a balance between memory constraints and performance.

However, DPP brings four primary challenges to fully unleash its potential:
\begin{itemize}[leftmargin=*]
\item {\textit{An Abstraction for Heterogeneous Chunks}. A unified abstraction is required to precisely characterize packed chunks, split chunks, and hybrid chunks. Such an abstraction is the foundation for cost modeling and optimization over heterogeneous micro-batches.}
\item {\textit{Precise Cost Estimation}. The system needs a cost model that makes heterogeneous PP optimization tractable by estimating computation, communication, and stage-wise memory footprint for heterogeneous chunks, so that the optimizer can balance workload to reduce pipeline bubbles while avoiding OOM errors.}
\item {\textit{Workload-Balanced Sequence Processing}. The system must decide how much to split long sequences and how aggressively to pack short sequences, constructing chunks that satisfy memory constraints without sacrificing computational efficiency.}
\item {\textit{Dynamic Pipeline Schedule}.
As variable-length long sequences may be split into a different number of slices, the rigid schedule of prior works fails to tackle the scheduling dependencies (Fig.~\ref{fig:tpp}) and apply to DPP.
A dynamic pipeline schedule is required to formulate the generated heterogeneous chunks into a pipeline at runtime, without violating the scheduling dependencies.}
\end{itemize}

The heterogeneous chunk abstraction further introduces a fundamental coupling:
\textbf{pipeline scheduling and gradient checkpointing can no longer be optimized independently}---they form a joint optimization problem rather than two separable knobs (\S~\ref{sec:overall_workflow}, \S~\ref{sec:chunk-level-ckpt}).
On the one hand, grouping variable-length sequences into separate 1F1B pipelines requires balancing the trade-off between pipeline launching overhead and recomputation cost (Fig.~\ref{fig:ckpt_insights}(a)).
On the other hand, within a 1F1B pipeline, heterogeneous chunk lengths necessitate \textit{\textbf{heterogeneous checkpointing strategy}}, where different chunks/micro-batches adopt distinct checkpointing configurations.
Prior works~\cite{herrmann2019optimal, sun2024adapipe, jeon2025graphpipe} assume homogeneous sequence length and derive checkpointing configurations shared across micro-batches, which become sub-optimal for heterogeneous chunks with different memory footprints and recomputation costs.
Moreover, the memory imbalance across pipeline stages further requires \textit{stage-aware} checkpointing decisions.
Therefore, effectively exploiting DPP requires jointly optimizing the runtime pipeline schedule and per-chunk, per-stage checkpointing configurations.

To tackle these challenges, we build \name, a distributed training system that integrates four techniques:
\begin{itemize}[leftmargin=*]
\item {\textit{A Unified System Abstraction for DPP (\S~\ref{sec:chunk_def}).}
We introduce a three-chunk abstraction, captured by a single mathematical representation that \textbf{subsumes} Batch-Level PP~\cite{DBLP:conf/ppopp/dapple} and Token-Level PP~\cite{sun2025mepipe,li2025slimpipe,sun2024seq1f1b} as special cases while additionally expressing their hybridity.}
\item {\textit{A Cost Model for Heterogeneous Chunks (\S~\ref{subsec:cost_model}).}
Based on the abstraction, we establish a sophisticated and effective cost model that estimates: 1) the computation and communication overhead of the heterogeneous micro-batch, 2) the stage-wise memory footprint of DPP, 3) the impact of gradient checkpointing on the time and memory overhead, with an error rate of less than 5\%.}
\item {\textit{A Workload-Balanced Sequence Processor (\S~\ref{sec:sequence_chunking}).}
Leveraging the cost model, we devise a sequence chunking algorithm that generates workload-balanced heterogeneous micro-batches.}
\item {\textit{A Chunk Scheduler for DPP-Induced Coupling (\S~\ref{sec:chunk_scheduler}).}
A \textit{co-optimization} methodology is devised to tackle the challenges of pipeline scheduling and gradient checkpointing.
The \ul{\textit{Stage-Aware Chunk-Level Adaptive Checkpointing}} helps improve the computational efficiency of DPP.}
\end{itemize}

This paper makes the following contributions:
\begin{itemize}[leftmargin=*]
    \item We propose \textit{Dynamic Pipeline Parallelism} (DPP), a generalized PP abstraction that transforms PP granularity from a static design choice into a runtime workload-adaptive optimization space.
    \item We identify a new coupling between pipeline scheduling and gradient checkpointing induced by heterogeneous chunks, and propose \textit{Stage-Aware Chunk-Level Adaptive Checkpointing} that co-optimizes the two.
    \item We develop \name and evaluate it across model sizes, datasets, and context lengths. \name achieves up to 1.69\texttimes\ speedup over FlexSP and 1.40\texttimes\ over MEPipe, with gains over MEPipe increasing to 1.63\texttimes\ at 192K context.
\end{itemize}

%% file: Main-Preliminary.tex
\begin{figure}
    \centering
\includegraphics[width=1\linewidth, keepaspectratio]{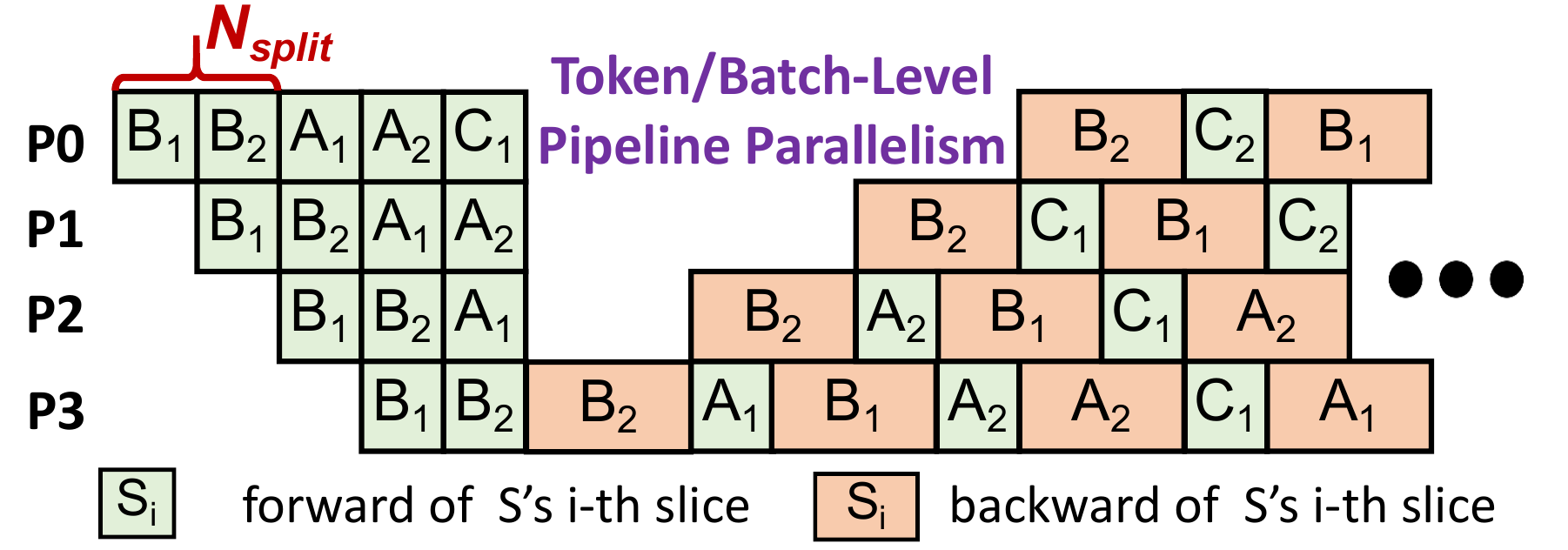}
    \caption{Illustration of DAPPLE ($N_{split}=1$) and MEPipe's ($N_{split}>1$) schedule, where sequences are divided uniformly into $N_{split}$ slices, forming \textit{homogeneous} micro-batches. There exists a scheduling dependency: \textbf{for each slice, the forward pass must be scheduled after its preceding slices, while the backward pass must be scheduled after its subsequent slices.}}
    \label{fig:tpp}
\end{figure}
\section{Preliminaries} \label{sec:background}
\subsection{Parallelism Strategies of Distributed LLM Training} \label{sec:parallel}
\begin{figure}
    \centering
\includegraphics[width=1\linewidth, keepaspectratio]{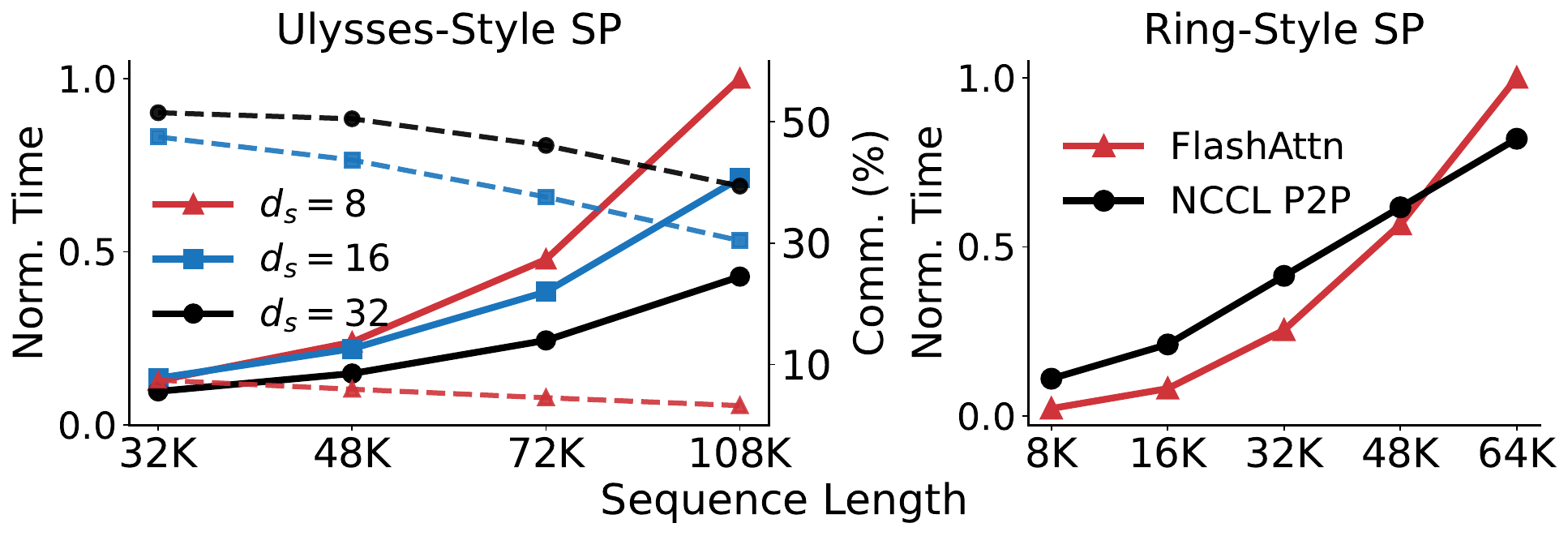}
    \caption{Communication Overhead of SP. The left plot presents the normalized total time (solid) and proportion of all-to-all communication (dashed) of USP under different context lengths and parallel degrees ($d_s$). The right plot shows the normalized kernel time of flash-attn and NCCL's P2P kernels w.r.t sequence length.}
    \label{fig:sp}
\end{figure}

\paragraph{Sequence Parallelism.} \label{sec:sp}
Sequence Parallelism (SP) partitions sequences into multiple slices, introducing communication for self-attention operations, based on which SP is typically categorized into two variants: Ulysses-style SP (USP) and Ring-style SP (RSP).
USP~\cite{DBLP:deepspeed-sp-Ulysses} introduces four all-to-all communications to transform query, key, value, and attention output between \textit{sequence-distributed} shaped $[\frac{S}{d_s}, H, D]$ and \textit{head-distributed} shaped $[S, \frac{H}{d_s}, D]$.
The inter-node all-to-all communications are costly, accounting for a large portion of end-to-end time (Fig.~\ref{fig:sp}, $d_s>8$).
In contrast, RSP~\cite{DBLP:ring-attn, DBLP:LightSeq, DBLP:striped_attn} performs a multi-step online self-attention, where the keys and values of different slices are exchanged via point-to-point (p2p) communication, which is overlapped with the attention operation.
However, the overlap is only achievable for sequences longer than 48K (Fig.~\ref{fig:sp}), which is minor in real-world datasets (Fig.~\ref{fig:moti} (b)).

\paragraph{Pipeline Parallelism.} \label{sec:tpp}
Pipeline Parallelism (PP) horizontally partitions a model into several stages that execute sequentially, requiring transmission of activations between two neighboring parts.
This transmission introduces \textbf{negligible} communication overhead as it occurs only once.
To enhance device occupancy, PP partitions training inputs into micro-batches, which categorizes PP into two variants: 1) batch-level PP~\cite{DBLP:conf/ppopp/dapple, pytorch_gpipe, li2021chimera, liu2023hanayo} that divides input samples, and 2) token-level PP~\cite{sun2025mepipe, li2021terapipe} that further splits a sequence into slices.
Various batch-level pipeline schedules~\cite{DBLP:conf/ppopp/dapple, li2021chimera, pytorch_gpipe, liu2023hanayo} have been proposed.
Fig.~\ref{fig:tpp} illustrates DAPPLE and MEPipe's schedule consisting of three distinct stages: \textit{warmup}, \textit{steady}, and \textit{cooldown}.
For the token-level PP's schedule, it is worth noting that an inter-micro-batch schedule dependency is introduced, as shown in Fig.~\ref{fig:tpp}. 
This is because the query of a token accesses only preceding tokens' keys and values in the forward pass. Consequently, gradients of the key and value of a token rely on those of subsequent tokens in the backward pass.
Token-level PP exhibits a lower memory footprint compared to batch-level PP but may introduce a performance degradation, as shown in Fig.~\ref{fig:moti}(a).

\noindent\paragraph{Fully Sharded Data Parallelism.} Fully sharded data parallelism (FSDP) partitions model states to alleviate the overhead of duplicate model states of traditional data parallelism.
Accordingly, gather and scatter communications are introduced to obtain complete parameters required for LLM execution and reduce gradients, respectively.
The gather and scatter communications in FSDP can overlap with computation.

\subsection{Sequence Packing}
Techniques such as \textit{padding} or \textit{packing} are used to deal with sequences of varying lengths. 
Padding pads or truncates sequences to the same length, employing an activation layout of $[B, max(S), D]$, introducing unnecessary computation overhead. 
In contrast, sequence packing~\cite{krell2021efficient}, which concatenates multiple input 
sequences into a single sequence and adopts a $[sum(S), D]$ layout, effectively eliminates redundant computation overhead.

\subsection{Gradient Checkpointing}
Gradient checkpointing trades computation for activation memory footprint reduction through freeing intermediate activations after the forward pass and recomputing them in the backward pass for gradient computation.

\subsection{Gradient Accumulation}
Gradient accumulation enables large effective batch sizes under limited memory by accumulating gradients over multiple micro-batches before updating model parameters, yielding the same optimization trajectory as processing the corresponding large batch directly.

%% file: Main-Method.tex
\section{\name} \label{sec:method}
\begin{figure}
    \centering
\includegraphics[width=1\linewidth, keepaspectratio]{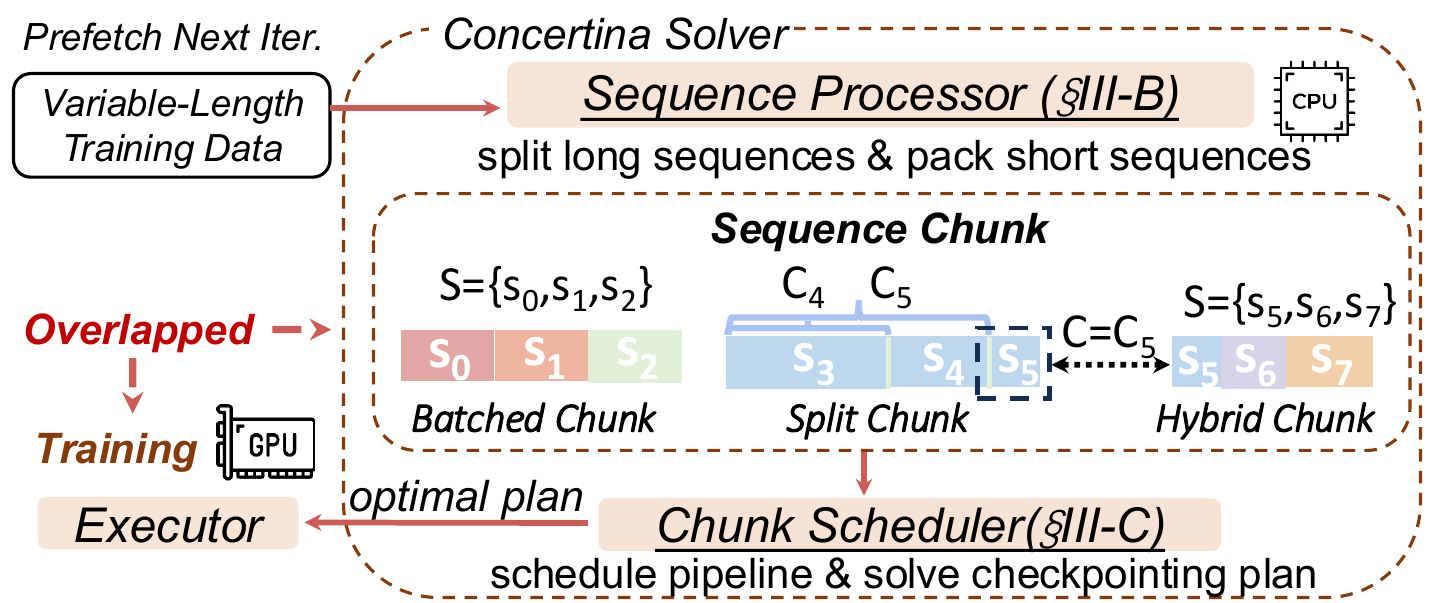}
    \caption{\name System Overview.}
    \label{fig:sys_overview}
\end{figure}

Fig.~\ref{fig:sys_overview} illustrates the overview of \name, which adopts a \ul{\textit{disaggregated}} architecture, where the solver solves the schedule plan on \textit{CPUs}, and the executor carries out training on \textit{GPUs}.
The current batch's training process \textbf{overlaps} with the next iteration's schedule solving, eliminating the solver's overhead.
Sequences sampled from the variable-length dataset are first processed by the solver's sequence processor (\S~\ref{sec:sequence_chunking}) into three types of \textit{chunks} (micro-batches of DPP) in a workload-balanced manner.
Afterward, the chunk scheduler (\S~\ref{sec:chunk_scheduler}) schedules the chunks into pipelines and solves the optimal gradient checkpointing plan for each chunk and each pipeline stage.
The tuned optimal execution plan is finally carried out by the executor to conduct LLM training.
\name employs a parallelism strategy of SP (equipped with ZeRO-3, degree $d_s$) and PP (degree $d_p$), where the cluster size is $N = d_s \times d_p$.

\subsection{Definition of \textit{Chunk}}
\label{sec:chunk_def}
We first introduce \textit{Chunk}, a key concept in \name that can be categorized into three types as illustrated in Fig.~\ref{fig:sys_overview}:
\begin{itemize}[leftmargin=*]
\item \textbf{Batched Chunk}. Short sequences are batched together, resulting in a chunk formulated as $S$, a set containing multiple slices.
\item \textbf{Split Chunk}. A long sequence is split into multiple slices, with the last one referred to as \textit{tail slice}. Due to the causal mask of self-attention, a split chunk's computation relies on keys and values of preceding slices. Therefore, we formulate a split chunk with its length $s$ and context length $C$, such as $\{0, \{s_3\}\}$ and $\{C_4, \{s_4\}\}$ in Fig.~\ref{fig:sys_overview}. 
\item \textbf{Hybrid Chunk}. Short sequences can be packed with a \textit{tail slice} $s_0$\footnote{Though it is the last slice of some sequence, it is also the first element in the slices set $S$ of this hybrid chunk. Therefore, we use the notation $s_0$ for a unified representation.} for the sake of workload balance, formulated as a context $C$ for $s_0$ and a set of slices $S$ ($s_0$ included), such as $\{C_5, \{s_5,s_6,s_7\}\}$ in Fig.~\ref{fig:sys_overview}. 
Notably, \textit{packing of two tail slices is avoided} as it forces co-scheduling of two long sequences\footnote{For instance, tail slice $A_3$ of sequence $A$ is packed with $B_2$ of sequence $B$ to $AB$. As a result, $A_1$, $A_2$ and $B_1$ must be scheduled before $AB$, introducing activation overhead of both $A$ and $B$.}, increasing memory overhead.
\end{itemize}

In summary, all three types of chunks can be expressed using a uniform representation $\{C,S\}$, where $S$ is the set containing the lengths of slices in the chunk and $C$ denotes the context length of $S$'s first slice $s_0$ (0 for a batched chunk).
Additionally, we introduce $I$ to denote whether a chunk is a split chunk, which is a frequently used notation in the later cost model.
We set $I=1$ only for split chunks, and $I=0$ for batched and hybrid chunks.
Estimations of computation and communication overhead as well as memory footprint are then introduced based on the concept of \textit{Chunk}.

\subsection{Cost Model} 
\label{subsec:cost_model}
In this section, we present the fundamental cost model employed by \name to depict the behavior of DPP.

\subsubsection{Computation and Communication Analysis}
As for computation, we assume a quadratic time complexity with respect to sequence length due to the self-attention operation.
Therefore, the computation time for processing chunk $\{C_k,S_k\}$ during both forward and backward passes is modeled as:
\begin{equation}
{
\begin{aligned}
\text{$T_{comp}$}(C_k,S_k) =  \frac{1}{N}(\alpha_1((C_k+s_0)^2-C_k^2)+\alpha_2s_0 \\
+ \sum_{s\in (S_k-\{s_0\})}(\alpha_1s^2 + \alpha_2s))+\frac{\beta_1}{d_p} 
\label{eq:time_chunk}
\end{aligned}
}
\end{equation}
\noindent where $\alpha_1$, $\alpha_2$, and $\beta_1$ are hyperparameters that characterize the model's computational behavior, which are obtained through offline profiling and regression fitting.

As for communication, $V$ (communication volume), $B_{\text{comm}}$ (bandwidth), $\beta_{\text{comm}}$ (latency) and $f$ (frequency) are utilized to model the overhead:
\begin{equation}
{
\begin{aligned}
\text{$T_{comm}$}(V,f) = (\frac{V}{B_{comm}}+\beta_{comm})\cdot f,  \label{eq:ori_comm}
\end{aligned}
}
\end{equation}
Specifically, Ulysses-style SP requires four All-to-All communications in each layer, resulting in the following overhead:
\begin{equation}
{
\begin{aligned}
T_{all2all}(S_k) = (\frac{e(D+D_{kv})\sum_{s\in S_k}s}{d_sB_{all2all}(d_s)}+2\cdot \beta_{all2all}(d_s))\frac{2L}{d_p}\label{eq:comm_a2a}
\end{aligned}
}
\end{equation}
where $D$, $D_{kv}$, and $L$ denote the hidden dimension, KV dimension, and number of layers, respectively, and $e$ is the size in bytes of an activation element (e.g., $e=2$ for bfloat16).
The total execution time for a chunk comprises both computation and communication components:
\begin{equation}
{
\begin{aligned}
\text{$T_{tot}$}(C_k,S_k) = T_{comp}(C_k,S_k) + T_{all2all}(S_k),  \label{eq:total_time}
\end{aligned}
}
\end{equation}
\noindent which is applicable in both forward and backward passes.

\subsubsection{Stage-Aware Memory Footprint Analysis}
We begin by analyzing the activation memory footprint of a chunk.
The activation memory footprint is \textit{proportional} to the number of tokens after employing flash-attn\cite{DBLP:flash_attn, DBLP:flash_attn_2}, which eliminates the need to materialize attention scores with $O(S^2)$ space complexity.
Notably, $M_{dkv}$, representing the overhead of keys and values' gradients for \textit{split chunk}, is included in our estimation based on two observations of TPP:
1) chunks of a sequence execute backward \textit{reversely}, with the last chunk executing backward first (see Fig.~\ref{fig:tpp}); 2) gradients for keys and values of all chunks are materialized \textit{simultaneously} during the backward pass of the last chunk, while being freed \textit{asynchronously until} the completion of its own backward pass.
Moreover, as current LLMs typically adopt a tokenizer with a large vocab size ($\ge$ 128K), the memory overhead for logits is non-negligible.
Consequently, the overall activation memory footprint is modeled as:
\begin{equation}
{
\begin{aligned}
\text{$M_{act}$}(p, S_k) = Act(p, S_k) + M_{dkv}(S_k),\quad\quad \\ Act(p, S_k)=(\frac{M_{token}}{N} + \frac{M_{logits}}{d_s}[p=d_p])\sum_{s\in S_k}s,\\
M_{dkv}(S_k) = I_k\cdot\frac{2eLD_{kv}}{N}\sum_{s\in S_k}s,\quad\quad\quad
\label{eq:activation_chunk}
\end{aligned}
}
\end{equation}
\noindent where $M_{\text{token}}$ and $M_{\text{logits}}$ are model-specific constants representing the activation and logits memory overhead per token, respectively.

The total memory footprint for the $p^{th}$ pipeline stage comprises the memory allocated for fixed model states $M_{ms}(p)$ (parameter, gradient and optimizer states) and the ever-changing activation memory $M_{act}(p,t)$:
\begin{equation}
{
\begin{aligned}
\text{$M_{tot}$}(p,t) = M_{ms}(p) + M_{act}(p,t) \label{eq:tot_act}
\end{aligned}
}
\end{equation}
The peak memory of the 1F1B pipeline occurs during the steady phase, where each pipeline stage $p$ maintains a \textit{constant} number of chunks that have not finished their backward passes, as shown in Fig.~\ref{fig:tpp}. Let $W_p(t)$, called the \textit{chunks window}, denote the set of these chunks at time $t$, satisfying: 
\begin{equation}
{
\label{eq:microbatch_window}
\begin{aligned}
|W_p(t)| = d_p-p+N_{split}
\end{aligned}
}
\end{equation}
which is equivalent to the number of micro-batches in the warmup phase. As activations of all chunks within $W_p(t)$ must be accommodated, the total memory footprint is modeled as:
\begin{equation}
{
\begin{aligned}
\text{$M_{tot}$}(p,t) = M_{ms}(p) + M_{act}(p, W_p(t)) \\ =M_{ms}(p)+\sum_{k\in W_p(t)}M_{act}(p, S_k)  \label{eq:total_mem}
\end{aligned}
}
\end{equation}

\subsubsection{Combining Gradient Checkpointing Together}\label{sec:ckpt_model}
Gradient checkpointing affects estimation for both memory footprint and time cost.
As in common practice in Megatron-LM~\cite{narayanan2021efficient}, \name applies checkpointing at layer granularity and lets $l_{ckpt}$ denote the checkpointed layers in each pipeline stage.

We first analyze how checkpointing affects memory footprint estimation.
The \textit{split chunk} exhibits a different behavior than other chunks: although gradient checkpointing is applied, its keys and values cannot be released, as they are needed by the subsequent slices to perform the self-attention operation. 
However, the other activations of a checkpointed layer can be ignored.
The phenomenon drives us to deal with keys and values individually in cost estimation.
Specifically, we include not only the checkpointed layer's input but also keys and values in the checkpointing memory overhead $M_{ckpt}$:
\begin{equation}
{
\begin{aligned}
M_{ckpt}(S_k)=\frac{e(D + 2\cdot I_kD_{kv})\cdot l_{ckpt}}{d_s}\sum_{s\in S_k}s
\label{eq:mem_ckpt_act}
\end{aligned}
}
\end{equation}
Moreover, checkpointing has no impact on $M_{dkv}$, and a chunk's activation footprint is further reformulated as:
\begin{equation}
{
\begin{aligned}
M_{act}(p, S_k) = M_{dkv}(S_k) + M_{ckpt}(S_k) +\quad\quad\\(\frac{L-l_{ckpt}\cdot d_p}{L}\cdot \frac{M_{token}}{N} +  \frac{M_{logits}}{d_s}[p=d_p])\sum_{s\in S_k}s,
\label{eq:mem_ckpt_tot}
\end{aligned}
}
\end{equation}
As for time cost estimation, checkpointing only affects the backward pass with a recomputation cost:
\begin{equation}
{
\begin{aligned}
    T_{ckpt}(C_k,S_k) = \frac{l_{ckpt}}{L\cdot d_s}\cdot T_{tot}(C_k,S_k)._{fwd}
\label{eq:chunk_recomp_cost}
\end{aligned}
}
\end{equation}

\subsection{Sequence Processor}
\label{sec:sequence_chunking}
The sequence processor ingests 
original, variable-length sequences sampled from the dataset and then organizes them into \textit{chunks}. 
There are two crucial issues to be addressed when designing the processing algorithm: 1) 
ensuring workload balance across chunks to avoid pipeline bubbles and 2) determining the proper granularity of a chunk that maximizes utilization while adhering to memory limits.
This section introduces a sketch of our resource-aware and workload-balanced chunking algorithm outlined in Alg.~\ref{alg:sequence_chunking}.
%



The algorithm consists of two steps: splitting long sequences to generate \textit{split chunks} and packing short sequences to form \textit{batched chunks} and \textit{hybrid chunks}. 
It takes the following inputs: the cost model $\mathcal{M}$; token capacity $\mathcal{C}$, i.e., maximum number of tokens the cluster can accommodate; sequence lengths $S$, i.e., the lengths of sequences sampled from dataset; and slice number $K$ that denotes the number of slices to split the longest sequence into, which is a hyper-parameter automatically tuned within the range [1, $d_p$ + 4] to obtain the optimal solution.

To begin with, a mesh is obtained by splitting the ingested longest sequence into $K$ workload-balanced slices (Line 1), which is based on $\mathcal{M}$'s time cost estimation $T_{tot}$ (Eq.~\ref{eq:total_time}) for the backward pass.
Then, long sequences are sharded by the mesh into multiple workload-balanced slices and a lower-cost \textit{tail slice} (Line 2). For instance, with a mesh \{8K,4K,2K\}, sequences with more than 12K tokens will be split into an 8K and 4K slice, followed by a variable-length remainder.

Afterward, a Best-Fit-Decreasing (BFD) algorithm is employed to pack the tail slices and short sequences that are not sharded by the mesh, resulting in \textit{batched chunk} and \textit{hybrid chunk} (Lines 3-15).
To balance the time cost of each chunk, a time threshold $\mathcal{T}_t$ indicating the maximum time cost of a chunk is set, initially equivalent to the workload of a slice of the mesh.
Notably, a token threshold $\mathcal{T}_m$ indicating the maximum number of tokens in a bucket is also set to balance memory footprint (determined according to $\mathcal{C}$ and $\mathcal{T}_t$; see our repository for a concrete expression).
To co-optimize the balance of time cost and chunk length, we prioritize putting the longest short sequence into the bucket with the minimum $\frac{tot\_time}{tot\_tokens}$ (Line 3, 9-12), where the metric indicates the proportion of long sequences in the bucket\footnote{Longer sequences exhibit a higher value of this metric because of the quadratic computation complexity w.r.t sequence length, we use this metric to avoid high chunk length variance due to aggregation of long sequences.}.
$\mathcal{T}_t$ is relaxed when $\mathcal{T}_m$ cannot be met in the BFD process (Line 14).

The sequence processor demonstrates a time complexity of $O(n^2|S|^2)$, introducing negligible overhead.


\begin{algorithm}[tb]
\small
\caption{Workload-Balanced Sequence Processing for Heterogeneous Chunks}
\label{alg:sequence_chunking}
\LinesNumbered
\KwIn{
    Cost model $\mathcal{M}$, token capacity $\mathcal{C}$, sequence lengths $S$, slice number $K$.
}
\KwOut{
    Chunks $\{(C_k, S_k) \mid k \le n\}$, where $n$ is the total number of generated chunks.
}
\(\text{mesh},\, \mathcal{T}_t,\, \mathcal{T}_m \leftarrow \mathcal{M}.\text{split}(\max(S), K)\)\;
\(split\_chunks,\, \text{tail\_slices},\, \text{short\_seqs} \leftarrow \text{split}(S, \text{mesh})\)\;
\tcp{\footnotesize a sorted_list with metric \(\frac{tot\_time}{tot\_tokens}\)}
\(\mathcal{B} \leftarrow \text{initialize\_buckets}(\text{tail\_slices})\)\;
\(\mathcal{M}.\text{descend\_sort\_by\_time}(\text{short\_seqs})\)\;

\While{\(\lnot\, \text{short\_seqs.is\_empty}()\)}{
    \(\text{short\_seq},\, \text{flag} \leftarrow \text{short\_seqs.pop}(0),\, \text{False}\)\;

    \If{\(\min_{b \in \mathcal{B}} b.\text{tot\_tokens} + \text{short\_seq.tokens} > \mathcal{T}_m\)}{
        \(\mathcal{B}.\text{create\_new\_bucket}(\text{short\_seq})\);\textbf{continue}\;
    }

    \For{\(\text{bucket} \in \mathcal{B}\)}{
        \If{\(\text{bucket.tot\_time} + \text{short\_seq.time} \le \mathcal{T}_t
        \And \text{bucket.tot\_tokens} + \text{short\_seq.tokens} \le \mathcal{T}_m\)}{
            \(\text{bucket.combine}(\text{short\_seq})\);\(\text{flag} \leftarrow \text{True}\)\;\textbf{break}\;
        }
    }

    \If{\(\lnot\, \text{flag}\)}{
        \(\mathcal{T}_t \leftarrow 
            \min_{b \in \mathcal{B}} b.\text{tot\_time} + \text{short\_seq.time}\);\textbf{goto} 9\;
    }
}

\(batched\_chunks,\, hybrid\_chunks \leftarrow \text{transform}(\mathcal{B})\)\;
\Return{\(split\_chunks \cup batched\_chunks \cup hybrid\_chunks\)}

\end{algorithm}

\subsection{Chunk Scheduler}
\label{sec:chunk_scheduler}
In this section, we first define our scheduling space through careful trade-off discussions and then present our two-level approach that jointly optimizes pipeline schedule (via \textit{sequence grouping}, \S~\ref{sec:overall_workflow}) and checkpointing configuration (via \textit{stage-aware chunk-level adaptive checkpointing}, \S~\ref{sec:chunk-level-ckpt}).

\subsubsection{Solving Space Definition}
\label{sec:solving_space_def}
It is non-trivial to solve a sophisticated schedule plan due to the vast optimization space: 1) There exist numerous basic scheduling patterns~\cite{DBLP:conf/ppopp/dapple, pytorch_gpipe, li2021chimera, qi2023zero}, each exhibiting distinct memory footprint and bubble characteristics. 2) The execution order of variable-length sequences needs to be determined. 3) A checkpointing mechanism tailored for DPP is required to efficiently integrate checkpointing.
We begin by introducing some key insights, based on which the solving space is then confined. 

To begin with, \textit{1F1B is taken as the basic schedule pattern.}
GPipe~\cite{pytorch_gpipe} features a \textit{forward-then-backward} schedule, which is not memory efficient because it needs to accommodate all micro-batches.
Chimera~\cite{li2021chimera} proposes bidirectional PP, but it duplicates model states and needs synchronization of gradients.
Other 1F1B schedule patterns~\cite{liu2023hanayo, qi2023zero, sun2025mepipe, li2021terapipe} are also proposed.
However, most improvements in these works stem from reducing the bubble of warmup and cooldown phases.
As the number of micro-batches increases, the benefits diminish because the steady phase with negligible bubbles dominates the total execution.
Most importantly, the number of micro-batches can be adjusted by controlling the splitting granularity of the sequence processor (hyperparameter $K$ of Alg.~\ref{alg:sequence_chunking}), which is optimized in the solving process.
To this end, the 1F1B schedule~\cite{sun2025mepipe} is taken as our basic schedule pattern, as we focus on training throughput.

\begin{figure}
    \centering
\includegraphics[width=1\linewidth, keepaspectratio]{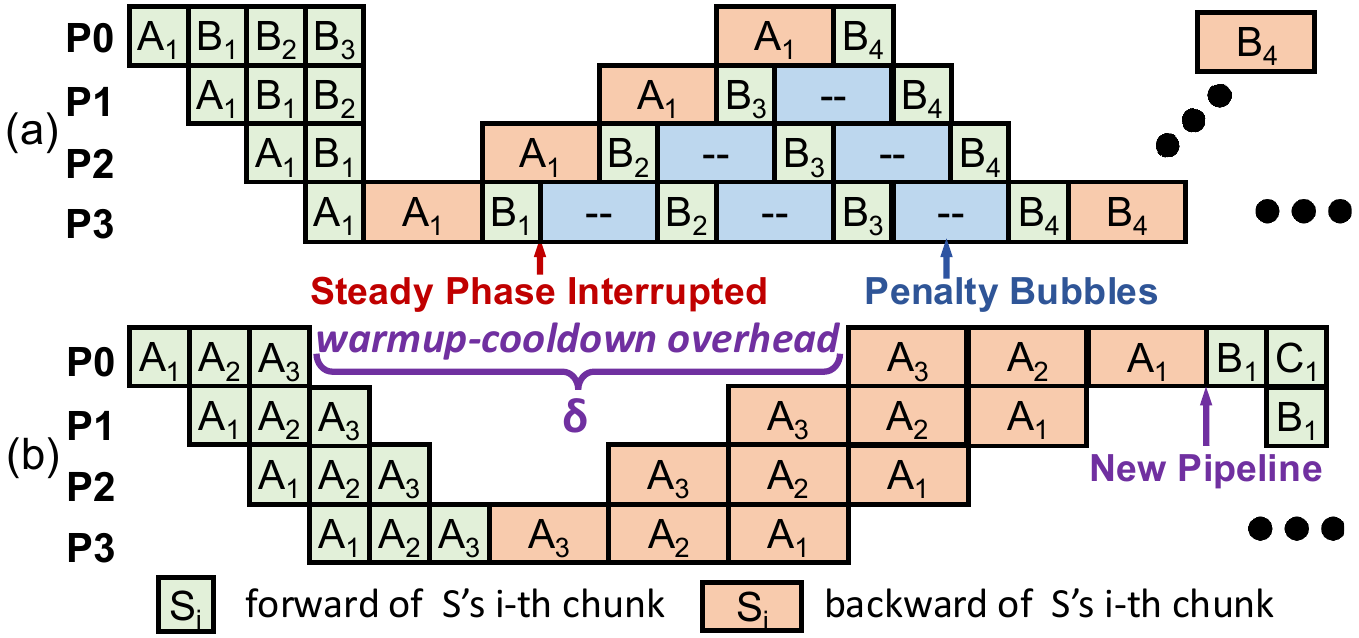}
    \caption{(a) Scheduling short sequence $A_1$ first interrupts the steady phase, as $B_4$'s backward pass is not available (see scheduling dependencies in Fig.~\ref{fig:tpp}). (b) If multiple 1F1B pipelines are scheduled, each will introduce an identical warmup-cooldown overhead $\delta$.}
    \label{fig:solving_space}
\end{figure}

Moreover, \textit{it is feasible to schedule one or multiple 1F1B pipelines, within which longer sequences are prioritized to execute.}
To begin with, the longest sequences should be scheduled first within a 1F1B pipeline.
As illustrated in Fig.~\ref{fig:solving_space}(a), sequence \textit{A} with one chunk is scheduled before sequence \textit{B} with three chunks.
We observe that the steady phase is interrupted due to the unavailability of a backward schedule, introducing severe penalty bubbles.
To address this, we prioritize longer sequences in a 1F1B pipeline and enforce this as a fundamental scheduling rule.
Moreover, scheduling of multiple 1F1B pipelines is necessary when sequences cannot be scheduled in a single 1F1B pipeline altogether.
As shown in Fig.~\ref{fig:solving_space}(b), sequences \textit{B} and \textit{C} cannot be co-scheduled with sequence \textit{A} due to the limited memory capacity, forcing a new pipeline to be scheduled.
Gradient accumulation is enabled among these 1F1B pipelines to ensure optimization consistency.
We do not exploit pipeline bubbles to overlap these 1F1B pipelines, as typically no more than two 1F1B pipelines are scheduled by our solver, offering marginal benefits but complicating the solving process.

\label{sec:moti_ckpt}
Furthermore, \textit{employing stage-aware chunk-level adaptive checkpointing}. 
Firstly, though sequence splitting effectively reduces the pipeline's memory footprint, gradient checkpointing remains \textit{indispensable}.
As indicated by Fig.~\ref{fig:moti}(a), for extremely long sequences, the sequence sharding granularity required to alleviate the memory overhead could cause performance degradation if checkpointing is naively disabled.
However, directly applying full checkpointing introduces unnecessary recomputation overhead. 
Furthermore, the memory footprint disparity across chunks and pipeline stages renders suboptimal performance of a uniform and static checkpointing strategy.
Specifically, different pipeline stages have varying requirements for checkpointing, and \textbf{applying checkpointing on longer chunks reduces more activation footprint at the same recomputation cost if chunks are workload-balanced}.
The trade-offs above inspire us to employ a stage-aware adaptive checkpointing strategy for each heterogeneous chunk.

In summary, we define our solving space as follows: for pipeline schedule, we group sequences and assign sequences of each group to a distinct 1F1B pipeline $P$, formulating a set $\mathcal{P}$; for checkpointing, we apply a customized checkpointing layer $ckpt(p,k)$ for each chunk $\{C_k, S_k\}$ at each pipeline stage $p$.
Let $\mathcal{S}[i]$ represent the set of sequences divided into $i$ chunks, the problem can be formulated as:
\begin{equation}
{
\begin{aligned}
    minimize \sum_{P\in\mathcal{P}} T(P)\label{eq:goal_raw}\\
    \text{s.t.} \bigcup_{P\in \mathcal{P}} \mathcal{S}_P = \mathcal{S}[:],
\end{aligned}
}
\end{equation}
\noindent where $\mathcal{S}_P$ represents the sequences scheduled in $P$ and $T(P)$ denotes the total execution time of $P$.
Thanks to the workload-balanced manner in which \name processes sequences, we ignore the pipeline bubbles in the steady phase and simplify Eq.~\ref{eq:goal_raw} as a summation of recomputation cost $T_{ckpt}(P)$ and warmup-cooldown overhead $\delta$:
\begin{equation}
\begin{aligned}
    minimize  \sum_{P\in\mathcal{P}} \delta + T_{ckpt}(P), \label{eq:opt_obj}
\end{aligned}
\end{equation}
\noindent where $\delta$ is a constant approximated as $(d_p-1)\cdot avg(T_{tot})$.
The optimal pipeline schedule $\mathcal{P}_0$ along with checkpointing configuration $ckpt_0(p,k)$, is co-optimized, with detailed methodology provided in \S~\ref{sec:overall_workflow} and \S~\ref{sec:chunk-level-ckpt}, respectively.

\subsubsection{Sequence Grouping}
\label{sec:overall_workflow}
\begin{figure}
    \centering
    \includegraphics[width=1\linewidth, keepaspectratio]{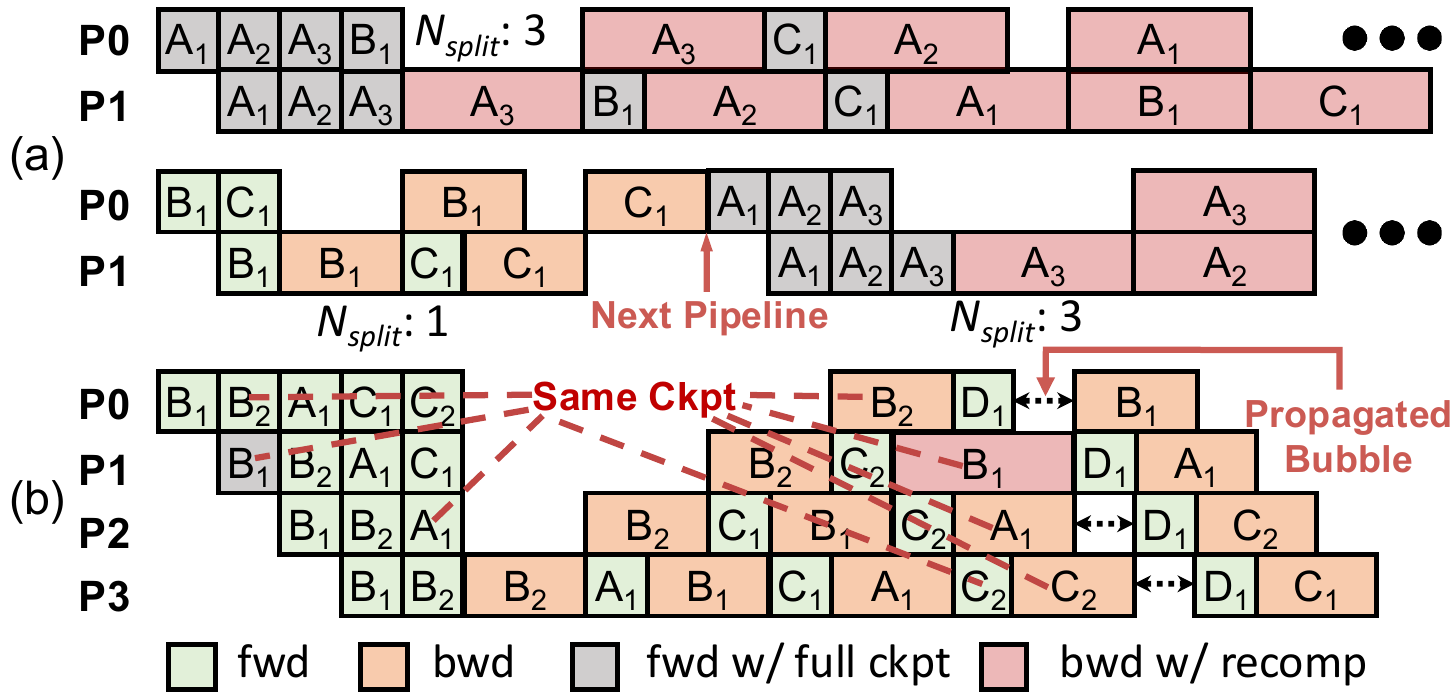}
    \caption{(a) Larger $N_{split}$ indicates scheduling more micro-batches in the warmup phase, requiring a tighter checkpointing configuration of $B_1$ and $C_1$. The bottom schedules another pipeline for $B_1$ and $C_1$, eliminating their recomputation overhead but increasing a warmup-cooldown overhead $\delta$. (b) The marked chunks $B_2$, $B_1$, $A_1$, and $C_2$ should share the same checkpointing configuration, as filling the propagated bubble does not change overall execution time.}
    \label{fig:ckpt_insights}
\end{figure}

An interference between the pipeline schedule and gradient checkpointing is observed in Fig.~\ref{fig:ckpt_insights}(a).
As indicated by Eqs.~\ref{eq:microbatch_window} and \ref{eq:total_mem}, the pipeline's memory footprint is related to $N_{split}$, i.e., the number of chunks the longest sequence in a sequence group is split into. 
Accordingly, when short sequences \textit{B} and \textit{C} are grouped with long sequence \textit{A}, they are forced to apply a tighter checkpointing setup than they are scheduled separately due to the enlarged $N_{split}$, introducing more recomputation overhead.
Therefore, 1) \textit{sequences of similar lengths should be grouped together} 2) \textit{scheduling more 1F1B pipelines can reduce recomputation cost $\sum_{P\in \mathcal{P}} T_{ckpt}(P)$ of Eq.~\ref{eq:opt_obj} at the cost of increased warmup-cooldown overhead $\delta\cdot |\mathcal{P}|$}, forming the trade-off when optimizing sequence grouping strategy.

We employ a dynamic programming method to resolve the optimal sequence grouping strategy.
Let $dp[i]$ represent the minimum cost to schedule sequences with at most $i$ chunks.
The state transition equation can be deduced as:
\begin{equation}
{
\begin{aligned}
dp[i+1]=\min_{0\leq k\leq i} \{dp[k]+ \delta + T_{ckpt}(P)\},
\label{dp:scheduler}
\end{aligned}
}
\end{equation}
\noindent where sequences in $\mathcal{S}[k+1:i+1]$ are scheduled by pipeline $P$ and $T_{ckpt}(P)$ is obtained by applying \textit{stage-aware chunk-level adaptive checkpointing} (\S~\ref{sec:chunk-level-ckpt}) on $P$. 
By tracking state transitions to $dp[N]$, we derive both the sequence grouping and its corresponding checkpointing configuration.
\subsubsection{Stage-Aware Chunk-Level Adaptive Checkpointing}
\label{sec:chunk-level-ckpt}
In this section, we elaborate on how we apply an optimal checkpointing configuration for a given 1F1B pipeline $P$ assigned $n$ chunks. As the Sequence Processor (\S~\ref{sec:sequence_chunking}) organizes sequences into chunks in a workload-balanced manner, we simplify the problem by assuming each chunk exhibits the same time cost but different lengths.

To begin with, we analyze the impact of checkpointing on PP and introduce a constraint on $ckpt(p,k)$.
As illustrated in Fig.~\ref{fig:ckpt_insights}(b), full checkpointing is applied to $B_1$ of the second stage.
We observe that checkpointing affects not only the second stage, introducing propagated bubbles of identical size in all the other stages.
A key insight is that \textbf{applying full checkpointing to the marked chunks $B_2$, $A_1$, and $C_2$ exploits the propagated bubble and maintains the total execution time unchanged}.
Let $ckpt'(p,k)$ represent the number of checkpointed layers for the chunk executing the $k^{th}$ backward pass in the $p^{th}$ pipeline stage (the execution order differs between forward and backward passes).
Based on this insight, we yield the following constraint:
\begin{equation}
{
\begin{aligned}
ckpt'(p,k)=ckpt'(p+i,k+i)=\mathcal{C}[k+d_p-p],
\label{eq:ckpt_prop}
\end{aligned}
}
\end{equation}
where $\mathcal{C}$ is a set containing $d_p-1+n$ independent integer variables.
Let $f2b[k]$ map the forward execution order to the backward execution order; we have:
\begin{equation}
{
\begin{aligned}
ckpt(p,k) = ckpt'(p,f2b[k]) = \mathcal{C}[f2b[k]+d_p-p]
\label{eq:ckpt_depend}
\end{aligned}
}
\end{equation}
This formulation reduces the number of optimization variables from $n\cdot d_p$ in $ckpt(p,k)$ to $d_p-1+n$ in $\mathcal{C}$, significantly reducing solving overhead.


Afterward, a solution based on ILP is introduced, as outlined in Alg.~\ref{alg:solve_ckpt_milp}. 
Fig.~\ref{fig:ckpt_insights}(b) reveals that recomputation cost $T_{ckpt}(P)$ is related to $\mathcal{C}$ with:
\begin{equation}
{
\begin{aligned}
    T_{ckpt}(P)=\hat{F} \cdot \sum_{c\in \mathcal{C}}c,
\label{eq:cost_recomp}
\end{aligned}
}
\end{equation}
\noindent where $\hat{F}$ denotes the estimated forward execution time of a model layer.
The optimal checkpointing strategy aims to \textit{minimize the recomputation cost $T_{ckpt}(P)$ (Eq.~\ref{eq:cost_recomp}) with peak memory $M_{tot}(p,t)$ (Eq.~\ref{eq:total_mem}) not exceeding hardware capacity limit $\mathcal{G}$}, formulated as:
\begin{equation}
{
\begin{aligned}
\label{eq:mem_cond}
    \argmin_{\mathcal{C} \in \mathbb{N}^{n+d_p-1}} \hat{F}\cdot\sum_{c\in \mathcal{C}}c\quad\quad\quad\quad\quad\quad\quad\quad\\
    \text{s.t.} \quad\quad M_{ms}(p)+\sum_{k\in W_p(t)}M_{act}(p, S_k) \leq \mathcal{G}\quad\forall (p,t),p\leq d_p\\
    c\leq \frac{L}{d_p},\forall c\in\mathcal{C} \quad\quad\quad\quad\quad\quad\quad\quad\quad
\end{aligned}
}    
\end{equation}
\noindent Combining Eqs.~\ref{eq:mem_ckpt_act},\ref{eq:mem_ckpt_tot},\ref{eq:ckpt_depend}, we derive a linearity:
\begin{equation}
{
\begin{aligned}
M_{act}(p, S_k) = \mathcal{I}_p[k] -\mathcal{F}[k]\cdot \mathcal{C}_{p,k},
\label{eq:mem_ckpt_tot_combine}
\end{aligned}
}
\end{equation}
where the coefficients ($\mathcal{I}_p[k]$, $\mathcal{F}[k]$) and $\mathcal{C}_{p,k}$ are defined explicitly in Alg.~\ref{alg:solve_ckpt_milp}.
To this end, constraint~\ref{eq:mem_cond} is further reformulated as a system of linear inequalities in terms of $\mathcal{C}$, and the ILP is finally expressed as:
\begin{equation}
{
\begin{aligned}
    \argmin_{\mathcal{C} \in \mathbb{N}^{n+d_p-1}} \sum_{c\in \mathcal{C}}c\quad\quad\quad\quad\quad\quad\quad\quad\quad\quad\\
    \text{s.t.} \sum_{k\in W_p(t)}\mathcal{I}_p[k]- \mathcal{F}[k]\cdot \mathcal{C}_{p,k}\leq \mathcal{G}-M_{ms}(p) \quad \forall (p,t),p\leq d_p\\
    c\leq \frac{L}{d_p},\forall c\in\mathcal{C} \quad\quad\quad\quad\quad\quad\quad\quad\quad\quad
\end{aligned}
}
\end{equation}
After optimizing $\mathcal{C}$ and $T_{ckpt}$, the optimal checkpointing configuration $ckpt_0(p,k)$ can be obtained by Eq.~\ref{eq:ckpt_depend}.

\begin{algorithm}[t]
\caption{Stage-Aware Chunk-Level Adaptive Checkpointing Solving Based on ILP}
\label{alg:solve_ckpt_milp}
\LinesNumbered
\KwIn{Chunks $\{S_k|k\leq n\}$, GPU memory capacity $\mathcal{G}$, chunks windows $W_p(t)$
}
\KwOut{Checkpointing configuration $\mathcal{C}$ and minimum recomputation cost $T_{ckpt}$}
\For{$k \leq n$}{
    $\mathcal{I}[k] \leftarrow (\frac{M_{token}}{N}+I_k\cdot\frac{2eLD_{kv}}{N})\sum_{s\in S_k}s$\;
    $\mathcal{F}[k] \leftarrow \left(\frac{M_{token}}{L\cdot d_s}-\frac{e(D + 2\cdot I_k D_{kv})}{d_s}\right)\sum_{s \in S_k} s$\;
}
$\mathcal{C} \leftarrow initialize\_ilp\_vars(n + d_p - 1,max=\frac{L}{d_p})$\;
\For{$p\leq d_p$}{
    \For{$W_p(t)\in W_p$}{
    $\mathcal{C}_{p,k}\leftarrow \mathcal{C}[d_p-p+f2b[k]]$\;
    $\mathcal{I}_p[k]\leftarrow \mathcal{I}[k] + [p=d_p]\frac{M_{logits}}{d_s}\sum_{s \in S_k} s$\;
    $cons.add(\sum_{k\in W_p(t)}(\mathcal{I}_p[k]-\mathcal{F}[k]\cdot \mathcal{C}_{p,k})\leq \mathcal{G})$\;
    }
}
$obj \leftarrow minimize \sum_{c \in \mathcal{C}} c$\;
$\mathcal{C}, T_{ckpt} \leftarrow solve\_ilp(cons, obj)$\;
\Return{$\mathcal{C}, T_{ckpt}$}
\end{algorithm}

%% file: Main-Implementation.tex
\section{Implementation} 
\label{sec:impl}
We implement \name in approximately 13K lines of code using Python, CUDA, and Triton~\cite{tillet2019triton}. The SCIP~\cite{BolusaniEtal2024OO} library is leveraged to solve the ILP problems.
Built on PyTorch, \name integrates the flash-attn~\cite{DBLP:flash_attn, DBLP:flash_attn_2} library for variable-length sequence packing and adopts NCCL~\cite{nccl} as the communication backend. Additionally, several key points in our implementation are highlighted as follows.

\noindent\paragraph{Tailored FSDP for Dynamic Pipeline Parallelism.}
FSDP operates orthogonally to SP and is commonly combined with SP to reduce model state memory overhead. Although PyTorch FSDP~\cite{DBLP:pytorch-fsdp} serves as the most widely-used implementation, its native version is not compatible with pipeline parallelism with gradient accumulation.
\name's runtime engine seamlessly integrates PyTorch FSDP with DPP, allowing ZeRO communication to be overlapped with computation when a dynamic pipeline schedule of DPP is adopted.

\noindent\paragraph{Fused In-Place Cross-Entropy.}
The peak memory usage during LLM training typically occurs at the beginning of the backward pass, i.e., when the cross-entropy loss starts its backward computation, where the logits' gradients as well as some intermediate tensors are materialized, introducing non-negligible estimation bias on peak memory.
We adopt Megatron-LM~\cite{megatron-lm}'s fused in-place cross-entropy operator to eliminate the materialization of other tensors, aligning the peak memory usage with our cost model.

%% file: Main-Experiment.tex
\begin{figure*}[t]
    \centering
    \includegraphics[width=0.95\linewidth, keepaspectratio]{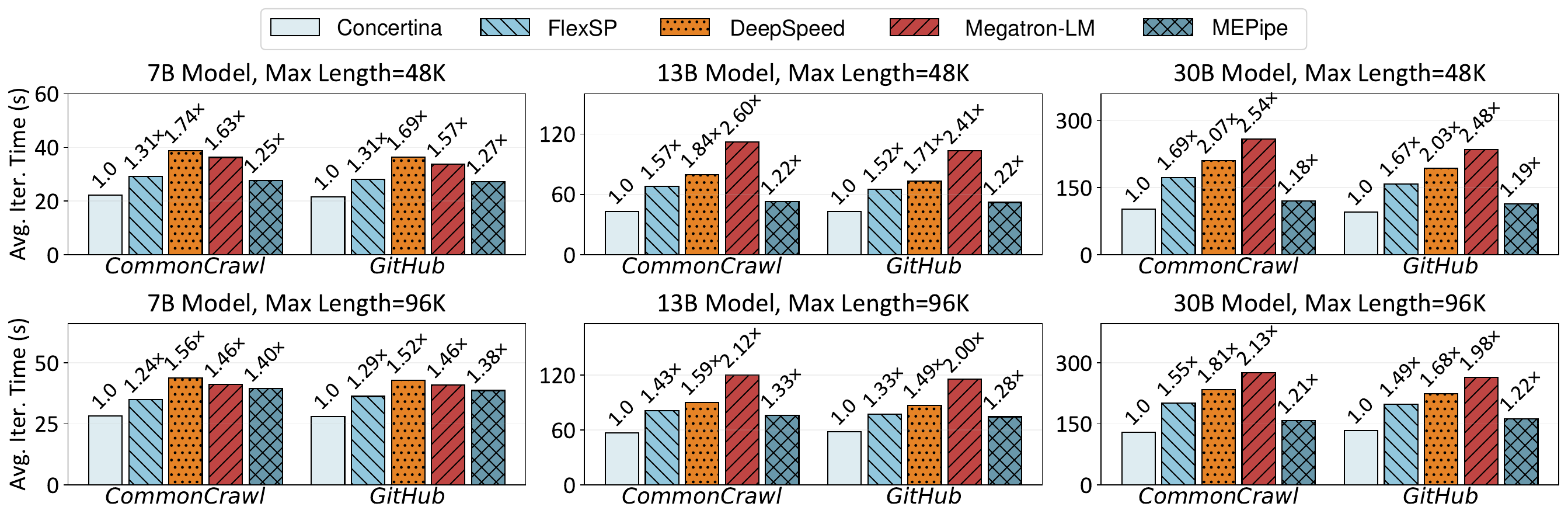}
    \caption{End-to-end Performance. Average end-to-end time of a training iteration under different settings of model sizes, context lengths, and datasets, with the speedup ratio of \name compared to baselines presented.}
    \label{fig:main_res}
\end{figure*}
\section{Experiments} \label{sec:experiments}
\subsection{Experiment Setup}
\paragraph{Environments.} 
Our testbed consists of four GPU servers, each equipped with 8 NVIDIA A800-80GB GPUs interconnected via NVLink (400 GB/s bandwidth). Inter-node communication is handled by a 400 Gb/s InfiniBand network. 
The software stack includes PyTorch 2.9.0 and CUDA 12.8.

\noindent\paragraph{Baseline Systems.}
We compare \name against four representative distributed training systems: 
\begin{itemize}[leftmargin=*]
\item {\textbf{Megatron-LM}. Megatron-LM is the current general-purpose SOTA featuring 4D parallelism comprising TP (equipped with Megatron-style SP), DP (ZeRO-1), CP, and PP.} 
\item {\textbf{DeepSpeed}. DeepSpeed integrates ZeRO of three stages and Ulysses-style SP, exhibiting superior performance on uniform-length long-context training workloads.}
\item {\textbf{FlexSP}. FlexSP extends DeepSpeed, representing the previous SOTA training system on variable-length corpora.}
\item {\textbf{MEPipe}. MEPipe is a representative PP baseline. Its original design ignores the skewed sequence-length distribution and divides sequences into a \textit{uniform} number of slices, leading to inefficient oversharding of short sequences. 
For a fair and stronger comparison, we evaluate an
enhanced MEPipe baseline that splits and packs sequences into fixed-size chunks, similar to ``w/o wbc'' in \S~\ref{sec:abla_study} but with a uniform and static checkpointing configuration for all micro-batches optimally tuned for each case.
This enhanced baseline also represents other uniform-splitting token-level PP systems, e.g., Seq1F1B~\cite{sun2024seq1f1b} and SlimPipe~\cite{li2025slimpipe}, strengthened for variable-length workloads.}
\end{itemize}

\noindent\paragraph{Workloads.}
We evaluate \name to train LLaMA2 architecture models (7B, 13B, 30B) on two widely used real-world datasets: \textit{CommonCrawl} and \textit{GitHub}.
The sequence length and token distribution of these two datasets are presented in Fig.~\ref{fig:moti} (b).
Excessively long sequences that exceed the context length are truncated.

\noindent\paragraph{Protocols}
The parallelism and gradient checkpointing configurations of all baselines are as follows:
\begin{itemize}[leftmargin=*]
\item \textbf{Parallelism Configuration}.
\name and MEPipe employ Ulysses-style SP within each node and PP across nodes; on our four-node testbed, this corresponds to SP degree 8 and PP degree 4.
For the baseline systems, we sweep feasible TP/CP/PP/DP/SP degrees and report the best-performing configuration for each workload.
For Megatron-LM, we enable TP, CP, PP, and DP with ZeRO-1; the TP degree is fixed to 8, and the CP degree is set to 2 for the 7B model and 4 for the 13B and 30B models, while the remaining parallelism dimensions are selected from the sweep.
For DeepSpeed, we use ZeRO-3 with Ulysses-style SP; the SP degree is set to 16 for the 7B model and 32 for the 13B and 30B models, with the remaining GPUs used for DP.
FlexSP determines its SP/DP configuration using its parallelism solver, and we use the fastest configuration selected by the solver.
 
\item \textbf{Gradient Checkpointing Configuration}. All systems use activation checkpointing configurations optimized for a 96K context length. For Megatron-LM, we checkpoint 10, 20, and 55 layers for 7B, 13B, and 30B models, respectively.
For DeepSpeed and FlexSP, we checkpoint 36 layers for the 30B model and none for the other model sizes. 
For MEPipe, consistent with MEPipe's single checkpointing policy for all micro-batches, we selected the checkpointing strategy with the fewest checkpointing layers without OOM error.
\end{itemize}

Evaluation metrics such as iteration time and token throughput are averaged over 20 training iterations.
Global batch size refers to the number of sequences in a training iteration.

\subsection{End-to-End Performance}
We evaluate the performance of \name by measuring the average end-to-end time of a training iteration with global batch size fixed to 512, as shown in Fig.~\ref{fig:main_res}. Experiments are conducted across various datasets, model sizes, and context lengths. 
Comprehensive results demonstrate that \name consistently outperforms baselines, achieving a maximum speedup of 1.69\texttimes\ compared to FlexSP, 2.07\texttimes\ compared to DeepSpeed, 2.60\texttimes\ compared to Megatron-LM, and 1.40\texttimes\ compared to MEPipe. 

The performance gains of \name on baseline systems except MEPipe primarily stem from the high communication efficiency of DPP. 
Specifically, DeepSpeed and FlexSP adopt Ulysses-style SP, where FSDP (ZeRO-3) is required to be applied on the whole cluster to shard the parameters and reduce gradients, resulting in frequent inter-node gather and scatter communications.
Moreover, the sequence parallelism pattern of DeepSpeed and Megatron-LM introduces costly inter-node communication overhead and harms training efficiency, which has been discussed in \S~\ref{sec:sp}.
In contrast, \name restricts SP and FSDP communication to intra-node by applying PP inter-node, significantly reducing the inter-node communication overhead.

FlexSP leverages heterogeneous sequence parallel groups to reduce the communication overhead of static Ulysses-style SP, where shorter sequences are scheduled with smaller SP groups with efficient intra-node communication. 
To this end, FlexSP accelerates DeepSpeed and Megatron-LM up to 1.33\texttimes\ and 1.66\texttimes\, respectively. 
However, this approach introduces workload imbalance across SP groups, and a longer sequence also necessitates being processed by a larger SP group, where the introduced inter-node communication overhead cannot be ignored.
This drawback is exacerbated when training larger models with limited resources. 
Correspondingly, the speedup of \name compared to FlexSP increases as model size scales, ranging from 1.31\texttimes\ to 1.69\texttimes\ on the \textit{CommonCrawl} dataset with a context length of 48K.

\begin{figure}
    \centering
\includegraphics[width=0.93\linewidth, keepaspectratio]{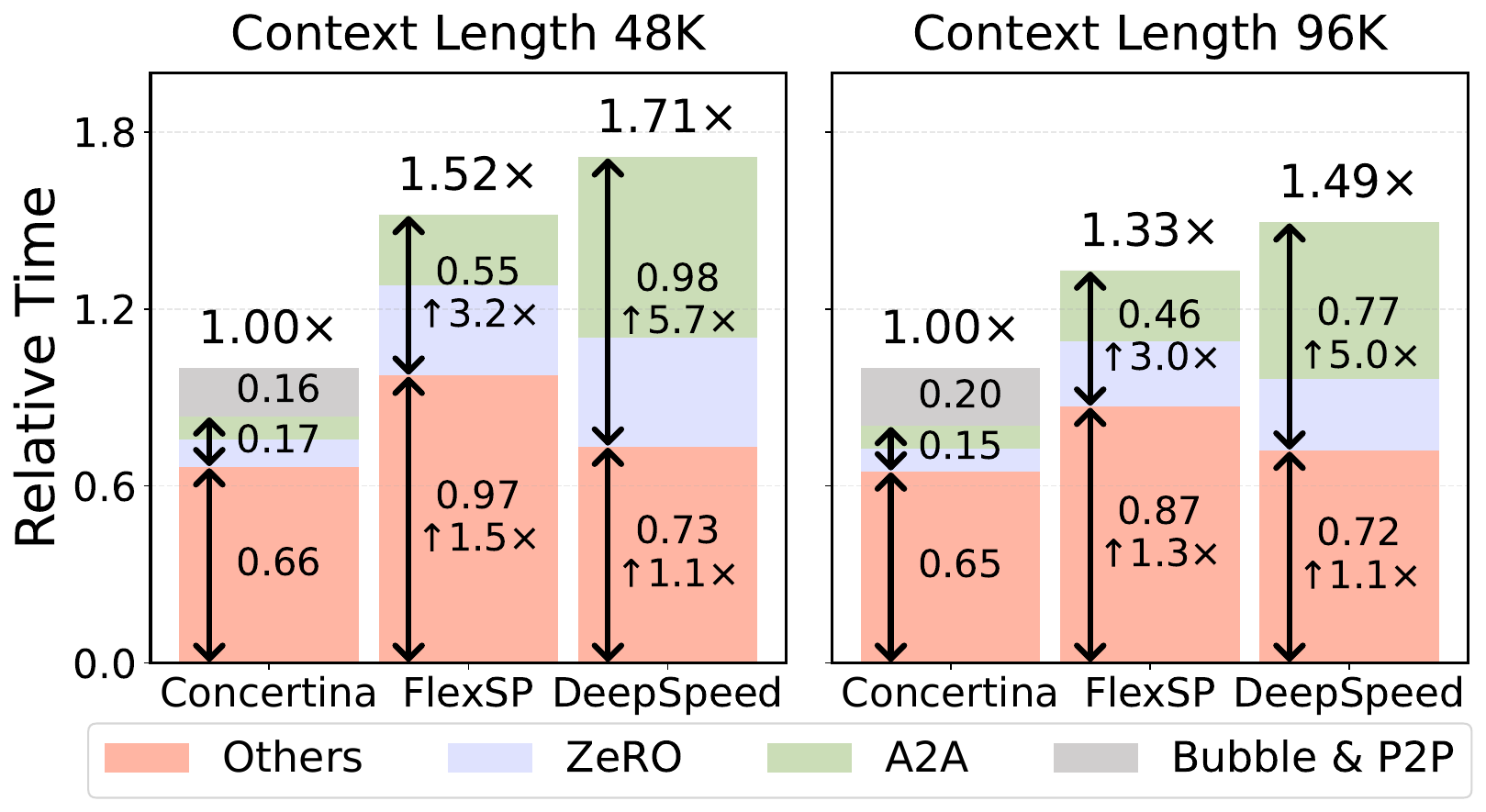}
    \caption{Case Study. End-to-end time breakdown of an iteration to train the 13B model with a fixed batch size of 512. The relative time and corresponding speedup of each component are indicated.}
    \label{fig:case_study}
\end{figure}

MEPipe suffers from pipeline bubbles resulting from workload imbalance across chunks.
As context length scales, the workload imbalance becomes more pronounced due to the enlarged variance of sequence lengths.
As a result, \name achieves a maximum speedup of 1.27\texttimes\ and 1.40\texttimes\ at a context length of 48K and 96K, respectively.
Moreover, MEPipe adopts a non-optimal and uniform checkpointing configuration to accommodate the longest sequence, introducing more unnecessary computation when handling relatively small models.
The adaptive chunk-level checkpointing pattern of \name reduces unnecessary recomputation overhead and further enhances training efficiency.


\subsection{Case Study}
To better understand \name's performance advantages in depth, we break down the end-to-end training time into several components: ``ZeRO'' (gather and scatter communications of ZeRO-3 that are not overlapped), ``A2A'' (All-to-All communication in Ulysses-style SP), ``Bubble \& P2P'' (time for PP's p2p communication and idle time of devices resulting from pipeline bubbles) and ``Others'' (computation, optimizer step, etc.).
The profiled time cost of each component is shown in Fig.~\ref{fig:case_study}.

To begin with, \name exhibits a similar performance in computation against DeepSpeed with a 1.1\texttimes\ improvement but outperforms FlexSP from 1.3\texttimes\ to 1.5\texttimes\, which is attributed to the unbalanced workload introduced by heterogeneous SP groups in FlexSP.
Furthermore, ``ZeRO'' and ``A2A'' overhead are the main bottlenecks of baselines.
However, these overheads account for only 17\% of the total end-to-end training time of \name.
By employing efficient intra-node communication, \name reduces these overheads significantly by up to 3.2\texttimes\ compared to FlexSP and 5.7\texttimes\ compared to DeepSpeed.
Last but not least, the bubble ratio is maintained at a relatively low level, less than 20\%, thanks to the workload-balanced chunking method and efficient pipeline schedule of \name.
The advantages above lead to an overall speedup of 1.52\texttimes\ and 1.71\texttimes\ compared to FlexSP and DeepSpeed, respectively.

\subsection{Scalability Study}
\begin{figure}
    \centering
    \includegraphics[width=1\linewidth, keepaspectratio]{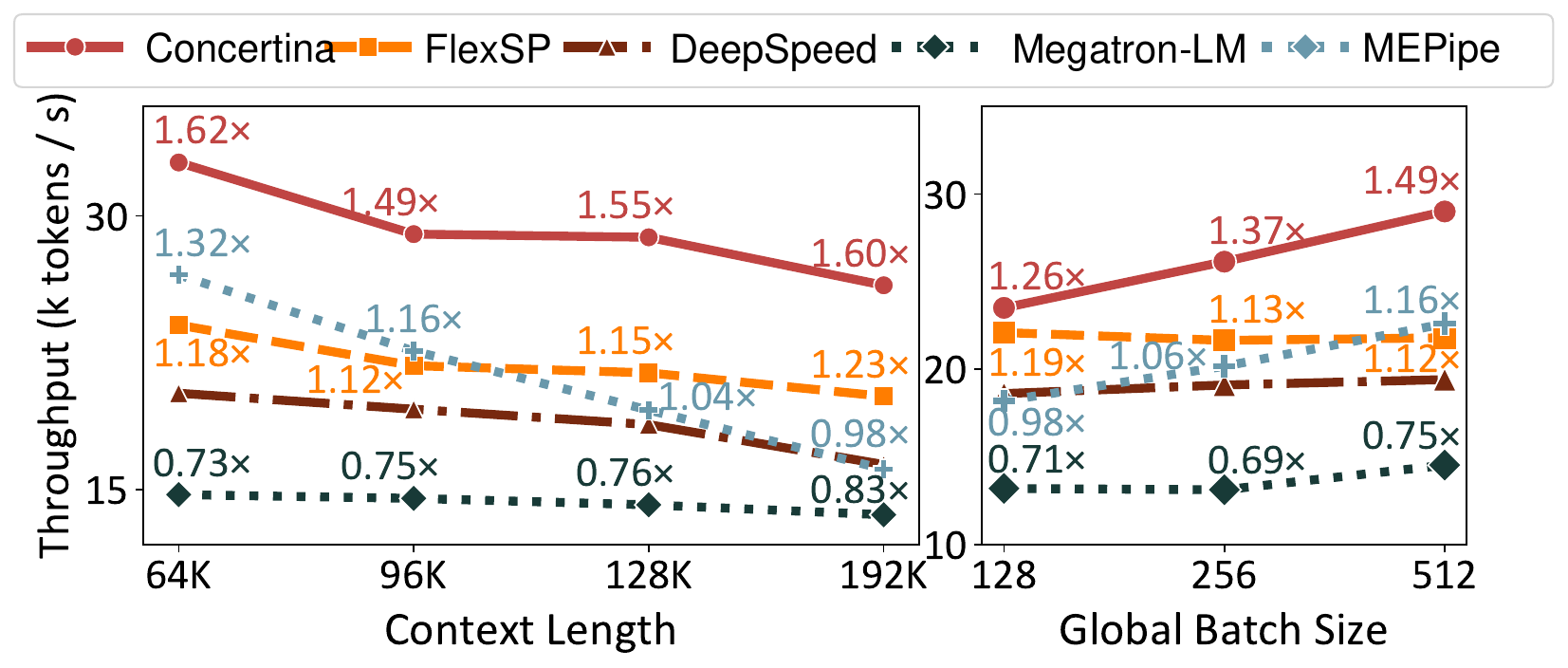}
    \caption{Scalability Study. Token throughput to train a 13B model under different context lengths and global batch sizes. Indicated improvements are normalized to DeepSpeed.}
    \label{fig:scalability}
\end{figure}
As shown in Fig.~\ref{fig:scalability}, token throughput under different settings of context length and batch size is measured to assess the scalability of \name.

\noindent\paragraph{Scalability w.r.t. context length.}
\name consistently achieves superior performance against baseline systems when context length is extended from 64K to 192K, achieving a speedup from 1.30\texttimes\ to 1.37\texttimes\ compared to FlexSP and from 1.23\texttimes\ to 1.63\texttimes\ compared to MEPipe.
As context length scales, the throughput of all systems tends to decrease, attributed to the increased computation overhead per token. Megatron-LM exhibits the least degradation because the quadratic-complexity self-attention operator is overlapped in CP's P2P kernel, resulting in similar processing time per token. 
On the contrary, MEPipe appears to be the most sensitive to context length, as the increasing variance of sequence lengths results in a more pronounced workload imbalance across chunks, leading to severe pipeline bubbles.

\begin{figure}
    \centering
\includegraphics[width=0.95\linewidth, keepaspectratio]{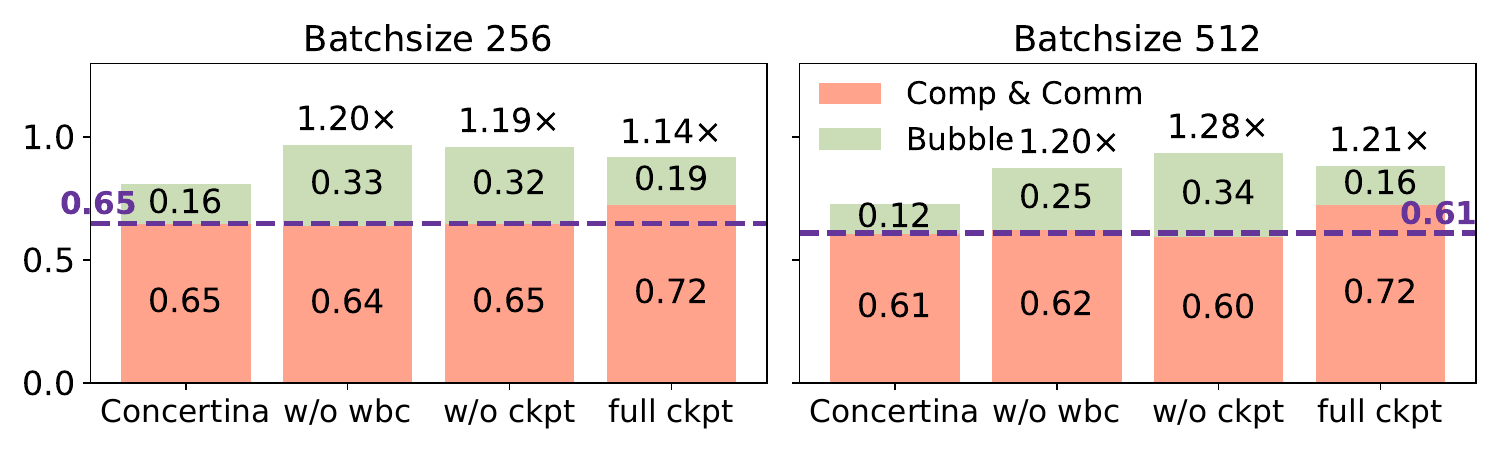}
    \caption{Ablation study. End-to-end time and bubble overhead to train a 13B model with a 64K context length. The times presented are all normalized to the end-to-end time of FlexSP.}
    \label{fig:ablation}
\end{figure}

\noindent\paragraph{Scalability w.r.t. global batch size.}
As global batch size ranges from 128 to 512, \name consistently outperforms baselines, and its performance exhibits a growing trend, with throughput improved by up to 1.18\texttimes.
In contrast, the throughput of baseline systems remains almost the same due to a similar computation overhead per token.
Benefiting from the lowered bubble ratio with more sequences, \name delivers a 1.33\texttimes\ and 1.49\texttimes\ throughput improvement compared to FlexSP and DeepSpeed, respectively.

\subsection{Ablation Study}
\label{sec:abla_study}
To validate the effectiveness of \name's key components, i.e., workload-balanced chunking and the co-optimization approach of pipeline scheduling and checkpointing, we compared \name with three ablated versions.
Specifically, ``w/o wbc'' denotes evenly splitting long sequences and packing short sequences into fixed-length chunks, ``w/o ckpt'' refers to disabling gradient checkpointing, and ``full ckpt'' represents applying full checkpointing.

As shown in Fig.~\ref{fig:ablation}, the variants exhibit distinct computation and pipeline bubble overhead.
Despite introducing no recomputation overhead, ``w/o ckpt'' brings limited computational benefits compared to \name due to the degradation of hardware utilization resulting from the finer granularity of a micro-batch. 
Moreover, the bubble ratios of all methods except ``w/o ckpt'' decrease as the global batch size scales.
This occurs as an increasing number of excessively long sequences forces scheduling of more 1F1B units, introducing severe warmup-cooldown overhead.
``w/o wbc'' suffers from bubble overhead caused by workload imbalance, while ``full ckpt'' incurs higher computation overhead due to suboptimal checkpointing configuration. 
Thanks to the co-optimization approach (\S~\ref{sec:chunk_scheduler}), \name consistently outperforms these variants with relatively low bubble ratio and computation overhead.

\begin{figure}
    \centering
\includegraphics[width=0.95\linewidth, keepaspectratio]{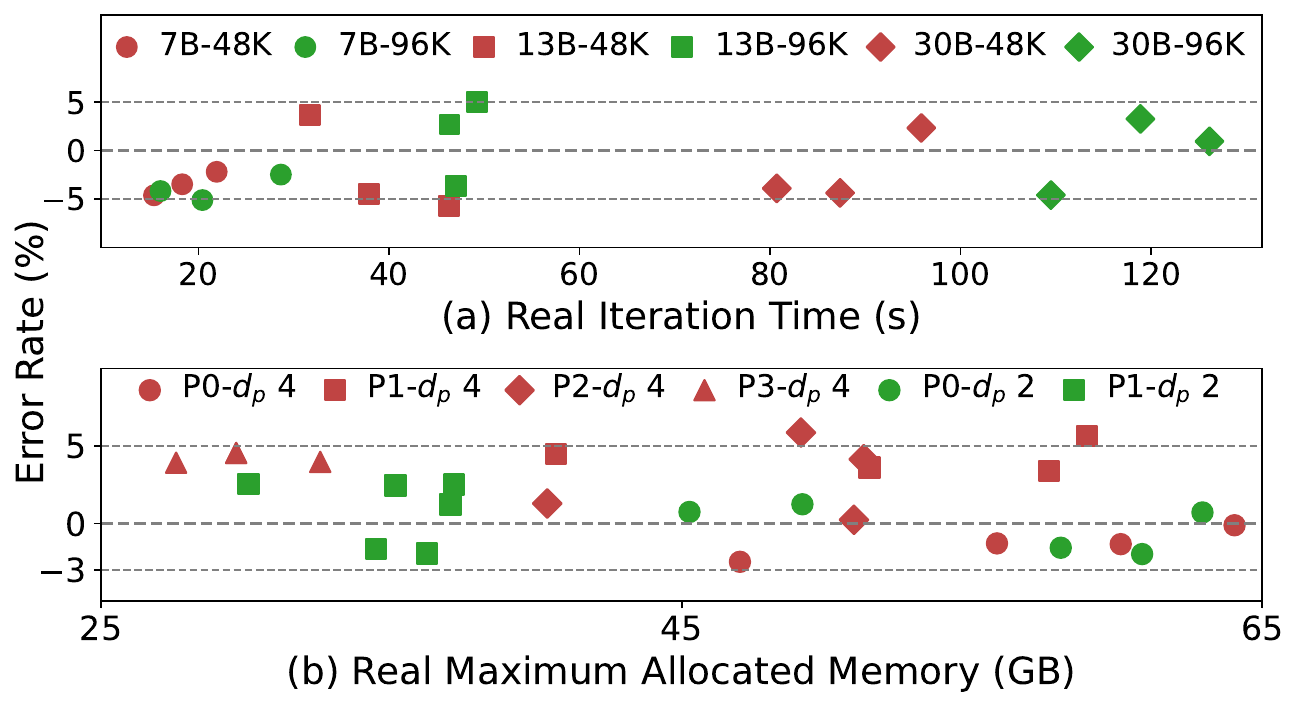}
    \caption{Accuracy of the Cost Model. The upper panel indicates the error rate of estimation on end-to-end time, while the bottom presents the error rate on the memory footprint of each pipeline stage. Experiments were conducted on a fixed cluster size of 32, i.e, $d_s \times d_p$.}
    \label{fig:cost_accu}
\end{figure}

\subsection{Solver Accuracy and Scalability}
We train a 13B model under a 64K context length to verify the solver's accuracy and light overhead.

\noindent\paragraph{Accuracy of Cost Model.}
\label{sec:accu_cost}
The deviations between the cost model's simulation and real profiled statistics are assessed under various settings of model sizes, context length, PP stages, and degrees.
As shown in Fig.~\ref{fig:cost_accu}, the error rates on both time cost and memory footprint are typically below 5\%, verifying the effectiveness of our cost model (\S~\ref{subsec:cost_model}).

\noindent\paragraph{Overhead and Scalability of Solver.}
\begin{figure}
    \centering
\includegraphics[width=1\linewidth, keepaspectratio]{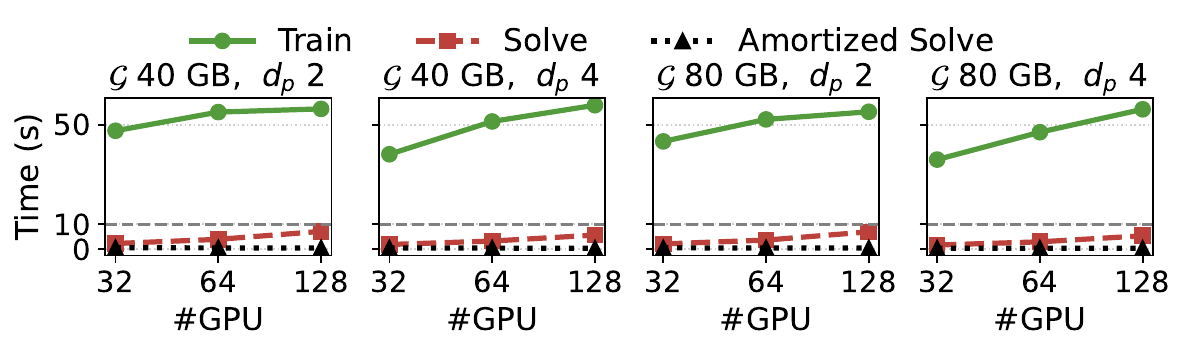}
\caption{Overhead and Scalability of Solver. The training time, solving time, and amortized solving time, i.e., solving time / (\#GPU / 8), under various cluster scales, hardware memory capacities ($\mathcal{G}$), and PP degrees ($d_p$).}
    \label{fig:solv_overhead}
\end{figure}
\label{sec:overhead_solver}
We conduct a simulation experiment to estimate the average training-iteration time and measure the solver's solve time within an optimality gap of \textbf{2\%} for the ILP problem (configurable in SCIP~\cite{BolusaniEtal2024OO}), and present the statistics in Fig.~\ref{fig:solv_overhead}.
The simulated training time is obtained using the validated cost model in Fig.~\ref{fig:cost_accu}.
The amortized solving time is also included due to the increased available CPU resources when deployed on larger-scale clusters.
The global batch size scales proportionally to cluster size, i.e., \#GPU, and is set to 512 initially for a cluster containing 32 GPUs.
The results indicate that the solver scales to larger clusters and that its overhead can be fully overlapped with the training process.

\subsection{Training Convergence Study}
\begin{figure}
    \centering
\includegraphics[width=1\linewidth, keepaspectratio]{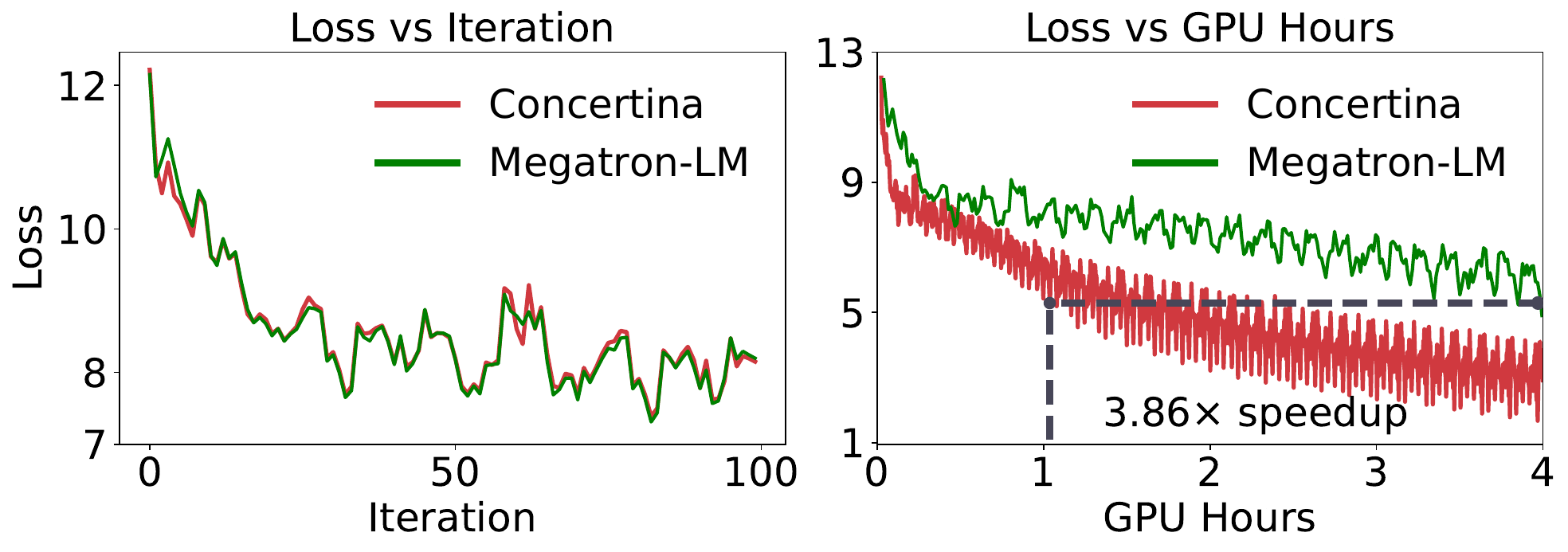}
\caption{Training convergence Study. Left: per-token loss over iterations, showing that \name follows the same optimization trajectory as Megatron-LM. Right: per-token loss over GPU hours.}
    \label{fig:loss}
\end{figure}

We randomly sampled 25,600 sequences from the \textit{GitHub} dataset and trained a 1B-parameter model from scratch using an AdamW optimizer.
We present the \textit{per-token loss} of \name compared with the reference implementation, Megatron-LM, with respect to iteration and GPU hours in Fig.~\ref{fig:loss}.
The results demonstrate that \name follows the same optimization trajectory as Megatron-LM.
\label{sec:training_convergence}

\subsection{Extension Study on MoE} \label{sec:moe_study}
We train a Qwen3-30B-A3B Mixture-of-Experts (MoE) model with PP degree 4 and Expert Parallelism (EP) degree 8 on the \textit{GitHub} dataset with varying context lengths and a batch size fixed to 512 to validate the generality of \name.
The attention modules employ an SP degree of 8 setup equipped with ZeRO-3, the same as the dense model.
We evaluate under balanced expert routing as workload-balance of EP is orthogonal to \name and has been well studied in prior MoE training systems~\cite{zhai2023smartmoe, nie2023flexmoe, he2022fastermoe}.

\noindent\paragraph{Accuracy of Cost Model.}
\begin{figure}
    \centering
\includegraphics[width=.95\linewidth, keepaspectratio]{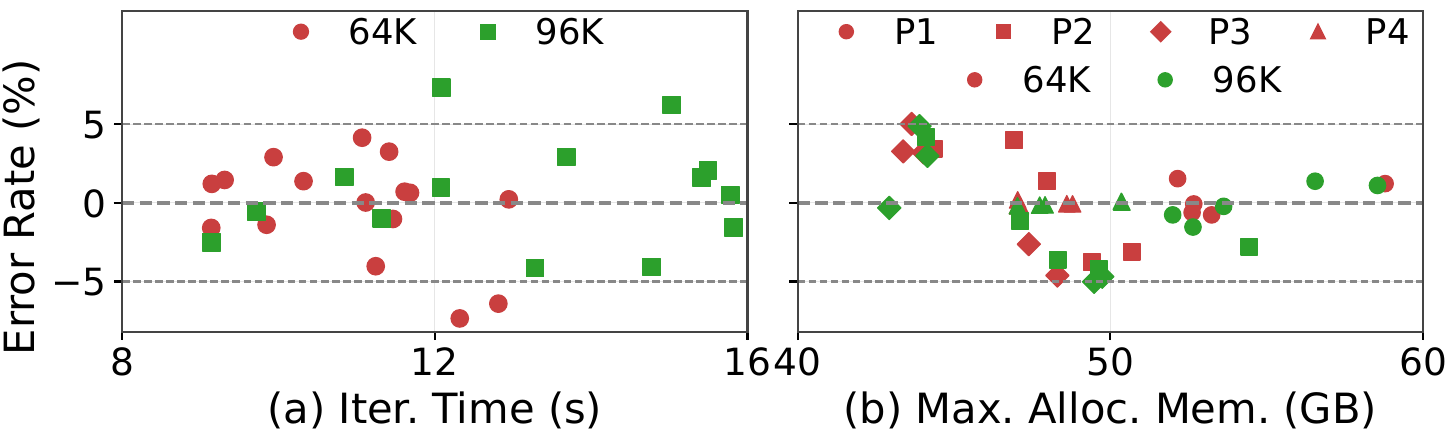}
    \caption{Accuracy of Cost Model for MoE. The predicted iteration time and maximum allocated memory, as well as the error rates, are presented on the left and right, respectively.}
    \label{fig:cost_accu_moe}
\end{figure}
Fig.~\ref{fig:cost_accu_moe} compares predicted and measured iteration time and stage-aware memory footprint under MoE workloads.
The average absolute error is 2.5\% for iteration time and 1.9\% for peak memory, with 95th-percentile errors of 6.8\% and 4.5\%, respectively.
These results validate the cost model's accuracy with MoE computation and expert-parallel communication, confirming that \name's assumptions on quadratic computation complexity and proportional activation memory remain applicable.

\begin{figure}
    \centering
\includegraphics[width=.95\linewidth, keepaspectratio]{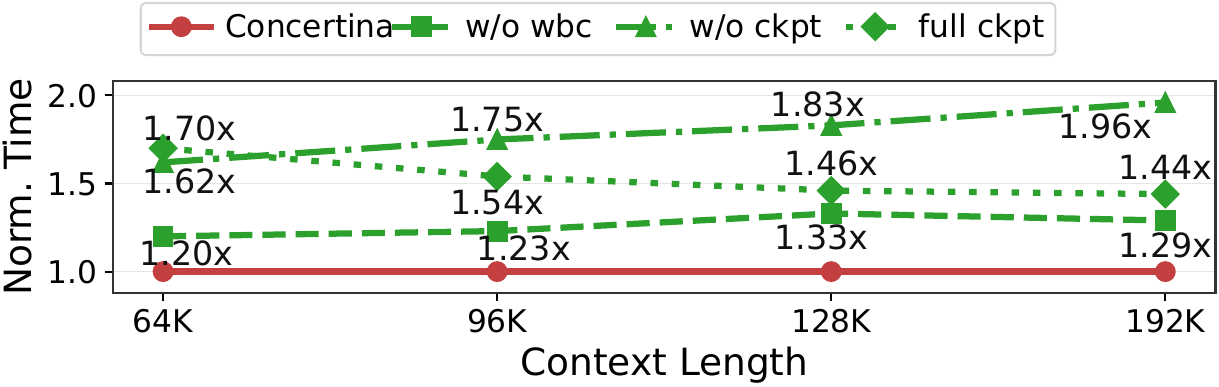}
    \caption{Ablation Study for MoE. The normalized times of the ablated variants w.r.t.\ \name are presented.}
    \label{fig:moe_perf}
\end{figure}

\noindent\paragraph{End-to-End Ablation.}
Fig.~\ref{fig:moe_perf} presents a scalability ablation on the \textit{GitHub} dataset, with all results normalized to \name.
\name consistently outperforms all variants across 64K--192K context lengths.
Removing workload-balanced chunking increases normalized time to 1.20--1.33\texttimes due to larger sequence-length variance.
Disabling checkpointing becomes increasingly inefficient at longer contexts (1.62--1.96\texttimes), while full checkpointing incurs 1.44--1.70\texttimes overhead due to redundant recomputation.
These results validate the effectiveness of workload-balanced chunking and stage-aware adaptive checkpointing for MoE training.

%% file: Main-Conclusion.tex
\section{Related Work} \label{sec:related}
We summarize the design space of \name compared with prior works in Tab.~\ref{tab:design_space}. 

\noindent\paragraph{Long Context Training.}
SP-based systems~\cite{DBLP:ring-attn, DBLP:striped_attn, DBLP:LightSeq}, do not consider PP and are orthogonal to \name.
Batch-level PP systems such as DAPPLE~\cite{DBLP:conf/ppopp/dapple}, GPipe~\cite{pytorch_gpipe}, DynaPipe~\cite{jiang2024dynapipe}, ByteScale~\cite{ge2025bytescale}, and WLB-LLM~\cite{wang2025wlb} pack or schedule variable-length batches, but remain in a packing-only regime.
Token-level PP systems such as MEPipe~\cite{sun2025mepipe}, TeraPipe~\cite{li2021terapipe}, Seq1F1B~\cite{sun2024seq1f1b}, and SlimPipe~\cite{li2025slimpipe} assume homogeneous sequence length and require sequences to be split into an identical number of slices.
In contrast, \name optimizes dynamic PP granularity, jointly choosing packed, split, and hybrid chunks.

\noindent\paragraph{Checkpointing Optimization.}
\name's optimization space differs fundamentally from AdaPipe~\cite{sun2024adapipe} and GraphPipe~\cite{jeon2025graphpipe}: these works apply a stage-wise checkpointing strategy, which is homogeneous to all micro-batches, while \name further discusses a distinct checkpointing strategy for each heterogeneous chunk, i.e., a runtime chunk-aware checkpointing strategy.

\begin{table}[t]
\small
\centering
\caption{Design-space comparison with prior systems.}
\label{tab:design_space}
\setlength{\tabcolsep}{2pt}
\begin{tabular}{@{}lcc@{}}
\toprule
\textbf{System} & \textbf{PP Granularity} & 
\textbf{Ckpt. Adaptivity} \\
\midrule
DAPPLE/GPipe & Batch & -- \\
MEPipe/SlimPipe & Token & -- \\
FlexSP/HotSPa & Non-PP & -- \\
\makecell[l]{DynaPipe/ByteScale\\WLB-LLM} & Batch & -- \\
AdaPipe/GraphPipe & Batch & Stage \\
\midrule
\textbf{\name} & \textbf{Dynamic} & 
\textbf{Chunk--Stage Joint} \\
\bottomrule
\end{tabular}
\end{table}

\section{Conclusion} \label{sec:conclusion}
In this paper, we propose \textit{Dynamic Pipeline Parallelism} (DPP) and \textit{Stage-Aware Chunk-Level Adaptive Checkpointing}.
We build \name, a distributed LLM training system that efficiently applies pipeline parallelism to variable-length heterogeneous long-context training workloads. 
Comprehensive evaluations across model sizes, datasets, and context lengths show that \name reduces communication overhead and improves training throughput by up to 1.69\texttimes\ over FlexSP and 1.40\texttimes\ over MEPipe.

%% file: main.bbl

\begin{thebibliography}{00}


\ifx \showCODEN    \undefined \def \showCODEN     #1{\unskip}     \fi
\ifx \showDOI      \undefined \def \showDOI       #1{#1}\fi
\ifx \showISBNx    \undefined \def \showISBNx     #1{\unskip}     \fi
\ifx \showISBNxiii \undefined \def \showISBNxiii  #1{\unskip}     \fi
\ifx \showISSN     \undefined \def \showISSN      #1{\unskip}     \fi
\ifx \showLCCN     \undefined \def \showLCCN      #1{\unskip}     \fi
\ifx \shownote     \undefined \def \shownote      #1{#1}          \fi
\ifx \showarticletitle \undefined \def \showarticletitle #1{#1}   \fi
\ifx \showURL      \undefined \def \showURL       {\relax}        \fi
\providecommand\bibfield[2]{#2}
\providecommand\bibinfo[2]{#2}
\providecommand\natexlab[1]{#1}
\providecommand\showeprint[2][]{arXiv:#2}

\bibitem[\protect\citeauthoryear{??}{ncc}{2021}]%
        {nccl}
 \bibinfo{year}{2021}\natexlab{}.
\newblock \bibinfo{title}{NVIDIA collective communications library (NCCL)}.
\newblock \bibinfo{howpublished}{\url{https://developer.nvidia.com/nccl}}.   (\bibinfo{year}{2021}).
\newblock


\bibitem[\protect\citeauthoryear{??}{pyt}{2021}]%
        {pytorch_gpipe}
 \bibinfo{year}{2021}\natexlab{}.
\newblock \bibinfo{title}{PyTorch GPipe}.
\newblock \bibinfo{howpublished}{\url{https://pytorch.org/docs/stable/pipeline.html}}.   (\bibinfo{year}{2021}).
\newblock


\bibitem[\protect\citeauthoryear{??}{lla}{2024}]%
        {llama-3}
 \bibinfo{year}{2024}\natexlab{}.
\newblock \bibinfo{title}{Introducing Meta Llama 3: The most capable openly available LLM to date}.
\newblock \bibinfo{howpublished}{\url{https://ai.meta.com/blog/meta-llama-3/}}.   (\bibinfo{year}{2024}).
\newblock


\bibitem[\protect\citeauthoryear{Achiam, Adler, Agarwal, Ahmad, Akkaya, Aleman, Almeida, Altenschmidt, Altman, Anadkat, Avila, Babuschkin, Balaji, Balcom, Baltescu, Bao, Bavarian, Belgum, Bello, Berdine, Bernadett-Shapiro, Berner, Bogdonoff, Boiko, Boyd, Brakman, Brockman, Brooks, Brundage, Button, Cai, Campbell, Cann, Carey, Carlson, Carmichael, Chan, Chang, Chantzis, Chen, Chen, Chen, Chen, Chen, Chess, Cho, Chu, Chung, Cummings, Currier, Dai, Decareaux, Degry, Deutsch, Deville, Dhar, Dohan, Dowling, Dunning, Ecoffet, Eleti, Eloundou, Farhi, Fedus, Felix, Fishman, Forte, Fulford, Gao, Georges, Gibson, Goel, Gogineni, Goh, Gontijo-Lopes, Gordon, Grafstein, Gray, Greene, Gross, Gu, Guo, Hallacy, Han, Harris, He, Heaton, Heidecke, Hesse, Hickey, Hickey, Hoeschele, Houghton, Hsu, Hu, Hu, Huizinga, Jain, Jain, Jang, Jiang, Jiang, Jin, Jin, Jomoto, Jonn, Jun, Kaftan, Kaiser, Kamali, Kanitscheider, Keskar, Khan, Kilpatrick, Kim, Kim, Kim, Kirchner, Kiros, Knight, Kokotajlo, Kondraciuk, Kondrich, Konstantinidis,
  Kosic, Krueger, Kuo, Lampe, Lan, Lee, Leike, Leung, Levy, Li, Lim, Lin, Lin, Litwin, Lopez, Lowe, Lue, Makanju, Malfacini, Manning, Markov, Markovski, Martin, Mayer, Mayne, McGrew, McKinney, McLeavey, McMillan, McNeil, Medina, Mehta, Menick, Metz, Mishchenko, Mishkin, Monaco, Morikawa, Mossing, Mu, Murati, Murk, Mély, Nair, Nakano, Nayak, Neelakantan, Ngo, Noh, Ouyang, O'Keefe, Pachocki, Paino, Palermo, Pantuliano, Parascandolo, Parish, Parparita, Passos, Pavlov, Peng, Perelman, Peres, Petrov, Pinto, Michael, Pokorny, Pokrass, Pong, Powell, Power, Power, Proehl, Puri, Radford, Rae, Ramesh, Raymond, Real, Rimbach, Ross, Rotsted, Roussez, Ryder, Saltarelli, Sanders, Santurkar, Sastry, Schmidt, Schnurr, Schulman, Selsam, Sheppard, Sherbakov, Shieh, Shoker, Shyam, Sidor, Sigler, Simens, Sitkin, Slama, Sohl, Sokolowsky, Song, Staudacher, Such, Summers, Sutskever, Tang, Tezak, Thompson, Tillet, Tootoonchian, Tseng, Tuggle, Turley, Tworek, Uribe, Vallone, Vijayvergiya, Voss, Wainwright, Wang, Wang, Wang, Ward,
  Wei, Weinmann, Welihinda, Welinder, Weng, Weng, Wiethoff, Willner, Winter, Wolrich, Wong, Workman, Wu, Wu, Wu, Xiao, Xu, Yoo, Yu, Yuan, Zaremba, Zellers, Zhang, Zhang, Zhao, Zheng, Zhuang, Zhuk, and Zoph}{Achiam et~al\mbox{.}}{2023}]%
        {gpt4}
\bibfield{author}{\bibinfo{person}{Josh Achiam}, \bibinfo{person}{Steven Adler}, \bibinfo{person}{Sandhini Agarwal}, \bibinfo{person}{Lama Ahmad}, \bibinfo{person}{Ilge Akkaya}, \bibinfo{person}{Florencia~Leoni Aleman}, \bibinfo{person}{Diogo Almeida}, \bibinfo{person}{Janko Altenschmidt}, \bibinfo{person}{Sam Altman}, \bibinfo{person}{Shyamal Anadkat}, \bibinfo{person}{Red Avila}, \bibinfo{person}{Igor Babuschkin}, \bibinfo{person}{Suchir Balaji}, \bibinfo{person}{Valerie Balcom}, \bibinfo{person}{Paul Baltescu}, \bibinfo{person}{Haiming Bao}, \bibinfo{person}{Mohammad Bavarian}, \bibinfo{person}{Jeff Belgum}, \bibinfo{person}{Irwan Bello}, \bibinfo{person}{Jake Berdine}, \bibinfo{person}{Gabriel Bernadett-Shapiro}, \bibinfo{person}{Christopher Berner}, \bibinfo{person}{Lenny Bogdonoff}, \bibinfo{person}{Oleg Boiko}, \bibinfo{person}{Madelaine Boyd}, \bibinfo{person}{Anna-Luisa Brakman}, \bibinfo{person}{Greg Brockman}, \bibinfo{person}{Tim Brooks}, \bibinfo{person}{Miles Brundage}, \bibinfo{person}{Kevin
  Button}, \bibinfo{person}{Trevor Cai}, \bibinfo{person}{Rosie Campbell}, \bibinfo{person}{Andrew Cann}, \bibinfo{person}{Brittany Carey}, \bibinfo{person}{Chelsea Carlson}, \bibinfo{person}{Rory Carmichael}, \bibinfo{person}{Brooke Chan}, \bibinfo{person}{Che Chang}, \bibinfo{person}{Fotis Chantzis}, \bibinfo{person}{Derek Chen}, \bibinfo{person}{Sully Chen}, \bibinfo{person}{Ruby Chen}, \bibinfo{person}{Jason Chen}, \bibinfo{person}{Mark Chen}, \bibinfo{person}{Ben Chess}, \bibinfo{person}{Chester Cho}, \bibinfo{person}{Casey Chu}, \bibinfo{person}{Hyung~Won Chung}, \bibinfo{person}{Dave Cummings}, \bibinfo{person}{Jeremiah Currier}, \bibinfo{person}{Yunxing Dai}, \bibinfo{person}{Cory Decareaux}, \bibinfo{person}{Thomas Degry}, \bibinfo{person}{Noah Deutsch}, \bibinfo{person}{Damien Deville}, \bibinfo{person}{Arka Dhar}, \bibinfo{person}{David Dohan}, \bibinfo{person}{Steve Dowling}, \bibinfo{person}{Sheila Dunning}, \bibinfo{person}{Adrien Ecoffet}, \bibinfo{person}{Atty Eleti}, \bibinfo{person}{Tyna
  Eloundou}, \bibinfo{person}{David Farhi}, \bibinfo{person}{Liam Fedus}, \bibinfo{person}{Niko Felix}, \bibinfo{person}{Simón~Posada Fishman}, \bibinfo{person}{Juston Forte}, \bibinfo{person}{Isabella Fulford}, \bibinfo{person}{Leo Gao}, \bibinfo{person}{Elie Georges}, \bibinfo{person}{Christian Gibson}, \bibinfo{person}{Vik Goel}, \bibinfo{person}{Tarun Gogineni}, \bibinfo{person}{Gabriel Goh}, \bibinfo{person}{Rapha Gontijo-Lopes}, \bibinfo{person}{Jonathan Gordon}, \bibinfo{person}{Morgan Grafstein}, \bibinfo{person}{Scott Gray}, \bibinfo{person}{Ryan Greene}, \bibinfo{person}{Joshua Gross}, \bibinfo{person}{Shixiang~Shane Gu}, \bibinfo{person}{Yufei Guo}, \bibinfo{person}{Chris Hallacy}, \bibinfo{person}{Jesse Han}, \bibinfo{person}{Jeff Harris}, \bibinfo{person}{Yuchen He}, \bibinfo{person}{Mike Heaton}, \bibinfo{person}{Johannes Heidecke}, \bibinfo{person}{Chris Hesse}, \bibinfo{person}{Alan Hickey}, \bibinfo{person}{Wade Hickey}, \bibinfo{person}{Peter Hoeschele}, \bibinfo{person}{Brandon Houghton},
  \bibinfo{person}{Kenny Hsu}, \bibinfo{person}{Shengli Hu}, \bibinfo{person}{Xin Hu}, \bibinfo{person}{Joost Huizinga}, \bibinfo{person}{Shantanu Jain}, \bibinfo{person}{Shawn Jain}, \bibinfo{person}{Joanne Jang}, \bibinfo{person}{Angela Jiang}, \bibinfo{person}{Roger Jiang}, \bibinfo{person}{Haozhun Jin}, \bibinfo{person}{Denny Jin}, \bibinfo{person}{Shino Jomoto}, \bibinfo{person}{Billie Jonn}, \bibinfo{person}{Heewoo Jun}, \bibinfo{person}{Tomer Kaftan}, \bibinfo{person}{Łukasz Kaiser}, \bibinfo{person}{Ali Kamali}, \bibinfo{person}{Ingmar Kanitscheider}, \bibinfo{person}{Nitish~Shirish Keskar}, \bibinfo{person}{Tabarak Khan}, \bibinfo{person}{Logan Kilpatrick}, \bibinfo{person}{Jong~Wook Kim}, \bibinfo{person}{Christina Kim}, \bibinfo{person}{Yongjik Kim}, \bibinfo{person}{Jan~Hendrik Kirchner}, \bibinfo{person}{Jamie Kiros}, \bibinfo{person}{Matt Knight}, \bibinfo{person}{Daniel Kokotajlo}, \bibinfo{person}{Łukasz Kondraciuk}, \bibinfo{person}{Andrew Kondrich}, \bibinfo{person}{Aris Konstantinidis},
  \bibinfo{person}{Kyle Kosic}, \bibinfo{person}{Gretchen Krueger}, \bibinfo{person}{Vishal Kuo}, \bibinfo{person}{Michael Lampe}, \bibinfo{person}{Ikai Lan}, \bibinfo{person}{Teddy Lee}, \bibinfo{person}{Jan Leike}, \bibinfo{person}{Jade Leung}, \bibinfo{person}{Daniel Levy}, \bibinfo{person}{Chak~Ming Li}, \bibinfo{person}{Rachel Lim}, \bibinfo{person}{Molly Lin}, \bibinfo{person}{Stephanie Lin}, \bibinfo{person}{Mateusz Litwin}, \bibinfo{person}{Theresa Lopez}, \bibinfo{person}{Ryan Lowe}, \bibinfo{person}{Patricia Lue}, \bibinfo{person}{Anna Makanju}, \bibinfo{person}{Kim Malfacini}, \bibinfo{person}{Sam Manning}, \bibinfo{person}{Todor Markov}, \bibinfo{person}{Yaniv Markovski}, \bibinfo{person}{Bianca Martin}, \bibinfo{person}{Katie Mayer}, \bibinfo{person}{Andrew Mayne}, \bibinfo{person}{Bob McGrew}, \bibinfo{person}{Scott~Mayer McKinney}, \bibinfo{person}{Christine McLeavey}, \bibinfo{person}{Paul McMillan}, \bibinfo{person}{Jake McNeil}, \bibinfo{person}{David Medina}, \bibinfo{person}{Aalok Mehta},
  \bibinfo{person}{Jacob Menick}, \bibinfo{person}{Luke Metz}, \bibinfo{person}{Andrey Mishchenko}, \bibinfo{person}{Pamela Mishkin}, \bibinfo{person}{Vinnie Monaco}, \bibinfo{person}{Evan Morikawa}, \bibinfo{person}{Daniel Mossing}, \bibinfo{person}{Tong Mu}, \bibinfo{person}{Mira Murati}, \bibinfo{person}{Oleg Murk}, \bibinfo{person}{David Mély}, \bibinfo{person}{Ashvin Nair}, \bibinfo{person}{Reiichiro Nakano}, \bibinfo{person}{Rajeev Nayak}, \bibinfo{person}{Arvind Neelakantan}, \bibinfo{person}{Richard Ngo}, \bibinfo{person}{Hyeonwoo Noh}, \bibinfo{person}{Long Ouyang}, \bibinfo{person}{Cullen O'Keefe}, \bibinfo{person}{Jakub Pachocki}, \bibinfo{person}{Alex Paino}, \bibinfo{person}{Joe Palermo}, \bibinfo{person}{Ashley Pantuliano}, \bibinfo{person}{Giambattista Parascandolo}, \bibinfo{person}{Joel Parish}, \bibinfo{person}{Emy Parparita}, \bibinfo{person}{Alex Passos}, \bibinfo{person}{Mikhail Pavlov}, \bibinfo{person}{Andrew Peng}, \bibinfo{person}{Adam Perelman}, \bibinfo{person}{Filipe de
  Avila~Belbute Peres}, \bibinfo{person}{Michael Petrov}, \bibinfo{person}{Henrique Ponde de~Oliveira Pinto}, \bibinfo{person}{Michael}, \bibinfo{person}{Pokorny}, \bibinfo{person}{Michelle Pokrass}, \bibinfo{person}{Vitchyr~H. Pong}, \bibinfo{person}{Tolly Powell}, \bibinfo{person}{Alethea Power}, \bibinfo{person}{Boris Power}, \bibinfo{person}{Elizabeth Proehl}, \bibinfo{person}{Raul Puri}, \bibinfo{person}{Alec Radford}, \bibinfo{person}{Jack Rae}, \bibinfo{person}{Aditya Ramesh}, \bibinfo{person}{Cameron Raymond}, \bibinfo{person}{Francis Real}, \bibinfo{person}{Kendra Rimbach}, \bibinfo{person}{Carl Ross}, \bibinfo{person}{Bob Rotsted}, \bibinfo{person}{Henri Roussez}, \bibinfo{person}{Nick Ryder}, \bibinfo{person}{Mario Saltarelli}, \bibinfo{person}{Ted Sanders}, \bibinfo{person}{Shibani Santurkar}, \bibinfo{person}{Girish Sastry}, \bibinfo{person}{Heather Schmidt}, \bibinfo{person}{David Schnurr}, \bibinfo{person}{John Schulman}, \bibinfo{person}{Daniel Selsam}, \bibinfo{person}{Kyla Sheppard},
  \bibinfo{person}{Toki Sherbakov}, \bibinfo{person}{Jessica Shieh}, \bibinfo{person}{Sarah Shoker}, \bibinfo{person}{Pranav Shyam}, \bibinfo{person}{Szymon Sidor}, \bibinfo{person}{Eric Sigler}, \bibinfo{person}{Maddie Simens}, \bibinfo{person}{Jordan Sitkin}, \bibinfo{person}{Katarina Slama}, \bibinfo{person}{Ian Sohl}, \bibinfo{person}{Benjamin Sokolowsky}, \bibinfo{person}{Yang Song}, \bibinfo{person}{Natalie Staudacher}, \bibinfo{person}{Felipe~Petroski Such}, \bibinfo{person}{Natalie Summers}, \bibinfo{person}{Ilya Sutskever}, \bibinfo{person}{Jie Tang}, \bibinfo{person}{Nikolas Tezak}, \bibinfo{person}{Madeleine~B. Thompson}, \bibinfo{person}{Phil Tillet}, \bibinfo{person}{Amin Tootoonchian}, \bibinfo{person}{Elizabeth Tseng}, \bibinfo{person}{Preston Tuggle}, \bibinfo{person}{Nick Turley}, \bibinfo{person}{Jerry Tworek}, \bibinfo{person}{Juan Felipe~Cerón Uribe}, \bibinfo{person}{Andrea Vallone}, \bibinfo{person}{Arun Vijayvergiya}, \bibinfo{person}{Chelsea Voss}, \bibinfo{person}{Carroll
  Wainwright}, \bibinfo{person}{Justin~Jay Wang}, \bibinfo{person}{Alvin Wang}, \bibinfo{person}{Ben Wang}, \bibinfo{person}{Jonathan Ward}, \bibinfo{person}{Jason Wei}, \bibinfo{person}{CJ Weinmann}, \bibinfo{person}{Akila Welihinda}, \bibinfo{person}{Peter Welinder}, \bibinfo{person}{Jiayi Weng}, \bibinfo{person}{Lilian Weng}, \bibinfo{person}{Matt Wiethoff}, \bibinfo{person}{Dave Willner}, \bibinfo{person}{Clemens Winter}, \bibinfo{person}{Samuel Wolrich}, \bibinfo{person}{Hannah Wong}, \bibinfo{person}{Lauren Workman}, \bibinfo{person}{Sherwin Wu}, \bibinfo{person}{Jeff Wu}, \bibinfo{person}{Michael Wu}, \bibinfo{person}{Kai Xiao}, \bibinfo{person}{Tao Xu}, \bibinfo{person}{Sarah Yoo}, \bibinfo{person}{Kevin Yu}, \bibinfo{person}{Qiming Yuan}, \bibinfo{person}{Wojciech Zaremba}, \bibinfo{person}{Rowan Zellers}, \bibinfo{person}{Chong Zhang}, \bibinfo{person}{Marvin Zhang}, \bibinfo{person}{Shengjia Zhao}, \bibinfo{person}{Tianhao Zheng}, \bibinfo{person}{Juntang Zhuang}, \bibinfo{person}{William Zhuk},
  {and} \bibinfo{person}{Barret Zoph}.} \bibinfo{year}{2023}\natexlab{}.
\newblock \showarticletitle{Gpt-4 technical report}.
\newblock \bibinfo{journal}{{\em arXiv preprint arXiv:2303.08774\/}} (\bibinfo{year}{2023}).
\newblock
\showURL{%
\url{https://doi.org/10.48550/arXiv.2303.08774}}


\bibitem[\protect\citeauthoryear{Bolusani, Besan{\c{c}}on, Bestuzheva, Chmiela, Dion{\'{i}}sio, Donkiewicz, van Doornmalen, Eifler, Ghannam, Gleixner, Graczyk, Halbig, Hedtke, Hoen, Hojny, van~der Hulst, Kamp, Koch, Kofler, Lentz, Manns, Mexi, M\"{u}hmer, Pfetsch, Schl{\"o}sser, Serrano, Shinano, Turner, Vigerske, Weninger, and Xu}{Bolusani et~al\mbox{.}}{2024}]%
        {BolusaniEtal2024OO}
\bibfield{author}{\bibinfo{person}{Suresh Bolusani}, \bibinfo{person}{Mathieu Besan{\c{c}}on}, \bibinfo{person}{Ksenia Bestuzheva}, \bibinfo{person}{Antonia Chmiela}, \bibinfo{person}{Jo{\~{a}}o Dion{\'{i}}sio}, \bibinfo{person}{Tim Donkiewicz}, \bibinfo{person}{Jasper van Doornmalen}, \bibinfo{person}{Leon Eifler}, \bibinfo{person}{Mohammed Ghannam}, \bibinfo{person}{Ambros Gleixner}, \bibinfo{person}{Christoph Graczyk}, \bibinfo{person}{Katrin Halbig}, \bibinfo{person}{Ivo Hedtke}, \bibinfo{person}{Alexander Hoen}, \bibinfo{person}{Christopher Hojny}, \bibinfo{person}{Rolf van~der Hulst}, \bibinfo{person}{Dominik Kamp}, \bibinfo{person}{Thorsten Koch}, \bibinfo{person}{Kevin Kofler}, \bibinfo{person}{Jurgen Lentz}, \bibinfo{person}{Julian Manns}, \bibinfo{person}{Gioni Mexi}, \bibinfo{person}{Erik M\"{u}hmer}, \bibinfo{person}{Marc~E. Pfetsch}, \bibinfo{person}{Franziska Schl{\"o}sser}, \bibinfo{person}{Felipe Serrano}, \bibinfo{person}{Yuji Shinano}, \bibinfo{person}{Mark Turner}, \bibinfo{person}{Stefan
  Vigerske}, \bibinfo{person}{Dieter Weninger}, {and} \bibinfo{person}{Lixing Xu}.} \bibinfo{year}{2024}\natexlab{}.
\newblock \bibinfo{booktitle}{{\em {The SCIP Optimization Suite 9.0}}}.
\newblock \bibinfo{type}{Technical Report}. \bibinfo{institution}{Optimization Online}.
\newblock
\showURL{%
\url{https://optimization-online.org/2024/02/the-scip-optimization-suite-9-0/}}


\bibitem[\protect\citeauthoryear{Brandon, Nrusimha, Qian, Ankner, Jin, Song, and Ragan{-}Kelley}{Brandon et~al\mbox{.}}{2023}]%
        {DBLP:striped_attn}
\bibfield{author}{\bibinfo{person}{William Brandon}, \bibinfo{person}{Aniruddha Nrusimha}, \bibinfo{person}{Kevin Qian}, \bibinfo{person}{Zachary Ankner}, \bibinfo{person}{Tian Jin}, \bibinfo{person}{Zhiye Song}, {and} \bibinfo{person}{Jonathan Ragan{-}Kelley}.} \bibinfo{year}{2023}\natexlab{}.
\newblock \showarticletitle{Striped Attention: Faster Ring Attention for Causal Transformers}.
\newblock \bibinfo{journal}{{\em CoRR\/}}  \bibinfo{volume}{abs/2311.09431} (\bibinfo{year}{2023}).
\newblock
\showDOI{%
\url{https://doi.org/10.48550/ARXIV.2311.09431}}
\showeprint{2311.09431}


\bibitem[\protect\citeauthoryear{Brown, Mann, Ryder, Subbiah, Kaplan, Dhariwal, Neelakantan, Shyam, Sastry, Askell, Agarwal, Herbert{-}Voss, Krueger, Henighan, Child, Ramesh, Ziegler, Wu, Winter, Hesse, Chen, Sigler, Litwin, Gray, Chess, Clark, Berner, McCandlish, Radford, Sutskever, and Amodei}{Brown et~al\mbox{.}}{2020}]%
        {DBLP:gpt3}
\bibfield{author}{\bibinfo{person}{Tom~B. Brown}, \bibinfo{person}{Benjamin Mann}, \bibinfo{person}{Nick Ryder}, \bibinfo{person}{Melanie Subbiah}, \bibinfo{person}{Jared Kaplan}, \bibinfo{person}{Prafulla Dhariwal}, \bibinfo{person}{Arvind Neelakantan}, \bibinfo{person}{Pranav Shyam}, \bibinfo{person}{Girish Sastry}, \bibinfo{person}{Amanda Askell}, \bibinfo{person}{Sandhini Agarwal}, \bibinfo{person}{Ariel Herbert{-}Voss}, \bibinfo{person}{Gretchen Krueger}, \bibinfo{person}{Tom Henighan}, \bibinfo{person}{Rewon Child}, \bibinfo{person}{Aditya Ramesh}, \bibinfo{person}{Daniel~M. Ziegler}, \bibinfo{person}{Jeffrey Wu}, \bibinfo{person}{Clemens Winter}, \bibinfo{person}{Christopher Hesse}, \bibinfo{person}{Mark Chen}, \bibinfo{person}{Eric Sigler}, \bibinfo{person}{Mateusz Litwin}, \bibinfo{person}{Scott Gray}, \bibinfo{person}{Benjamin Chess}, \bibinfo{person}{Jack Clark}, \bibinfo{person}{Christopher Berner}, \bibinfo{person}{Sam McCandlish}, \bibinfo{person}{Alec Radford}, \bibinfo{person}{Ilya Sutskever},
  {and} \bibinfo{person}{Dario Amodei}.} \bibinfo{year}{2020}\natexlab{}.
\newblock \showarticletitle{Language Models are Few-Shot Learners}. In \bibinfo{booktitle}{{\em NeurIPS}}.
\newblock
\showURL{%
\url{https://dl.acm.org/doi/abs/10.5555/3495724.3495883}}


\bibitem[\protect\citeauthoryear{Dao}{Dao}{2023}]%
        {DBLP:flash_attn_2}
\bibfield{author}{\bibinfo{person}{Tri Dao}.} \bibinfo{year}{2023}\natexlab{}.
\newblock \showarticletitle{FlashAttention-2: Faster Attention with Better Parallelism and Work Partitioning}.
\newblock \bibinfo{journal}{{\em CoRR\/}}  \bibinfo{volume}{abs/2307.08691} (\bibinfo{year}{2023}).
\newblock
\showDOI{%
\url{https://doi.org/10.48550/ARXIV.2307.08691}}
\showeprint{2307.08691}


\bibitem[\protect\citeauthoryear{Dao, Fu, Ermon, Rudra, and R{\'{e}}}{Dao et~al\mbox{.}}{2022}]%
        {DBLP:flash_attn}
\bibfield{author}{\bibinfo{person}{Tri Dao}, \bibinfo{person}{Daniel~Y. Fu}, \bibinfo{person}{Stefano Ermon}, \bibinfo{person}{Atri Rudra}, {and} \bibinfo{person}{Christopher R{\'{e}}}.} \bibinfo{year}{2022}\natexlab{}.
\newblock \showarticletitle{FlashAttention: Fast and Memory-Efficient Exact Attention with IO-Awareness}. In \bibinfo{booktitle}{{\em Advances in Neural Information Processing Systems 35: Annual Conference on Neural Information Processing Systems 2022, NeurIPS 2022, New Orleans, LA, USA, November 28 - December 9, 2022}}, \bibfield{editor}{\bibinfo{person}{Sanmi Koyejo}, \bibinfo{person}{S.~Mohamed}, \bibinfo{person}{A.~Agarwal}, \bibinfo{person}{Danielle Belgrave}, \bibinfo{person}{K.~Cho}, {and} \bibinfo{person}{A.~Oh}} (Eds.).
\newblock
\showURL{%
\url{https://dl.acm.org/doi/10.5555/3600270.3601459}}


\bibitem[\protect\citeauthoryear{DeepSeek-AI, Liu, Feng, Wang, Wang, Liu, Zhao, Dengr, Ruan, Dai, Guo, Yang, Chen, Ji, Li, Lin, Luo, Hao, Chen, Li, Zhang, Xu, Yang, Zhang, Ding, Xin, Gao, Li, Qu, Cai, Liang, Guo, Ni, Li, Chen, Yuan, Qiu, Song, Dong, Gao, Guan, Wang, Zhang, Xu, Xia, Zhao, Zhang, Li, Wang, Zhang, Zhang, Tang, Li, Tian, Huang, Wang, Zhang, Zhu, Chen, Du, Chen, Jin, Ge, Pan, Xu, Chen, Li, Lu, Zhou, Chen, Wu, Ye, Ma, Wang, Zhou, Yu, Zhou, Zheng, Wang, Pei, Yuan, Sun, Xiao, Zeng, An, Liu, Liang, Gao, Zhang, Li, Jin, Wang, Bi, Liu, Wang, Shen, Chen, Chen, Nie, Sun, Wang, Liu, Xie, Yu, Song, Zhou, Yang, Lu, Su, Wu, Li, Wei, Zhu, Xu, Huang, Li, Zhao, Sun, Li, Wang, Zheng, Zhang, Xiong, Zhao, He, Tang, Piao, Dong, Tan, Liu, Wang, Guo, Zhu, Wang, Zou, Zha, Ma, Yan, You, Liu, Ren, Ren, Sha, Fu, Huang, Zhang, Xie, Hao, Shao, Wen, Xu, Zhang, Li, Wang, Gu, Li, and Xie}{DeepSeek-AI et~al\mbox{.}}{2024}]%
        {deepseekai2024deepseekv2strongeconomicalefficient}
\bibfield{author}{\bibinfo{person}{DeepSeek-AI}, \bibinfo{person}{Aixin Liu}, \bibinfo{person}{Bei Feng}, \bibinfo{person}{Bin Wang}, \bibinfo{person}{Bingxuan Wang}, \bibinfo{person}{Bo Liu}, \bibinfo{person}{Chenggang Zhao}, \bibinfo{person}{Chengqi Dengr}, \bibinfo{person}{Chong Ruan}, \bibinfo{person}{Damai Dai}, \bibinfo{person}{Daya Guo}, \bibinfo{person}{Dejian Yang}, \bibinfo{person}{Deli Chen}, \bibinfo{person}{Dongjie Ji}, \bibinfo{person}{Erhang Li}, \bibinfo{person}{Fangyun Lin}, \bibinfo{person}{Fuli Luo}, \bibinfo{person}{Guangbo Hao}, \bibinfo{person}{Guanting Chen}, \bibinfo{person}{Guowei Li}, \bibinfo{person}{H. Zhang}, \bibinfo{person}{Hanwei Xu}, \bibinfo{person}{Hao Yang}, \bibinfo{person}{Haowei Zhang}, \bibinfo{person}{Honghui Ding}, \bibinfo{person}{Huajian Xin}, \bibinfo{person}{Huazuo Gao}, \bibinfo{person}{Hui Li}, \bibinfo{person}{Hui Qu}, \bibinfo{person}{J.~L. Cai}, \bibinfo{person}{Jian Liang}, \bibinfo{person}{Jianzhong Guo}, \bibinfo{person}{Jiaqi Ni}, \bibinfo{person}{Jiashi
  Li}, \bibinfo{person}{Jin Chen}, \bibinfo{person}{Jingyang Yuan}, \bibinfo{person}{Junjie Qiu}, \bibinfo{person}{Junxiao Song}, \bibinfo{person}{Kai Dong}, \bibinfo{person}{Kaige Gao}, \bibinfo{person}{Kang Guan}, \bibinfo{person}{Lean Wang}, \bibinfo{person}{Lecong Zhang}, \bibinfo{person}{Lei Xu}, \bibinfo{person}{Leyi Xia}, \bibinfo{person}{Liang Zhao}, \bibinfo{person}{Liyue Zhang}, \bibinfo{person}{Meng Li}, \bibinfo{person}{Miaojun Wang}, \bibinfo{person}{Mingchuan Zhang}, \bibinfo{person}{Minghua Zhang}, \bibinfo{person}{Minghui Tang}, \bibinfo{person}{Mingming Li}, \bibinfo{person}{Ning Tian}, \bibinfo{person}{Panpan Huang}, \bibinfo{person}{Peiyi Wang}, \bibinfo{person}{Peng Zhang}, \bibinfo{person}{Qihao Zhu}, \bibinfo{person}{Qinyu Chen}, \bibinfo{person}{Qiushi Du}, \bibinfo{person}{R.~J. Chen}, \bibinfo{person}{R.~L. Jin}, \bibinfo{person}{Ruiqi Ge}, \bibinfo{person}{Ruizhe Pan}, \bibinfo{person}{Runxin Xu}, \bibinfo{person}{Ruyi Chen}, \bibinfo{person}{S.~S. Li}, \bibinfo{person}{Shanghao Lu},
  \bibinfo{person}{Shangyan Zhou}, \bibinfo{person}{Shanhuang Chen}, \bibinfo{person}{Shaoqing Wu}, \bibinfo{person}{Shengfeng Ye}, \bibinfo{person}{Shirong Ma}, \bibinfo{person}{Shiyu Wang}, \bibinfo{person}{Shuang Zhou}, \bibinfo{person}{Shuiping Yu}, \bibinfo{person}{Shunfeng Zhou}, \bibinfo{person}{Size Zheng}, \bibinfo{person}{T. Wang}, \bibinfo{person}{Tian Pei}, \bibinfo{person}{Tian Yuan}, \bibinfo{person}{Tianyu Sun}, \bibinfo{person}{W.~L. Xiao}, \bibinfo{person}{Wangding Zeng}, \bibinfo{person}{Wei An}, \bibinfo{person}{Wen Liu}, \bibinfo{person}{Wenfeng Liang}, \bibinfo{person}{Wenjun Gao}, \bibinfo{person}{Wentao Zhang}, \bibinfo{person}{X.~Q. Li}, \bibinfo{person}{Xiangyue Jin}, \bibinfo{person}{Xianzu Wang}, \bibinfo{person}{Xiao Bi}, \bibinfo{person}{Xiaodong Liu}, \bibinfo{person}{Xiaohan Wang}, \bibinfo{person}{Xiaojin Shen}, \bibinfo{person}{Xiaokang Chen}, \bibinfo{person}{Xiaosha Chen}, \bibinfo{person}{Xiaotao Nie}, \bibinfo{person}{Xiaowen Sun}, \bibinfo{person}{Xiaoxiang Wang},
  \bibinfo{person}{Xin Liu}, \bibinfo{person}{Xin Xie}, \bibinfo{person}{Xingkai Yu}, \bibinfo{person}{Xinnan Song}, \bibinfo{person}{Xinyi Zhou}, \bibinfo{person}{Xinyu Yang}, \bibinfo{person}{Xuan Lu}, \bibinfo{person}{Xuecheng Su}, \bibinfo{person}{Y. Wu}, \bibinfo{person}{Y.~K. Li}, \bibinfo{person}{Y.~X. Wei}, \bibinfo{person}{Y.~X. Zhu}, \bibinfo{person}{Yanhong Xu}, \bibinfo{person}{Yanping Huang}, \bibinfo{person}{Yao Li}, \bibinfo{person}{Yao Zhao}, \bibinfo{person}{Yaofeng Sun}, \bibinfo{person}{Yaohui Li}, \bibinfo{person}{Yaohui Wang}, \bibinfo{person}{Yi Zheng}, \bibinfo{person}{Yichao Zhang}, \bibinfo{person}{Yiliang Xiong}, \bibinfo{person}{Yilong Zhao}, \bibinfo{person}{Ying He}, \bibinfo{person}{Ying Tang}, \bibinfo{person}{Yishi Piao}, \bibinfo{person}{Yixin Dong}, \bibinfo{person}{Yixuan Tan}, \bibinfo{person}{Yiyuan Liu}, \bibinfo{person}{Yongji Wang}, \bibinfo{person}{Yongqiang Guo}, \bibinfo{person}{Yuchen Zhu}, \bibinfo{person}{Yuduan Wang}, \bibinfo{person}{Yuheng Zou},
  \bibinfo{person}{Yukun Zha}, \bibinfo{person}{Yunxian Ma}, \bibinfo{person}{Yuting Yan}, \bibinfo{person}{Yuxiang You}, \bibinfo{person}{Yuxuan Liu}, \bibinfo{person}{Z.~Z. Ren}, \bibinfo{person}{Zehui Ren}, \bibinfo{person}{Zhangli Sha}, \bibinfo{person}{Zhe Fu}, \bibinfo{person}{Zhen Huang}, \bibinfo{person}{Zhen Zhang}, \bibinfo{person}{Zhenda Xie}, \bibinfo{person}{Zhewen Hao}, \bibinfo{person}{Zhihong Shao}, \bibinfo{person}{Zhiniu Wen}, \bibinfo{person}{Zhipeng Xu}, \bibinfo{person}{Zhongyu Zhang}, \bibinfo{person}{Zhuoshu Li}, \bibinfo{person}{Zihan Wang}, \bibinfo{person}{Zihui Gu}, \bibinfo{person}{Zilin Li}, {and} \bibinfo{person}{Ziwei Xie}.} \bibinfo{year}{2024}\natexlab{}.
\newblock \bibinfo{title}{DeepSeek-V2: A Strong, Economical, and Efficient Mixture-of-Experts Language Model}.
\newblock   (\bibinfo{year}{2024}).
\newblock
\showeprint[arxiv]{cs.CL/2405.04434}
\showURL{%
\url{https://doi.org/10.48550/arXiv.2405.04434}}


\bibitem[\protect\citeauthoryear{Fan, Rong, Meng, Cao, Wang, Zheng, Wu, Long, Yang, Xia, Diao, Liu, and Lin}{Fan et~al\mbox{.}}{2021}]%
        {DBLP:conf/ppopp/dapple}
\bibfield{author}{\bibinfo{person}{Shiqing Fan}, \bibinfo{person}{Yi Rong}, \bibinfo{person}{Chen Meng}, \bibinfo{person}{Zongyan Cao}, \bibinfo{person}{Siyu Wang}, \bibinfo{person}{Zhen Zheng}, \bibinfo{person}{Chuan Wu}, \bibinfo{person}{Guoping Long}, \bibinfo{person}{Jun Yang}, \bibinfo{person}{Lixue Xia}, \bibinfo{person}{Lansong Diao}, \bibinfo{person}{Xiaoyong Liu}, {and} \bibinfo{person}{Wei Lin}.} \bibinfo{year}{2021}\natexlab{}.
\newblock \showarticletitle{{DAPPLE:} a pipelined data parallel approach for training large models}. In \bibinfo{booktitle}{{\em PPoPP}}. \bibinfo{publisher}{{ACM}}, \bibinfo{pages}{431--445}.
\newblock
\showURL{%
\url{https://dl.acm.org/doi/10.1145/3437801.3441593}}


\bibitem[\protect\citeauthoryear{Ge, Feng, Huang, Fu, Nie, Zuo, Lin, Cui, and Liu}{Ge et~al\mbox{.}}{2025}]%
        {ge2025bytescale}
\bibfield{author}{\bibinfo{person}{Hao Ge}, \bibinfo{person}{Junda Feng}, \bibinfo{person}{Qi Huang}, \bibinfo{person}{Fangcheng Fu}, \bibinfo{person}{Xiaonan Nie}, \bibinfo{person}{Lei Zuo}, \bibinfo{person}{Haibin Lin}, \bibinfo{person}{Bin Cui}, {and} \bibinfo{person}{Xin Liu}.} \bibinfo{year}{2025}\natexlab{}.
\newblock \showarticletitle{ByteScale: Efficient Scaling of LLM Training with a 2048K Context Length on More Than 12,000 GPUs}.
\newblock \bibinfo{journal}{{\em arXiv preprint arXiv:2502.21231\/}} (\bibinfo{year}{2025}).
\newblock
\showURL{%
\url{https://doi.org/10.48550/arXiv.2502.21231}}


\bibitem[\protect\citeauthoryear{Ge, Fu, Li, Wang, Lin, Wang, Nie, Zhang, Miao, and Cui}{Ge et~al\mbox{.}}{2024}]%
        {ge2024enabling}
\bibfield{author}{\bibinfo{person}{Hao Ge}, \bibinfo{person}{Fangcheng Fu}, \bibinfo{person}{Haoyang Li}, \bibinfo{person}{Xuanyu Wang}, \bibinfo{person}{Sheng Lin}, \bibinfo{person}{Yujie Wang}, \bibinfo{person}{Xiaonan Nie}, \bibinfo{person}{Hailin Zhang}, \bibinfo{person}{Xupeng Miao}, {and} \bibinfo{person}{Bin Cui}.} \bibinfo{year}{2024}\natexlab{}.
\newblock \showarticletitle{Enabling Parallelism Hot Switching for Efficient Training of Large Language Models}. In \bibinfo{booktitle}{{\em Proceedings of the ACM SIGOPS 30th Symposium on Operating Systems Principles}}. \bibinfo{pages}{178--194}.
\newblock
\showURL{%
\url{https://dl.acm.org/doi/10.1145/3694715.3695969}}


\bibitem[\protect\citeauthoryear{Grattafiori, Dubey, Jauhri, Pandey, Kadian, Al-Dahle, Letman, Mathur, Schelten, Vaughan, Yang, Fan, Goyal, Hartshorn, Yang, Mitra, Sravankumar, Korenev, Hinsvark, Rao, Zhang, Rodriguez, Gregerson, Spataru, Roziere, Biron, Tang, Chern, Caucheteux, Nayak, Bi, Marra, McConnell, Keller, Touret, Wu, Wong, Ferrer, Nikolaidis, Allonsius, Song, Pintz, Livshits, Wyatt, Esiobu, Choudhary, Mahajan, Garcia-Olano, Perino, Hupkes, Lakomkin, AlBadawy, Lobanova, Dinan, Smith, Radenovic, Guzmán, Zhang, Synnaeve, Lee, Anderson, Thattai, Nail, Mialon, Pang, Cucurell, Nguyen, Korevaar, Xu, Touvron, Zarov, Ibarra, Kloumann, Misra, Evtimov, Zhang, Copet, Lee, Geffert, Vranes, Park, Mahadeokar, Shah, Linde, Billock, Hong, Lee, Fu, Chi, Huang, Liu, Wang, Yu, Bitton, Spisak, Park, Rocca, Johnstun, Saxe, Jia, Alwala, Prasad, Upasani, Plawiak, Li, Heafield, Stone, El-Arini, Iyer, Malik, Chiu, Bhalla, Lakhotia, Rantala-Yeary, Maaten, Chen, Tan, Jenkins, Martin, Madaan, Malo, Blecher, Landzaat, Oliveira,
  Muzzi, Pasupuleti, Singh, Paluri, Kardas, Tsimpoukelli, Oldham, Rita, Pavlova, Kambadur, Lewis, Si, Singh, Hassan, Goyal, Torabi, Bashlykov, Bogoychev, Chatterji, Zhang, Duchenne, Çelebi, Alrassy, Zhang, Li, Vasic, Weng, Bhargava, Dubal, Krishnan, Koura, Xu, He, Dong, Srinivasan, Ganapathy, Calderer, Cabral, Stojnic, Raileanu, Maheswari, Girdhar, Patel, Sauvestre, Polidoro, Sumbaly, Taylor, Silva, Hou, Wang, Hosseini, Chennabasappa, Singh, Bell, Kim, Edunov, Nie, Narang, Raparthy, Shen, Wan, Bhosale, Zhang, Vandenhende, Batra, Whitman, Sootla, Collot, Gururangan, Borodinsky, Herman, Fowler, Sheasha, Georgiou, Scialom, Speckbacher, Mihaylov, Xiao, Karn, Goswami, Gupta, Ramanathan, Kerkez, Gonguet, Do, Vogeti, Albiero, Petrovic, Chu, Xiong, Fu, Meers, Martinet, Wang, Wang, Tan, Xia, Xie, Jia, Wang, Goldschlag, Gaur, Babaei, Wen, Song, Zhang, Li, Mao, Coudert, Yan, Chen, Papakipos, Singh, Srivastava, Jain, Kelsey, Shajnfeld, Gangidi, Victoria, Goldstand, Menon, Sharma, Boesenberg, Baevski, Feinstein, Kallet,
  Sangani, Teo, Yunus, Lupu, Alvarado, Caples, Gu, Ho, Poulton, Ryan, Ramchandani, Dong, Franco, Goyal, Saraf, Chowdhury, Gabriel, Bharambe, Eisenman, Yazdan, James, Maurer, Leonhardi, Huang, Loyd, Paola, Paranjape, Liu, Wu, Ni, Hancock, Wasti, Spence, Stojkovic, Gamido, Montalvo, Parker, Burton, Mejia, Liu, Wang, Kim, Zhou, Hu, Chu, Cai, Tindal, Feichtenhofer, Gao, Civin, Beaty, Kreymer, Li, Adkins, Xu, Testuggine, David, Parikh, Liskovich, Foss, Wang, Le, Holland, Dowling, Jamil, Montgomery, Presani, Hahn, Wood, Le, Brinkman, Arcaute, Dunbar, Smothers, Sun, Kreuk, Tian, Kokkinos, Ozgenel, Caggioni, Kanayet, Seide, Florez, Schwarz, Badeer, Swee, Halpern, Herman, Sizov, Guangyi, Zhang, Lakshminarayanan, Inan, Shojanazeri, Zou, Wang, Zha, Habeeb, Rudolph, Suk, Aspegren, Goldman, Zhan, Damlaj, Molybog, Tufanov, Leontiadis, Veliche, Gat, Weissman, Geboski, Kohli, Lam, Asher, Gaya, Marcus, Tang, Chan, Zhen, Reizenstein, Teboul, Zhong, Jin, Yang, Cummings, Carvill, Shepard, McPhie, Torres, Ginsburg, Wang, Wu, U,
  Saxena, Khandelwal, Zand, Matosich, Veeraraghavan, Michelena, Li, Jagadeesh, Huang, Chawla, Huang, Chen, Garg, A, Silva, Bell, Zhang, Guo, Yu, Moshkovich, Wehrstedt, Khabsa, Avalani, Bhatt, Mankus, Hasson, Lennie, Reso, Groshev, Naumov, Lathi, Keneally, Liu, Seltzer, Valko, Restrepo, Patel, Vyatskov, Samvelyan, Clark, Macey, Wang, Hermoso, Metanat, Rastegari, Bansal, Santhanam, Parks, White, Bawa, Singhal, Egebo, Usunier, Mehta, Laptev, Dong, Cheng, Chernoguz, Hart, Salpekar, Kalinli, Kent, Parekh, Saab, Balaji, Rittner, Bontrager, Roux, Dollar, Zvyagina, Ratanchandani, Yuvraj, Liang, Alao, Rodriguez, Ayub, Murthy, Nayani, Mitra, Parthasarathy, Li, Hogan, Battey, Wang, Howes, Rinott, Mehta, Siby, Bondu, Datta, Chugh, Hunt, Dhillon, Sidorov, Pan, Mahajan, Verma, Yamamoto, Ramaswamy, Lindsay, Lindsay, Feng, Lin, Zha, Patil, Shankar, Zhang, Zhang, Wang, Agarwal, Sajuyigbe, Chintala, Max, Chen, Kehoe, Satterfield, Govindaprasad, Gupta, Deng, Cho, Virk, Subramanian, Choudhury, Goldman, Remez, Glaser, Best,
  Koehler, Robinson, Li, Zhang, Matthews, Chou, Shaked, Vontimitta, Ajayi, Montanez, Mohan, Kumar, Mangla, Ionescu, Poenaru, Mihailescu, Ivanov, Li, Wang, Jiang, Bouaziz, Constable, Tang, Wu, Wang, Wu, Gao, Kleinman, Chen, Hu, Jia, Qi, Li, Zhang, Zhang, Adi, Nam, Yu, Wang, Zhao, Hao, Qian, Li, He, Rait, DeVito, Rosnbrick, Wen, Yang, Zhao, and Ma}{Grattafiori et~al\mbox{.}}{2024}]%
        {llama3}
\bibfield{author}{\bibinfo{person}{Aaron Grattafiori}, \bibinfo{person}{Abhimanyu Dubey}, \bibinfo{person}{Abhinav Jauhri}, \bibinfo{person}{Abhinav Pandey}, \bibinfo{person}{Abhishek Kadian}, \bibinfo{person}{Ahmad Al-Dahle}, \bibinfo{person}{Aiesha Letman}, \bibinfo{person}{Akhil Mathur}, \bibinfo{person}{Alan Schelten}, \bibinfo{person}{Alex Vaughan}, \bibinfo{person}{Amy Yang}, \bibinfo{person}{Angela Fan}, \bibinfo{person}{Anirudh Goyal}, \bibinfo{person}{Anthony Hartshorn}, \bibinfo{person}{Aobo Yang}, \bibinfo{person}{Archi Mitra}, \bibinfo{person}{Archie Sravankumar}, \bibinfo{person}{Artem Korenev}, \bibinfo{person}{Arthur Hinsvark}, \bibinfo{person}{Arun Rao}, \bibinfo{person}{Aston Zhang}, \bibinfo{person}{Aurelien Rodriguez}, \bibinfo{person}{Austen Gregerson}, \bibinfo{person}{Ava Spataru}, \bibinfo{person}{Baptiste Roziere}, \bibinfo{person}{Bethany Biron}, \bibinfo{person}{Binh Tang}, \bibinfo{person}{Bobbie Chern}, \bibinfo{person}{Charlotte Caucheteux}, \bibinfo{person}{Chaya Nayak},
  \bibinfo{person}{Chloe Bi}, \bibinfo{person}{Chris Marra}, \bibinfo{person}{Chris McConnell}, \bibinfo{person}{Christian Keller}, \bibinfo{person}{Christophe Touret}, \bibinfo{person}{Chunyang Wu}, \bibinfo{person}{Corinne Wong}, \bibinfo{person}{Cristian~Canton Ferrer}, \bibinfo{person}{Cyrus Nikolaidis}, \bibinfo{person}{Damien Allonsius}, \bibinfo{person}{Daniel Song}, \bibinfo{person}{Danielle Pintz}, \bibinfo{person}{Danny Livshits}, \bibinfo{person}{Danny Wyatt}, \bibinfo{person}{David Esiobu}, \bibinfo{person}{Dhruv Choudhary}, \bibinfo{person}{Dhruv Mahajan}, \bibinfo{person}{Diego Garcia-Olano}, \bibinfo{person}{Diego Perino}, \bibinfo{person}{Dieuwke Hupkes}, \bibinfo{person}{Egor Lakomkin}, \bibinfo{person}{Ehab AlBadawy}, \bibinfo{person}{Elina Lobanova}, \bibinfo{person}{Emily Dinan}, \bibinfo{person}{Eric~Michael Smith}, \bibinfo{person}{Filip Radenovic}, \bibinfo{person}{Francisco Guzmán}, \bibinfo{person}{Frank Zhang}, \bibinfo{person}{Gabriel Synnaeve}, \bibinfo{person}{Gabrielle Lee},
  \bibinfo{person}{Georgia~Lewis Anderson}, \bibinfo{person}{Govind Thattai}, \bibinfo{person}{Graeme Nail}, \bibinfo{person}{Gregoire Mialon}, \bibinfo{person}{Guan Pang}, \bibinfo{person}{Guillem Cucurell}, \bibinfo{person}{Hailey Nguyen}, \bibinfo{person}{Hannah Korevaar}, \bibinfo{person}{Hu Xu}, \bibinfo{person}{Hugo Touvron}, \bibinfo{person}{Iliyan Zarov}, \bibinfo{person}{Imanol~Arrieta Ibarra}, \bibinfo{person}{Isabel Kloumann}, \bibinfo{person}{Ishan Misra}, \bibinfo{person}{Ivan Evtimov}, \bibinfo{person}{Jack Zhang}, \bibinfo{person}{Jade Copet}, \bibinfo{person}{Jaewon Lee}, \bibinfo{person}{Jan Geffert}, \bibinfo{person}{Jana Vranes}, \bibinfo{person}{Jason Park}, \bibinfo{person}{Jay Mahadeokar}, \bibinfo{person}{Jeet Shah}, \bibinfo{person}{Jelmer van~der Linde}, \bibinfo{person}{Jennifer Billock}, \bibinfo{person}{Jenny Hong}, \bibinfo{person}{Jenya Lee}, \bibinfo{person}{Jeremy Fu}, \bibinfo{person}{Jianfeng Chi}, \bibinfo{person}{Jianyu Huang}, \bibinfo{person}{Jiawen Liu},
  \bibinfo{person}{Jie Wang}, \bibinfo{person}{Jiecao Yu}, \bibinfo{person}{Joanna Bitton}, \bibinfo{person}{Joe Spisak}, \bibinfo{person}{Jongsoo Park}, \bibinfo{person}{Joseph Rocca}, \bibinfo{person}{Joshua Johnstun}, \bibinfo{person}{Joshua Saxe}, \bibinfo{person}{Junteng Jia}, \bibinfo{person}{Kalyan~Vasuden Alwala}, \bibinfo{person}{Karthik Prasad}, \bibinfo{person}{Kartikeya Upasani}, \bibinfo{person}{Kate Plawiak}, \bibinfo{person}{Ke Li}, \bibinfo{person}{Kenneth Heafield}, \bibinfo{person}{Kevin Stone}, \bibinfo{person}{Khalid El-Arini}, \bibinfo{person}{Krithika Iyer}, \bibinfo{person}{Kshitiz Malik}, \bibinfo{person}{Kuenley Chiu}, \bibinfo{person}{Kunal Bhalla}, \bibinfo{person}{Kushal Lakhotia}, \bibinfo{person}{Lauren Rantala-Yeary}, \bibinfo{person}{Laurens van~der Maaten}, \bibinfo{person}{Lawrence Chen}, \bibinfo{person}{Liang Tan}, \bibinfo{person}{Liz Jenkins}, \bibinfo{person}{Louis Martin}, \bibinfo{person}{Lovish Madaan}, \bibinfo{person}{Lubo Malo}, \bibinfo{person}{Lukas Blecher},
  \bibinfo{person}{Lukas Landzaat}, \bibinfo{person}{Luke~de Oliveira}, \bibinfo{person}{Madeline Muzzi}, \bibinfo{person}{Mahesh Pasupuleti}, \bibinfo{person}{Mannat Singh}, \bibinfo{person}{Manohar Paluri}, \bibinfo{person}{Marcin Kardas}, \bibinfo{person}{Maria Tsimpoukelli}, \bibinfo{person}{Mathew Oldham}, \bibinfo{person}{Mathieu Rita}, \bibinfo{person}{Maya Pavlova}, \bibinfo{person}{Melanie Kambadur}, \bibinfo{person}{Mike Lewis}, \bibinfo{person}{Min Si}, \bibinfo{person}{Mitesh~Kumar Singh}, \bibinfo{person}{Mona Hassan}, \bibinfo{person}{Naman Goyal}, \bibinfo{person}{Narjes Torabi}, \bibinfo{person}{Nikolay Bashlykov}, \bibinfo{person}{Nikolay Bogoychev}, \bibinfo{person}{Niladri Chatterji}, \bibinfo{person}{Ning Zhang}, \bibinfo{person}{Olivier Duchenne}, \bibinfo{person}{Onur Çelebi}, \bibinfo{person}{Patrick Alrassy}, \bibinfo{person}{Pengchuan Zhang}, \bibinfo{person}{Pengwei Li}, \bibinfo{person}{Petar Vasic}, \bibinfo{person}{Peter Weng}, \bibinfo{person}{Prajjwal Bhargava},
  \bibinfo{person}{Pratik Dubal}, \bibinfo{person}{Praveen Krishnan}, \bibinfo{person}{Punit~Singh Koura}, \bibinfo{person}{Puxin Xu}, \bibinfo{person}{Qing He}, \bibinfo{person}{Qingxiao Dong}, \bibinfo{person}{Ragavan Srinivasan}, \bibinfo{person}{Raj Ganapathy}, \bibinfo{person}{Ramon Calderer}, \bibinfo{person}{Ricardo~Silveira Cabral}, \bibinfo{person}{Robert Stojnic}, \bibinfo{person}{Roberta Raileanu}, \bibinfo{person}{Rohan Maheswari}, \bibinfo{person}{Rohit Girdhar}, \bibinfo{person}{Rohit Patel}, \bibinfo{person}{Romain Sauvestre}, \bibinfo{person}{Ronnie Polidoro}, \bibinfo{person}{Roshan Sumbaly}, \bibinfo{person}{Ross Taylor}, \bibinfo{person}{Ruan Silva}, \bibinfo{person}{Rui Hou}, \bibinfo{person}{Rui Wang}, \bibinfo{person}{Saghar Hosseini}, \bibinfo{person}{Sahana Chennabasappa}, \bibinfo{person}{Sanjay Singh}, \bibinfo{person}{Sean Bell}, \bibinfo{person}{Seohyun~Sonia Kim}, \bibinfo{person}{Sergey Edunov}, \bibinfo{person}{Shaoliang Nie}, \bibinfo{person}{Sharan Narang},
  \bibinfo{person}{Sharath Raparthy}, \bibinfo{person}{Sheng Shen}, \bibinfo{person}{Shengye Wan}, \bibinfo{person}{Shruti Bhosale}, \bibinfo{person}{Shun Zhang}, \bibinfo{person}{Simon Vandenhende}, \bibinfo{person}{Soumya Batra}, \bibinfo{person}{Spencer Whitman}, \bibinfo{person}{Sten Sootla}, \bibinfo{person}{Stephane Collot}, \bibinfo{person}{Suchin Gururangan}, \bibinfo{person}{Sydney Borodinsky}, \bibinfo{person}{Tamar Herman}, \bibinfo{person}{Tara Fowler}, \bibinfo{person}{Tarek Sheasha}, \bibinfo{person}{Thomas Georgiou}, \bibinfo{person}{Thomas Scialom}, \bibinfo{person}{Tobias Speckbacher}, \bibinfo{person}{Todor Mihaylov}, \bibinfo{person}{Tong Xiao}, \bibinfo{person}{Ujjwal Karn}, \bibinfo{person}{Vedanuj Goswami}, \bibinfo{person}{Vibhor Gupta}, \bibinfo{person}{Vignesh Ramanathan}, \bibinfo{person}{Viktor Kerkez}, \bibinfo{person}{Vincent Gonguet}, \bibinfo{person}{Virginie Do}, \bibinfo{person}{Vish Vogeti}, \bibinfo{person}{Vítor Albiero}, \bibinfo{person}{Vladan Petrovic},
  \bibinfo{person}{Weiwei Chu}, \bibinfo{person}{Wenhan Xiong}, \bibinfo{person}{Wenyin Fu}, \bibinfo{person}{Whitney Meers}, \bibinfo{person}{Xavier Martinet}, \bibinfo{person}{Xiaodong Wang}, \bibinfo{person}{Xiaofang Wang}, \bibinfo{person}{Xiaoqing~Ellen Tan}, \bibinfo{person}{Xide Xia}, \bibinfo{person}{Xinfeng Xie}, \bibinfo{person}{Xuchao Jia}, \bibinfo{person}{Xuewei Wang}, \bibinfo{person}{Yaelle Goldschlag}, \bibinfo{person}{Yashesh Gaur}, \bibinfo{person}{Yasmine Babaei}, \bibinfo{person}{Yi Wen}, \bibinfo{person}{Yiwen Song}, \bibinfo{person}{Yuchen Zhang}, \bibinfo{person}{Yue Li}, \bibinfo{person}{Yuning Mao}, \bibinfo{person}{Zacharie~Delpierre Coudert}, \bibinfo{person}{Zheng Yan}, \bibinfo{person}{Zhengxing Chen}, \bibinfo{person}{Zoe Papakipos}, \bibinfo{person}{Aaditya Singh}, \bibinfo{person}{Aayushi Srivastava}, \bibinfo{person}{Abha Jain}, \bibinfo{person}{Adam Kelsey}, \bibinfo{person}{Adam Shajnfeld}, \bibinfo{person}{Adithya Gangidi}, \bibinfo{person}{Adolfo Victoria},
  \bibinfo{person}{Ahuva Goldstand}, \bibinfo{person}{Ajay Menon}, \bibinfo{person}{Ajay Sharma}, \bibinfo{person}{Alex Boesenberg}, \bibinfo{person}{Alexei Baevski}, \bibinfo{person}{Allie Feinstein}, \bibinfo{person}{Amanda Kallet}, \bibinfo{person}{Amit Sangani}, \bibinfo{person}{Amos Teo}, \bibinfo{person}{Anam Yunus}, \bibinfo{person}{Andrei Lupu}, \bibinfo{person}{Andres Alvarado}, \bibinfo{person}{Andrew Caples}, \bibinfo{person}{Andrew Gu}, \bibinfo{person}{Andrew Ho}, \bibinfo{person}{Andrew Poulton}, \bibinfo{person}{Andrew Ryan}, \bibinfo{person}{Ankit Ramchandani}, \bibinfo{person}{Annie Dong}, \bibinfo{person}{Annie Franco}, \bibinfo{person}{Anuj Goyal}, \bibinfo{person}{Aparajita Saraf}, \bibinfo{person}{Arkabandhu Chowdhury}, \bibinfo{person}{Ashley Gabriel}, \bibinfo{person}{Ashwin Bharambe}, \bibinfo{person}{Assaf Eisenman}, \bibinfo{person}{Azadeh Yazdan}, \bibinfo{person}{Beau James}, \bibinfo{person}{Ben Maurer}, \bibinfo{person}{Benjamin Leonhardi}, \bibinfo{person}{Bernie Huang},
  \bibinfo{person}{Beth Loyd}, \bibinfo{person}{Beto~De Paola}, \bibinfo{person}{Bhargavi Paranjape}, \bibinfo{person}{Bing Liu}, \bibinfo{person}{Bo Wu}, \bibinfo{person}{Boyu Ni}, \bibinfo{person}{Braden Hancock}, \bibinfo{person}{Bram Wasti}, \bibinfo{person}{Brandon Spence}, \bibinfo{person}{Brani Stojkovic}, \bibinfo{person}{Brian Gamido}, \bibinfo{person}{Britt Montalvo}, \bibinfo{person}{Carl Parker}, \bibinfo{person}{Carly Burton}, \bibinfo{person}{Catalina Mejia}, \bibinfo{person}{Ce Liu}, \bibinfo{person}{Changhan Wang}, \bibinfo{person}{Changkyu Kim}, \bibinfo{person}{Chao Zhou}, \bibinfo{person}{Chester Hu}, \bibinfo{person}{Ching-Hsiang Chu}, \bibinfo{person}{Chris Cai}, \bibinfo{person}{Chris Tindal}, \bibinfo{person}{Christoph Feichtenhofer}, \bibinfo{person}{Cynthia Gao}, \bibinfo{person}{Damon Civin}, \bibinfo{person}{Dana Beaty}, \bibinfo{person}{Daniel Kreymer}, \bibinfo{person}{Daniel Li}, \bibinfo{person}{David Adkins}, \bibinfo{person}{David Xu}, \bibinfo{person}{Davide Testuggine},
  \bibinfo{person}{Delia David}, \bibinfo{person}{Devi Parikh}, \bibinfo{person}{Diana Liskovich}, \bibinfo{person}{Didem Foss}, \bibinfo{person}{Dingkang Wang}, \bibinfo{person}{Duc Le}, \bibinfo{person}{Dustin Holland}, \bibinfo{person}{Edward Dowling}, \bibinfo{person}{Eissa Jamil}, \bibinfo{person}{Elaine Montgomery}, \bibinfo{person}{Eleonora Presani}, \bibinfo{person}{Emily Hahn}, \bibinfo{person}{Emily Wood}, \bibinfo{person}{Eric-Tuan Le}, \bibinfo{person}{Erik Brinkman}, \bibinfo{person}{Esteban Arcaute}, \bibinfo{person}{Evan Dunbar}, \bibinfo{person}{Evan Smothers}, \bibinfo{person}{Fei Sun}, \bibinfo{person}{Felix Kreuk}, \bibinfo{person}{Feng Tian}, \bibinfo{person}{Filippos Kokkinos}, \bibinfo{person}{Firat Ozgenel}, \bibinfo{person}{Francesco Caggioni}, \bibinfo{person}{Frank Kanayet}, \bibinfo{person}{Frank Seide}, \bibinfo{person}{Gabriela~Medina Florez}, \bibinfo{person}{Gabriella Schwarz}, \bibinfo{person}{Gada Badeer}, \bibinfo{person}{Georgia Swee}, \bibinfo{person}{Gil Halpern},
  \bibinfo{person}{Grant Herman}, \bibinfo{person}{Grigory Sizov}, \bibinfo{person}{Guangyi}, \bibinfo{person}{Zhang}, \bibinfo{person}{Guna Lakshminarayanan}, \bibinfo{person}{Hakan Inan}, \bibinfo{person}{Hamid Shojanazeri}, \bibinfo{person}{Han Zou}, \bibinfo{person}{Hannah Wang}, \bibinfo{person}{Hanwen Zha}, \bibinfo{person}{Haroun Habeeb}, \bibinfo{person}{Harrison Rudolph}, \bibinfo{person}{Helen Suk}, \bibinfo{person}{Henry Aspegren}, \bibinfo{person}{Hunter Goldman}, \bibinfo{person}{Hongyuan Zhan}, \bibinfo{person}{Ibrahim Damlaj}, \bibinfo{person}{Igor Molybog}, \bibinfo{person}{Igor Tufanov}, \bibinfo{person}{Ilias Leontiadis}, \bibinfo{person}{Irina-Elena Veliche}, \bibinfo{person}{Itai Gat}, \bibinfo{person}{Jake Weissman}, \bibinfo{person}{James Geboski}, \bibinfo{person}{James Kohli}, \bibinfo{person}{Janice Lam}, \bibinfo{person}{Japhet Asher}, \bibinfo{person}{Jean-Baptiste Gaya}, \bibinfo{person}{Jeff Marcus}, \bibinfo{person}{Jeff Tang}, \bibinfo{person}{Jennifer Chan},
  \bibinfo{person}{Jenny Zhen}, \bibinfo{person}{Jeremy Reizenstein}, \bibinfo{person}{Jeremy Teboul}, \bibinfo{person}{Jessica Zhong}, \bibinfo{person}{Jian Jin}, \bibinfo{person}{Jingyi Yang}, \bibinfo{person}{Joe Cummings}, \bibinfo{person}{Jon Carvill}, \bibinfo{person}{Jon Shepard}, \bibinfo{person}{Jonathan McPhie}, \bibinfo{person}{Jonathan Torres}, \bibinfo{person}{Josh Ginsburg}, \bibinfo{person}{Junjie Wang}, \bibinfo{person}{Kai Wu}, \bibinfo{person}{Kam~Hou U}, \bibinfo{person}{Karan Saxena}, \bibinfo{person}{Kartikay Khandelwal}, \bibinfo{person}{Katayoun Zand}, \bibinfo{person}{Kathy Matosich}, \bibinfo{person}{Kaushik Veeraraghavan}, \bibinfo{person}{Kelly Michelena}, \bibinfo{person}{Keqian Li}, \bibinfo{person}{Kiran Jagadeesh}, \bibinfo{person}{Kun Huang}, \bibinfo{person}{Kunal Chawla}, \bibinfo{person}{Kyle Huang}, \bibinfo{person}{Lailin Chen}, \bibinfo{person}{Lakshya Garg}, \bibinfo{person}{Lavender A}, \bibinfo{person}{Leandro Silva}, \bibinfo{person}{Lee Bell}, \bibinfo{person}{Lei
  Zhang}, \bibinfo{person}{Liangpeng Guo}, \bibinfo{person}{Licheng Yu}, \bibinfo{person}{Liron Moshkovich}, \bibinfo{person}{Luca Wehrstedt}, \bibinfo{person}{Madian Khabsa}, \bibinfo{person}{Manav Avalani}, \bibinfo{person}{Manish Bhatt}, \bibinfo{person}{Martynas Mankus}, \bibinfo{person}{Matan Hasson}, \bibinfo{person}{Matthew Lennie}, \bibinfo{person}{Matthias Reso}, \bibinfo{person}{Maxim Groshev}, \bibinfo{person}{Maxim Naumov}, \bibinfo{person}{Maya Lathi}, \bibinfo{person}{Meghan Keneally}, \bibinfo{person}{Miao Liu}, \bibinfo{person}{Michael~L. Seltzer}, \bibinfo{person}{Michal Valko}, \bibinfo{person}{Michelle Restrepo}, \bibinfo{person}{Mihir Patel}, \bibinfo{person}{Mik Vyatskov}, \bibinfo{person}{Mikayel Samvelyan}, \bibinfo{person}{Mike Clark}, \bibinfo{person}{Mike Macey}, \bibinfo{person}{Mike Wang}, \bibinfo{person}{Miquel~Jubert Hermoso}, \bibinfo{person}{Mo Metanat}, \bibinfo{person}{Mohammad Rastegari}, \bibinfo{person}{Munish Bansal}, \bibinfo{person}{Nandhini Santhanam},
  \bibinfo{person}{Natascha Parks}, \bibinfo{person}{Natasha White}, \bibinfo{person}{Navyata Bawa}, \bibinfo{person}{Nayan Singhal}, \bibinfo{person}{Nick Egebo}, \bibinfo{person}{Nicolas Usunier}, \bibinfo{person}{Nikhil Mehta}, \bibinfo{person}{Nikolay~Pavlovich Laptev}, \bibinfo{person}{Ning Dong}, \bibinfo{person}{Norman Cheng}, \bibinfo{person}{Oleg Chernoguz}, \bibinfo{person}{Olivia Hart}, \bibinfo{person}{Omkar Salpekar}, \bibinfo{person}{Ozlem Kalinli}, \bibinfo{person}{Parkin Kent}, \bibinfo{person}{Parth Parekh}, \bibinfo{person}{Paul Saab}, \bibinfo{person}{Pavan Balaji}, \bibinfo{person}{Pedro Rittner}, \bibinfo{person}{Philip Bontrager}, \bibinfo{person}{Pierre Roux}, \bibinfo{person}{Piotr Dollar}, \bibinfo{person}{Polina Zvyagina}, \bibinfo{person}{Prashant Ratanchandani}, \bibinfo{person}{Pritish Yuvraj}, \bibinfo{person}{Qian Liang}, \bibinfo{person}{Rachad Alao}, \bibinfo{person}{Rachel Rodriguez}, \bibinfo{person}{Rafi Ayub}, \bibinfo{person}{Raghotham Murthy}, \bibinfo{person}{Raghu
  Nayani}, \bibinfo{person}{Rahul Mitra}, \bibinfo{person}{Rangaprabhu Parthasarathy}, \bibinfo{person}{Raymond Li}, \bibinfo{person}{Rebekkah Hogan}, \bibinfo{person}{Robin Battey}, \bibinfo{person}{Rocky Wang}, \bibinfo{person}{Russ Howes}, \bibinfo{person}{Ruty Rinott}, \bibinfo{person}{Sachin Mehta}, \bibinfo{person}{Sachin Siby}, \bibinfo{person}{Sai~Jayesh Bondu}, \bibinfo{person}{Samyak Datta}, \bibinfo{person}{Sara Chugh}, \bibinfo{person}{Sara Hunt}, \bibinfo{person}{Sargun Dhillon}, \bibinfo{person}{Sasha Sidorov}, \bibinfo{person}{Satadru Pan}, \bibinfo{person}{Saurabh Mahajan}, \bibinfo{person}{Saurabh Verma}, \bibinfo{person}{Seiji Yamamoto}, \bibinfo{person}{Sharadh Ramaswamy}, \bibinfo{person}{Shaun Lindsay}, \bibinfo{person}{Shaun Lindsay}, \bibinfo{person}{Sheng Feng}, \bibinfo{person}{Shenghao Lin}, \bibinfo{person}{Shengxin~Cindy Zha}, \bibinfo{person}{Shishir Patil}, \bibinfo{person}{Shiva Shankar}, \bibinfo{person}{Shuqiang Zhang}, \bibinfo{person}{Shuqiang Zhang}, \bibinfo{person}{Sinong
  Wang}, \bibinfo{person}{Sneha Agarwal}, \bibinfo{person}{Soji Sajuyigbe}, \bibinfo{person}{Soumith Chintala}, \bibinfo{person}{Stephanie Max}, \bibinfo{person}{Stephen Chen}, \bibinfo{person}{Steve Kehoe}, \bibinfo{person}{Steve Satterfield}, \bibinfo{person}{Sudarshan Govindaprasad}, \bibinfo{person}{Sumit Gupta}, \bibinfo{person}{Summer Deng}, \bibinfo{person}{Sungmin Cho}, \bibinfo{person}{Sunny Virk}, \bibinfo{person}{Suraj Subramanian}, \bibinfo{person}{Sy Choudhury}, \bibinfo{person}{Sydney Goldman}, \bibinfo{person}{Tal Remez}, \bibinfo{person}{Tamar Glaser}, \bibinfo{person}{Tamara Best}, \bibinfo{person}{Thilo Koehler}, \bibinfo{person}{Thomas Robinson}, \bibinfo{person}{Tianhe Li}, \bibinfo{person}{Tianjun Zhang}, \bibinfo{person}{Tim Matthews}, \bibinfo{person}{Timothy Chou}, \bibinfo{person}{Tzook Shaked}, \bibinfo{person}{Varun Vontimitta}, \bibinfo{person}{Victoria Ajayi}, \bibinfo{person}{Victoria Montanez}, \bibinfo{person}{Vijai Mohan}, \bibinfo{person}{Vinay~Satish Kumar},
  \bibinfo{person}{Vishal Mangla}, \bibinfo{person}{Vlad Ionescu}, \bibinfo{person}{Vlad Poenaru}, \bibinfo{person}{Vlad~Tiberiu Mihailescu}, \bibinfo{person}{Vladimir Ivanov}, \bibinfo{person}{Wei Li}, \bibinfo{person}{Wenchen Wang}, \bibinfo{person}{Wenwen Jiang}, \bibinfo{person}{Wes Bouaziz}, \bibinfo{person}{Will Constable}, \bibinfo{person}{Xiaocheng Tang}, \bibinfo{person}{Xiaojian Wu}, \bibinfo{person}{Xiaolan Wang}, \bibinfo{person}{Xilun Wu}, \bibinfo{person}{Xinbo Gao}, \bibinfo{person}{Yaniv Kleinman}, \bibinfo{person}{Yanjun Chen}, \bibinfo{person}{Ye Hu}, \bibinfo{person}{Ye Jia}, \bibinfo{person}{Ye Qi}, \bibinfo{person}{Yenda Li}, \bibinfo{person}{Yilin Zhang}, \bibinfo{person}{Ying Zhang}, \bibinfo{person}{Yossi Adi}, \bibinfo{person}{Youngjin Nam}, \bibinfo{person}{Yu}, \bibinfo{person}{Wang}, \bibinfo{person}{Yu Zhao}, \bibinfo{person}{Yuchen Hao}, \bibinfo{person}{Yundi Qian}, \bibinfo{person}{Yunlu Li}, \bibinfo{person}{Yuzi He}, \bibinfo{person}{Zach Rait}, \bibinfo{person}{Zachary
  DeVito}, \bibinfo{person}{Zef Rosnbrick}, \bibinfo{person}{Zhaoduo Wen}, \bibinfo{person}{Zhenyu Yang}, \bibinfo{person}{Zhiwei Zhao}, {and} \bibinfo{person}{Zhiyu Ma}.} \bibinfo{year}{2024}\natexlab{}.
\newblock \showarticletitle{The llama 3 herd of models}.
\newblock \bibinfo{journal}{{\em arXiv preprint arXiv:2407.21783\/}} (\bibinfo{year}{2024}).
\newblock
\showURL{%
\url{https://doi.org/10.48550/arXiv.2407.21783}}


\bibitem[\protect\citeauthoryear{He, Zhai, Antunes, Wang, Luo, Shi, and Li}{He et~al\mbox{.}}{2022}]%
        {he2022fastermoe}
\bibfield{author}{\bibinfo{person}{Jiaao He}, \bibinfo{person}{Jidong Zhai}, \bibinfo{person}{Tiago Antunes}, \bibinfo{person}{Haojie Wang}, \bibinfo{person}{Fuwen Luo}, \bibinfo{person}{Shangfeng Shi}, {and} \bibinfo{person}{Qin Li}.} \bibinfo{year}{2022}\natexlab{}.
\newblock \showarticletitle{Fastermoe: modeling and optimizing training of large-scale dynamic pre-trained models}. In \bibinfo{booktitle}{{\em Proceedings of the 27th ACM SIGPLAN Symposium on Principles and Practice of Parallel Programming}}. \bibinfo{pages}{120--134}.
\newblock


\bibitem[\protect\citeauthoryear{Herrmann, Beaumont, Eyraud-Dubois, Hermann, Joly, and Shilova}{Herrmann et~al\mbox{.}}{2019}]%
        {herrmann2019optimal}
\bibfield{author}{\bibinfo{person}{Julien Herrmann}, \bibinfo{person}{Olivier Beaumont}, \bibinfo{person}{Lionel Eyraud-Dubois}, \bibinfo{person}{Julien Hermann}, \bibinfo{person}{Alexis Joly}, {and} \bibinfo{person}{Alena Shilova}.} \bibinfo{year}{2019}\natexlab{}.
\newblock \showarticletitle{Optimal checkpointing for heterogeneous chains: how to train deep neural networks with limited memory}.
\newblock \bibinfo{journal}{{\em arXiv preprint arXiv:1911.13214\/}} (\bibinfo{year}{2019}).
\newblock
\showURL{%
\url{https://doi.org/10.48550/arXiv.1911.13214}}


\bibitem[\protect\citeauthoryear{Jacobs, Tanaka, Zhang, Zhang, Song, Rajbhandari, and He}{Jacobs et~al\mbox{.}}{2023}]%
        {DBLP:deepspeed-sp-Ulysses}
\bibfield{author}{\bibinfo{person}{Sam~Ade Jacobs}, \bibinfo{person}{Masahiro Tanaka}, \bibinfo{person}{Chengming Zhang}, \bibinfo{person}{Minjia Zhang}, \bibinfo{person}{Shuaiwen~Leon Song}, \bibinfo{person}{Samyam Rajbhandari}, {and} \bibinfo{person}{Yuxiong He}.} \bibinfo{year}{2023}\natexlab{}.
\newblock \showarticletitle{DeepSpeed Ulysses: System Optimizations for Enabling Training of Extreme Long Sequence Transformer Models}.
\newblock \bibinfo{journal}{{\em CoRR\/}}  \bibinfo{volume}{abs/2309.14509} (\bibinfo{year}{2023}).
\newblock
\showDOI{%
\url{https://doi.org/10.48550/ARXIV.2309.14509}}
\showeprint{2309.14509}


\bibitem[\protect\citeauthoryear{Jeon, Wu, Cao, Kim, Park, Aggarwal, Unger, Arfeen, Liao, and Miao}{Jeon et~al\mbox{.}}{2025}]%
        {jeon2025graphpipe}
\bibfield{author}{\bibinfo{person}{Byungsoo Jeon}, \bibinfo{person}{Mengdi Wu}, \bibinfo{person}{Shiyi Cao}, \bibinfo{person}{Sunghyun Kim}, \bibinfo{person}{Sunghyun Park}, \bibinfo{person}{Neeraj Aggarwal}, \bibinfo{person}{Colin Unger}, \bibinfo{person}{Daiyaan Arfeen}, \bibinfo{person}{Peiyuan Liao}, {and} \bibinfo{person}{Xupeng Miao}.} \bibinfo{year}{2025}\natexlab{}.
\newblock \showarticletitle{Graphpipe: Improving performance and scalability of dnn training with graph pipeline parallelism}. In \bibinfo{booktitle}{{\em Proceedings of the 30th ACM International Conference on Architectural Support for Programming Languages and Operating Systems, Volume 1}}. \bibinfo{pages}{557--571}.
\newblock


\bibitem[\protect\citeauthoryear{Jiang, Jia, Zheng, Wang, and Wu}{Jiang et~al\mbox{.}}{2024}]%
        {jiang2024dynapipe}
\bibfield{author}{\bibinfo{person}{Chenyu Jiang}, \bibinfo{person}{Zhen Jia}, \bibinfo{person}{Shuai Zheng}, \bibinfo{person}{Yida Wang}, {and} \bibinfo{person}{Chuan Wu}.} \bibinfo{year}{2024}\natexlab{}.
\newblock \showarticletitle{DynaPipe: Optimizing multi-task training through dynamic pipelines}. In \bibinfo{booktitle}{{\em Proceedings of the Nineteenth European Conference on Computer Systems}}. \bibinfo{pages}{542--559}.
\newblock


\bibitem[\protect\citeauthoryear{Krell, Kosec, Perez, and Fitzgibbon}{Krell et~al\mbox{.}}{2021}]%
        {krell2021efficient}
\bibfield{author}{\bibinfo{person}{Mario~Michael Krell}, \bibinfo{person}{Matej Kosec}, \bibinfo{person}{Sergio~P Perez}, {and} \bibinfo{person}{Andrew Fitzgibbon}.} \bibinfo{year}{2021}\natexlab{}.
\newblock \showarticletitle{Efficient sequence packing without cross-contamination: Accelerating large language models without impacting performance}.
\newblock \bibinfo{journal}{{\em arXiv preprint arXiv:2107.02027\/}} (\bibinfo{year}{2021}).
\newblock
\showURL{%
\url{https://doi.org/10.48550/arXiv.2107.02027}}


\bibitem[\protect\citeauthoryear{Li, Gong, Yang, Shan, Liu, Zhu, Zhang, Guo, Chen, Li, Jiao, Li, Zhang, Sun, Dong, Zhu, Zhuang, Song, Zhu, Han, Li, Xie, Xu, Yan, Zhang, Xiao, Kang, Han, Wang, Yu, Feng, Zheng, Chai, Xing, Ju, Chi, Zhang, Huang, Niu, Li, Zhao, Yang, Xu, Wang, Wang, Li, Leng, Shi, Yu, Li, Zhu, Huang, Liang, Sun, Sun, Cheng, Li, Song, Su, Han, Zhang, Hou, Min, Zou, Shen, Gong, Zhu, Zhou, Zhong, Hu, Fan, Yu, Yang, Li, Huang, Li, Huang, Xu, Mao, Li, Li, Tao, Ying, Cong, Qin, Fan, Yu, Jiang, and Wu}{Li et~al\mbox{.}}{2025a}]%
        {li2025minimax}
\bibfield{author}{\bibinfo{person}{Aonian Li}, \bibinfo{person}{Bangwei Gong}, \bibinfo{person}{Bo Yang}, \bibinfo{person}{Boji Shan}, \bibinfo{person}{Chang Liu}, \bibinfo{person}{Cheng Zhu}, \bibinfo{person}{Chunhao Zhang}, \bibinfo{person}{Congchao Guo}, \bibinfo{person}{Da Chen}, \bibinfo{person}{Dong Li}, \bibinfo{person}{Enwei Jiao}, \bibinfo{person}{Gengxin Li}, \bibinfo{person}{Guojun Zhang}, \bibinfo{person}{Haohai Sun}, \bibinfo{person}{Houze Dong}, \bibinfo{person}{Jiadai Zhu}, \bibinfo{person}{Jiaqi Zhuang}, \bibinfo{person}{Jiayuan Song}, \bibinfo{person}{Jin Zhu}, \bibinfo{person}{Jingtao Han}, \bibinfo{person}{Jingyang Li}, \bibinfo{person}{Junbin Xie}, \bibinfo{person}{Junhao Xu}, \bibinfo{person}{Junjie Yan}, \bibinfo{person}{Kaishun Zhang}, \bibinfo{person}{Kecheng Xiao}, \bibinfo{person}{Kexi Kang}, \bibinfo{person}{Le Han}, \bibinfo{person}{Leyang Wang}, \bibinfo{person}{Lianfei Yu}, \bibinfo{person}{Liheng Feng}, \bibinfo{person}{Lin Zheng}, \bibinfo{person}{Linbo Chai},
  \bibinfo{person}{Long Xing}, \bibinfo{person}{Meizhi Ju}, \bibinfo{person}{Mingyuan Chi}, \bibinfo{person}{Mozhi Zhang}, \bibinfo{person}{Peikai Huang}, \bibinfo{person}{Pengcheng Niu}, \bibinfo{person}{Pengfei Li}, \bibinfo{person}{Pengyu Zhao}, \bibinfo{person}{Qi Yang}, \bibinfo{person}{Qidi Xu}, \bibinfo{person}{Qiexiang Wang}, \bibinfo{person}{Qin Wang}, \bibinfo{person}{Qiuhui Li}, \bibinfo{person}{Ruitao Leng}, \bibinfo{person}{Shengmin Shi}, \bibinfo{person}{Shuqi Yu}, \bibinfo{person}{Sichen Li}, \bibinfo{person}{Songquan Zhu}, \bibinfo{person}{Tao Huang}, \bibinfo{person}{Tianrun Liang}, \bibinfo{person}{Weigao Sun}, \bibinfo{person}{Weixuan Sun}, \bibinfo{person}{Weiyu Cheng}, \bibinfo{person}{Wenkai Li}, \bibinfo{person}{Xiangjun Song}, \bibinfo{person}{Xiao Su}, \bibinfo{person}{Xiaodong Han}, \bibinfo{person}{Xinjie Zhang}, \bibinfo{person}{Xinzhu Hou}, \bibinfo{person}{Xu Min}, \bibinfo{person}{Xun Zou}, \bibinfo{person}{Xuyang Shen}, \bibinfo{person}{Yan Gong}, \bibinfo{person}{Yingjie Zhu},
  \bibinfo{person}{Yipeng Zhou}, \bibinfo{person}{Yiran Zhong}, \bibinfo{person}{Yongyi Hu}, \bibinfo{person}{Yuanxiang Fan}, \bibinfo{person}{Yue Yu}, \bibinfo{person}{Yufeng Yang}, \bibinfo{person}{Yuhao Li}, \bibinfo{person}{Yunan Huang}, \bibinfo{person}{Yunji Li}, \bibinfo{person}{Yunpeng Huang}, \bibinfo{person}{Yunzhi Xu}, \bibinfo{person}{Yuxin Mao}, \bibinfo{person}{Zehan Li}, \bibinfo{person}{Zekang Li}, \bibinfo{person}{Zewei Tao}, \bibinfo{person}{Zewen Ying}, \bibinfo{person}{Zhaoyang Cong}, \bibinfo{person}{Zhen Qin}, \bibinfo{person}{Zhenhua Fan}, \bibinfo{person}{Zhihang Yu}, \bibinfo{person}{Zhuo Jiang}, {and} \bibinfo{person}{Zijia Wu}.} \bibinfo{year}{2025}\natexlab{a}.
\newblock \showarticletitle{Minimax-01: Scaling foundation models with lightning attention}.
\newblock \bibinfo{journal}{{\em arXiv preprint arXiv:2501.08313\/}} (\bibinfo{year}{2025}).
\newblock
\showURL{%
\url{https://doi.org/10.48550/arXiv.2501.08313}}


\bibitem[\protect\citeauthoryear{Li, Shao, Xie, Xing, Gonzalez, Stoica, Ma, and Zhang}{Li et~al\mbox{.}}{2023}]%
        {DBLP:LightSeq}
\bibfield{author}{\bibinfo{person}{Dacheng Li}, \bibinfo{person}{Rulin Shao}, \bibinfo{person}{Anze Xie}, \bibinfo{person}{Eric~P. Xing}, \bibinfo{person}{Joseph~E. Gonzalez}, \bibinfo{person}{Ion Stoica}, \bibinfo{person}{Xuezhe Ma}, {and} \bibinfo{person}{Hao Zhang}.} \bibinfo{year}{2023}\natexlab{}.
\newblock \showarticletitle{LightSeq: Sequence Level Parallelism for Distributed Training of Long Context Transformers}.
\newblock \bibinfo{journal}{{\em CoRR\/}}  \bibinfo{volume}{abs/2310.03294} (\bibinfo{year}{2023}).
\newblock
\showDOI{%
\url{https://doi.org/10.48550/ARXIV.2310.03294}}
\showeprint{2310.03294}


\bibitem[\protect\citeauthoryear{Li and Hoefler}{Li and Hoefler}{2021}]%
        {li2021chimera}
\bibfield{author}{\bibinfo{person}{Shigang Li} {and} \bibinfo{person}{Torsten Hoefler}.} \bibinfo{year}{2021}\natexlab{}.
\newblock \showarticletitle{Chimera: efficiently training large-scale neural networks with bidirectional pipelines}. In \bibinfo{booktitle}{{\em Proceedings of the International Conference for High Performance Computing, Networking, Storage and Analysis}}. \bibinfo{pages}{1--14}.
\newblock
\showURL{%
\url{https://dl.acm.org/doi/10.1145/3458817.3476145}}


\bibitem[\protect\citeauthoryear{Li, Liu, Zhang, Yuan, Chen, and Song}{Li et~al\mbox{.}}{2025b}]%
        {li2025slimpipe}
\bibfield{author}{\bibinfo{person}{Zhouyang Li}, \bibinfo{person}{Yuliang Liu}, \bibinfo{person}{Wei Zhang}, \bibinfo{person}{Tailing Yuan}, \bibinfo{person}{Bin Chen}, {and} \bibinfo{person}{Chengru Song}.} \bibinfo{year}{2025}\natexlab{b}.
\newblock \showarticletitle{SlimPipe: Memory-Thrifty and Efficient Pipeline Parallelism for Long-Context LLM Training}. In \bibinfo{booktitle}{{\em Proceedings of the International Conference for High Performance Computing, Networking, Storage and Analysis}}. \bibinfo{pages}{1409--1428}.
\newblock


\bibitem[\protect\citeauthoryear{Li, Zhuang, Guo, Zhuo, Zhang, Song, and Stoica}{Li et~al\mbox{.}}{2021}]%
        {li2021terapipe}
\bibfield{author}{\bibinfo{person}{Zhuohan Li}, \bibinfo{person}{Siyuan Zhuang}, \bibinfo{person}{Shiyuan Guo}, \bibinfo{person}{Danyang Zhuo}, \bibinfo{person}{Hao Zhang}, \bibinfo{person}{Dawn Song}, {and} \bibinfo{person}{Ion Stoica}.} \bibinfo{year}{2021}\natexlab{}.
\newblock \showarticletitle{Terapipe: Token-level pipeline parallelism for training large-scale language models}. In \bibinfo{booktitle}{{\em International Conference on Machine Learning}}. PMLR, \bibinfo{pages}{6543--6552}.
\newblock
\showURL{%
\url{https://proceedings.mlr.press/v139/li21y.html}}


\bibitem[\protect\citeauthoryear{Liu, Feng, Xue, Wang, Wu, Lu, Zhao, Deng, Zhang, Ruan, Dai, Guo, Yang, Chen, Ji, Li, Lin, Dai, Luo, Hao, Chen, Li, Zhang, Bao, Xu, Wang, Zhang, Ding, Xin, Gao, Li, Qu, Cai, Liang, Guo, Ni, Li, Wang, Chen, Chen, Yuan, Qiu, Li, Song, Dong, Hu, Gao, Guan, Huang, Yu, Wang, Zhang, Xu, Xia, Zhao, Wang, Zhang, Li, Wang, Zhang, Zhang, Tang, Li, Tian, Huang, Wang, Zhang, Wang, Zhu, Chen, Du, Chen, Jin, Ge, Zhang, Pan, Wang, Xu, Zhang, Chen, Li, Lu, Zhou, Chen, Wu, Ye, Ye, Ma, Wang, Zhou, Yu, Zhou, Pan, Wang, Yun, Pei, Sun, Xiao, Zeng, Zhao, An, Liu, Liang, Gao, Yu, Zhang, Li, Jin, Wang, Bi, Liu, Wang, Shen, Chen, Zhang, Chen, Nie, Sun, Wang, Cheng, Liu, Xie, Liu, Yu, Song, Shan, Zhou, Yang, Li, Su, Lin, Li, Wang, Wei, Zhu, Zhang, Xu, Xu, Huang, Li, Zhao, Sun, Li, Wang, Yu, Zheng, Zhang, Shi, Xiong, He, Tang, Piao, Wang, Tan, Ma, Liu, Guo, Wu, Ou, Zhu, Wang, Gong, Zou, He, Zha, Xiong, Ma, Yan, Luo, You, Liu, Zhou, Wu, Ren, Ren, Sha, Fu, Xu, Huang, Zhang, Xie, Zhang, Hao, Gou, Ma, Yan,
  Shao, Xu, Wu, Zhang, Li, Gu, Zhu, Liu, Li, Xie, Song, Gao, and Pan}{Liu et~al\mbox{.}}{2024}]%
        {liu2024deepseek}
\bibfield{author}{\bibinfo{person}{Aixin Liu}, \bibinfo{person}{Bei Feng}, \bibinfo{person}{Bing Xue}, \bibinfo{person}{Bingxuan Wang}, \bibinfo{person}{Bochao Wu}, \bibinfo{person}{Chengda Lu}, \bibinfo{person}{Chenggang Zhao}, \bibinfo{person}{Chengqi Deng}, \bibinfo{person}{Chenyu Zhang}, \bibinfo{person}{Chong Ruan}, \bibinfo{person}{Damai Dai}, \bibinfo{person}{Daya Guo}, \bibinfo{person}{Dejian Yang}, \bibinfo{person}{Deli Chen}, \bibinfo{person}{Dongjie Ji}, \bibinfo{person}{Erhang Li}, \bibinfo{person}{Fangyun Lin}, \bibinfo{person}{Fucong Dai}, \bibinfo{person}{Fuli Luo}, \bibinfo{person}{Guangbo Hao}, \bibinfo{person}{Guanting Chen}, \bibinfo{person}{Guowei Li}, \bibinfo{person}{H. Zhang}, \bibinfo{person}{Han Bao}, \bibinfo{person}{Hanwei Xu}, \bibinfo{person}{Haocheng Wang}, \bibinfo{person}{Haowei Zhang}, \bibinfo{person}{Honghui Ding}, \bibinfo{person}{Huajian Xin}, \bibinfo{person}{Huazuo Gao}, \bibinfo{person}{Hui Li}, \bibinfo{person}{Hui Qu}, \bibinfo{person}{J.~L. Cai},
  \bibinfo{person}{Jian Liang}, \bibinfo{person}{Jianzhong Guo}, \bibinfo{person}{Jiaqi Ni}, \bibinfo{person}{Jiashi Li}, \bibinfo{person}{Jiawei Wang}, \bibinfo{person}{Jin Chen}, \bibinfo{person}{Jingchang Chen}, \bibinfo{person}{Jingyang Yuan}, \bibinfo{person}{Junjie Qiu}, \bibinfo{person}{Junlong Li}, \bibinfo{person}{Junxiao Song}, \bibinfo{person}{Kai Dong}, \bibinfo{person}{Kai Hu}, \bibinfo{person}{Kaige Gao}, \bibinfo{person}{Kang Guan}, \bibinfo{person}{Kexin Huang}, \bibinfo{person}{Kuai Yu}, \bibinfo{person}{Lean Wang}, \bibinfo{person}{Lecong Zhang}, \bibinfo{person}{Lei Xu}, \bibinfo{person}{Leyi Xia}, \bibinfo{person}{Liang Zhao}, \bibinfo{person}{Litong Wang}, \bibinfo{person}{Liyue Zhang}, \bibinfo{person}{Meng Li}, \bibinfo{person}{Miaojun Wang}, \bibinfo{person}{Mingchuan Zhang}, \bibinfo{person}{Minghua Zhang}, \bibinfo{person}{Minghui Tang}, \bibinfo{person}{Mingming Li}, \bibinfo{person}{Ning Tian}, \bibinfo{person}{Panpan Huang}, \bibinfo{person}{Peiyi Wang}, \bibinfo{person}{Peng
  Zhang}, \bibinfo{person}{Qiancheng Wang}, \bibinfo{person}{Qihao Zhu}, \bibinfo{person}{Qinyu Chen}, \bibinfo{person}{Qiushi Du}, \bibinfo{person}{R.~J. Chen}, \bibinfo{person}{R.~L. Jin}, \bibinfo{person}{Ruiqi Ge}, \bibinfo{person}{Ruisong Zhang}, \bibinfo{person}{Ruizhe Pan}, \bibinfo{person}{Runji Wang}, \bibinfo{person}{Runxin Xu}, \bibinfo{person}{Ruoyu Zhang}, \bibinfo{person}{Ruyi Chen}, \bibinfo{person}{S.~S. Li}, \bibinfo{person}{Shanghao Lu}, \bibinfo{person}{Shangyan Zhou}, \bibinfo{person}{Shanhuang Chen}, \bibinfo{person}{Shaoqing Wu}, \bibinfo{person}{Shengfeng Ye}, \bibinfo{person}{Shengfeng Ye}, \bibinfo{person}{Shirong Ma}, \bibinfo{person}{Shiyu Wang}, \bibinfo{person}{Shuang Zhou}, \bibinfo{person}{Shuiping Yu}, \bibinfo{person}{Shunfeng Zhou}, \bibinfo{person}{Shuting Pan}, \bibinfo{person}{T. Wang}, \bibinfo{person}{Tao Yun}, \bibinfo{person}{Tian Pei}, \bibinfo{person}{Tianyu Sun}, \bibinfo{person}{W.~L. Xiao}, \bibinfo{person}{Wangding Zeng}, \bibinfo{person}{Wanjia Zhao},
  \bibinfo{person}{Wei An}, \bibinfo{person}{Wen Liu}, \bibinfo{person}{Wenfeng Liang}, \bibinfo{person}{Wenjun Gao}, \bibinfo{person}{Wenqin Yu}, \bibinfo{person}{Wentao Zhang}, \bibinfo{person}{X.~Q. Li}, \bibinfo{person}{Xiangyue Jin}, \bibinfo{person}{Xianzu Wang}, \bibinfo{person}{Xiao Bi}, \bibinfo{person}{Xiaodong Liu}, \bibinfo{person}{Xiaohan Wang}, \bibinfo{person}{Xiaojin Shen}, \bibinfo{person}{Xiaokang Chen}, \bibinfo{person}{Xiaokang Zhang}, \bibinfo{person}{Xiaosha Chen}, \bibinfo{person}{Xiaotao Nie}, \bibinfo{person}{Xiaowen Sun}, \bibinfo{person}{Xiaoxiang Wang}, \bibinfo{person}{Xin Cheng}, \bibinfo{person}{Xin Liu}, \bibinfo{person}{Xin Xie}, \bibinfo{person}{Xingchao Liu}, \bibinfo{person}{Xingkai Yu}, \bibinfo{person}{Xinnan Song}, \bibinfo{person}{Xinxia Shan}, \bibinfo{person}{Xinyi Zhou}, \bibinfo{person}{Xinyu Yang}, \bibinfo{person}{Xinyuan Li}, \bibinfo{person}{Xuecheng Su}, \bibinfo{person}{Xuheng Lin}, \bibinfo{person}{Y.~K. Li}, \bibinfo{person}{Y.~Q. Wang},
  \bibinfo{person}{Y.~X. Wei}, \bibinfo{person}{Y.~X. Zhu}, \bibinfo{person}{Yang Zhang}, \bibinfo{person}{Yanhong Xu}, \bibinfo{person}{Yanhong Xu}, \bibinfo{person}{Yanping Huang}, \bibinfo{person}{Yao Li}, \bibinfo{person}{Yao Zhao}, \bibinfo{person}{Yaofeng Sun}, \bibinfo{person}{Yaohui Li}, \bibinfo{person}{Yaohui Wang}, \bibinfo{person}{Yi Yu}, \bibinfo{person}{Yi Zheng}, \bibinfo{person}{Yichao Zhang}, \bibinfo{person}{Yifan Shi}, \bibinfo{person}{Yiliang Xiong}, \bibinfo{person}{Ying He}, \bibinfo{person}{Ying Tang}, \bibinfo{person}{Yishi Piao}, \bibinfo{person}{Yisong Wang}, \bibinfo{person}{Yixuan Tan}, \bibinfo{person}{Yiyang Ma}, \bibinfo{person}{Yiyuan Liu}, \bibinfo{person}{Yongqiang Guo}, \bibinfo{person}{Yu Wu}, \bibinfo{person}{Yuan Ou}, \bibinfo{person}{Yuchen Zhu}, \bibinfo{person}{Yuduan Wang}, \bibinfo{person}{Yue Gong}, \bibinfo{person}{Yuheng Zou}, \bibinfo{person}{Yujia He}, \bibinfo{person}{Yukun Zha}, \bibinfo{person}{Yunfan Xiong}, \bibinfo{person}{Yunxian Ma},
  \bibinfo{person}{Yuting Yan}, \bibinfo{person}{Yuxiang Luo}, \bibinfo{person}{Yuxiang You}, \bibinfo{person}{Yuxuan Liu}, \bibinfo{person}{Yuyang Zhou}, \bibinfo{person}{Z.~F. Wu}, \bibinfo{person}{Z.~Z. Ren}, \bibinfo{person}{Zehui Ren}, \bibinfo{person}{Zhangli Sha}, \bibinfo{person}{Zhe Fu}, \bibinfo{person}{Zhean Xu}, \bibinfo{person}{Zhen Huang}, \bibinfo{person}{Zhen Zhang}, \bibinfo{person}{Zhenda Xie}, \bibinfo{person}{Zhengyan Zhang}, \bibinfo{person}{Zhewen Hao}, \bibinfo{person}{Zhibin Gou}, \bibinfo{person}{Zhicheng Ma}, \bibinfo{person}{Zhigang Yan}, \bibinfo{person}{Zhihong Shao}, \bibinfo{person}{Zhipeng Xu}, \bibinfo{person}{Zhiyu Wu}, \bibinfo{person}{Zhongyu Zhang}, \bibinfo{person}{Zhuoshu Li}, \bibinfo{person}{Zihui Gu}, \bibinfo{person}{Zijia Zhu}, \bibinfo{person}{Zijun Liu}, \bibinfo{person}{Zilin Li}, \bibinfo{person}{Ziwei Xie}, \bibinfo{person}{Ziyang Song}, \bibinfo{person}{Ziyi Gao}, {and} \bibinfo{person}{Zizheng Pan}.} \bibinfo{year}{2024}\natexlab{}.
\newblock \showarticletitle{Deepseek-v3 technical report}.
\newblock \bibinfo{journal}{{\em arXiv preprint arXiv:2412.19437\/}} (\bibinfo{year}{2024}).
\newblock
\showURL{%
\url{https://doi.org/10.48550/arXiv.2412.19437}}


\bibitem[\protect\citeauthoryear{Liu, Zaharia, and Abbeel}{Liu et~al\mbox{.}}{2023}]%
        {DBLP:ring-attn}
\bibfield{author}{\bibinfo{person}{Hao Liu}, \bibinfo{person}{Matei Zaharia}, {and} \bibinfo{person}{Pieter Abbeel}.} \bibinfo{year}{2023}\natexlab{}.
\newblock \showarticletitle{Ring Attention with Blockwise Transformers for Near-Infinite Context}.
\newblock \bibinfo{journal}{{\em CoRR\/}}  \bibinfo{volume}{abs/2310.01889} (\bibinfo{year}{2023}).
\newblock
\showDOI{%
\url{https://doi.org/10.48550/ARXIV.2310.01889}}
\showeprint{2310.01889}


\bibitem[\protect\citeauthoryear{Liu, Cheng, Zhou, and You}{Liu et~al\mbox{.}}{2023}]%
        {liu2023hanayo}
\bibfield{author}{\bibinfo{person}{Ziming Liu}, \bibinfo{person}{Shenggan Cheng}, \bibinfo{person}{Haotian Zhou}, {and} \bibinfo{person}{Yang You}.} \bibinfo{year}{2023}\natexlab{}.
\newblock \showarticletitle{Hanayo: Harnessing wave-like pipeline parallelism for enhanced large model training efficiency}. In \bibinfo{booktitle}{{\em Proceedings of the International Conference for High Performance Computing, Networking, Storage and Analysis}}. \bibinfo{pages}{1--13}.
\newblock
\showURL{%
\url{https://dl.acm.org/doi/10.1145/3581784.3607073}}


\bibitem[\protect\citeauthoryear{Narayanan, Shoeybi, Casper, LeGresley, Patwary, Korthikanti, Vainbrand, Kashinkunti, Bernauer, Catanzaro, Phanishayee, and Zaharia}{Narayanan et~al\mbox{.}}{2021}]%
        {narayanan2021efficient}
\bibfield{author}{\bibinfo{person}{Deepak Narayanan}, \bibinfo{person}{Mohammad Shoeybi}, \bibinfo{person}{Jared Casper}, \bibinfo{person}{Patrick LeGresley}, \bibinfo{person}{Mostofa Patwary}, \bibinfo{person}{Vijay Korthikanti}, \bibinfo{person}{Dmitri Vainbrand}, \bibinfo{person}{Prethvi Kashinkunti}, \bibinfo{person}{Julie Bernauer}, \bibinfo{person}{Bryan Catanzaro}, \bibinfo{person}{Amar Phanishayee}, {and} \bibinfo{person}{Matei Zaharia}.} \bibinfo{year}{2021}\natexlab{}.
\newblock \showarticletitle{Efficient large-scale language model training on gpu clusters using megatron-lm}. In \bibinfo{booktitle}{{\em Proceedings of the International Conference for High Performance Computing, Networking, Storage and Analysis}}. \bibinfo{pages}{1--15}.
\newblock
\showURL{%
\url{https://dl.acm.org/doi/10.1145/3458817.3476209}}


\bibitem[\protect\citeauthoryear{Nie, Miao, Wang, Yang, Xue, Ma, Cao, and Cui}{Nie et~al\mbox{.}}{2023}]%
        {nie2023flexmoe}
\bibfield{author}{\bibinfo{person}{Xiaonan Nie}, \bibinfo{person}{Xupeng Miao}, \bibinfo{person}{Zilong Wang}, \bibinfo{person}{Zichao Yang}, \bibinfo{person}{Jilong Xue}, \bibinfo{person}{Lingxiao Ma}, \bibinfo{person}{Gang Cao}, {and} \bibinfo{person}{Bin Cui}.} \bibinfo{year}{2023}\natexlab{}.
\newblock \showarticletitle{Flexmoe: Scaling large-scale sparse pre-trained model training via dynamic device placement}.
\newblock \bibinfo{journal}{{\em Proceedings of the ACM on Management of Data\/}} \bibinfo{volume}{1}, \bibinfo{number}{1} (\bibinfo{year}{2023}), \bibinfo{pages}{1--19}.
\newblock


\bibitem[\protect\citeauthoryear{Qi, Wan, Huang, and Lin}{Qi et~al\mbox{.}}{2023}]%
        {qi2023zero}
\bibfield{author}{\bibinfo{person}{Penghui Qi}, \bibinfo{person}{Xinyi Wan}, \bibinfo{person}{Guangxing Huang}, {and} \bibinfo{person}{Min Lin}.} \bibinfo{year}{2023}\natexlab{}.
\newblock \showarticletitle{Zero bubble pipeline parallelism}.
\newblock \bibinfo{journal}{{\em arXiv preprint arXiv:2401.10241\/}} (\bibinfo{year}{2023}).
\newblock
\showURL{%
\url{https://doi.org/10.48550/arXiv.2401.10241}}


\bibitem[\protect\citeauthoryear{Shoeybi, Patwary, Puri, LeGresley, Casper, and Catanzaro}{Shoeybi et~al\mbox{.}}{2019}]%
        {megatron-lm}
\bibfield{author}{\bibinfo{person}{Mohammad Shoeybi}, \bibinfo{person}{Mostofa Patwary}, \bibinfo{person}{Raul Puri}, \bibinfo{person}{Patrick LeGresley}, \bibinfo{person}{Jared Casper}, {and} \bibinfo{person}{Bryan Catanzaro}.} \bibinfo{year}{2019}\natexlab{}.
\newblock \showarticletitle{Megatron-LM: Training Multi-Billion Parameter Language Models Using Model Parallelism}.
\newblock \bibinfo{journal}{{\em arXiv preprint arXiv:1909.08053\/}} (\bibinfo{year}{2019}).
\newblock
\showURL{%
\url{https://doi.org/10.48550/arXiv.1909.08053}}


\bibitem[\protect\citeauthoryear{Sun, Zhao, Han, Yang, Zhang, Liu, Shi, and Sun}{Sun et~al\mbox{.}}{2024}]%
        {sun2024seq1f1b}
\bibfield{author}{\bibinfo{person}{Ao Sun}, \bibinfo{person}{Weilin Zhao}, \bibinfo{person}{Xu Han}, \bibinfo{person}{Cheng Yang}, \bibinfo{person}{Xinrong Zhang}, \bibinfo{person}{Zhiyuan Liu}, \bibinfo{person}{Chuan Shi}, {and} \bibinfo{person}{Maosong Sun}.} \bibinfo{year}{2024}\natexlab{}.
\newblock \showarticletitle{Seq1f1b: Efficient sequence-level pipeline parallelism for large language model training}.
\newblock \bibinfo{journal}{{\em arXiv preprint arXiv:2406.03488\/}} (\bibinfo{year}{2024}).
\newblock
\showURL{%
\url{https://doi.org/10.48550/arXiv.2406.03488}}


\bibitem[\protect\citeauthoryear{Sun, Cao, Wang, Feng, Chen, Wang, and Chen}{Sun et~al\mbox{.}}{2024}]%
        {sun2024adapipe}
\bibfield{author}{\bibinfo{person}{Zhenbo Sun}, \bibinfo{person}{Huanqi Cao}, \bibinfo{person}{Yuanwei Wang}, \bibinfo{person}{Guanyu Feng}, \bibinfo{person}{Shengqi Chen}, \bibinfo{person}{Haojie Wang}, {and} \bibinfo{person}{Wenguang Chen}.} \bibinfo{year}{2024}\natexlab{}.
\newblock \showarticletitle{Adapipe: Optimizing pipeline parallelism with adaptive recomputation and partitioning}. In \bibinfo{booktitle}{{\em Proceedings of the 29th ACM International Conference on Architectural Support for Programming Languages and Operating Systems, Volume 3}}. \bibinfo{pages}{86--100}.
\newblock
\showURL{%
\url{https://dl.acm.org/doi/10.1145/3620666.3651359}}


\bibitem[\protect\citeauthoryear{Sun, Chen, Wang, Sha, Feng, and Chen}{Sun et~al\mbox{.}}{2025}]%
        {sun2025mepipe}
\bibfield{author}{\bibinfo{person}{Zhenbo Sun}, \bibinfo{person}{Shengqi Chen}, \bibinfo{person}{Yuanwei Wang}, \bibinfo{person}{Jian Sha}, \bibinfo{person}{Guanyu Feng}, {and} \bibinfo{person}{Wenguang Chen}.} \bibinfo{year}{2025}\natexlab{}.
\newblock \showarticletitle{Mepipe: Democratizing llm training with memory-efficient slice-level pipeline scheduling on cost-effective accelerators}. In \bibinfo{booktitle}{{\em Proceedings of the Twentieth European Conference on Computer Systems}}. \bibinfo{pages}{1263--1278}.
\newblock


\bibitem[\protect\citeauthoryear{Tillet, Kung, and Cox}{Tillet et~al\mbox{.}}{2019}]%
        {tillet2019triton}
\bibfield{author}{\bibinfo{person}{Philippe Tillet}, \bibinfo{person}{Hsiang-Tsung Kung}, {and} \bibinfo{person}{David Cox}.} \bibinfo{year}{2019}\natexlab{}.
\newblock \showarticletitle{Triton: an intermediate language and compiler for tiled neural network computations}. In \bibinfo{booktitle}{{\em Proceedings of the 3rd ACM SIGPLAN International Workshop on Machine Learning and Programming Languages}}. \bibinfo{pages}{10--19}.
\newblock
\showURL{%
\url{https://dl.acm.org/doi/10.1145/3315508.3329973}}


\bibitem[\protect\citeauthoryear{Touvron, Martin, Stone, Albert, Almahairi, Babaei, Bashlykov, Batra, Bhargava, Bhosale, Bikel, Blecher, Canton{-}Ferrer, Chen, Cucurull, Esiobu, Fernandes, Fu, Fu, Fuller, Gao, Goswami, Goyal, Hartshorn, Hosseini, Hou, Inan, Kardas, Kerkez, Khabsa, Kloumann, Korenev, Koura, Lachaux, Lavril, Lee, Liskovich, Lu, Mao, Martinet, Mihaylov, Mishra, Molybog, Nie, Poulton, Reizenstein, Rungta, Saladi, Schelten, Silva, Smith, Subramanian, Tan, Tang, Taylor, Williams, Kuan, Xu, Yan, Zarov, Zhang, Fan, Kambadur, Narang, Rodriguez, Stojnic, Edunov, and Scialom}{Touvron et~al\mbox{.}}{2023}]%
        {DBLP:llama2}
\bibfield{author}{\bibinfo{person}{Hugo Touvron}, \bibinfo{person}{Louis Martin}, \bibinfo{person}{Kevin Stone}, \bibinfo{person}{Peter Albert}, \bibinfo{person}{Amjad Almahairi}, \bibinfo{person}{Yasmine Babaei}, \bibinfo{person}{Nikolay Bashlykov}, \bibinfo{person}{Soumya Batra}, \bibinfo{person}{Prajjwal Bhargava}, \bibinfo{person}{Shruti Bhosale}, \bibinfo{person}{Dan Bikel}, \bibinfo{person}{Lukas Blecher}, \bibinfo{person}{Cristian Canton{-}Ferrer}, \bibinfo{person}{Moya Chen}, \bibinfo{person}{Guillem Cucurull}, \bibinfo{person}{David Esiobu}, \bibinfo{person}{Jude Fernandes}, \bibinfo{person}{Jeremy Fu}, \bibinfo{person}{Wenyin Fu}, \bibinfo{person}{Brian Fuller}, \bibinfo{person}{Cynthia Gao}, \bibinfo{person}{Vedanuj Goswami}, \bibinfo{person}{Naman Goyal}, \bibinfo{person}{Anthony Hartshorn}, \bibinfo{person}{Saghar Hosseini}, \bibinfo{person}{Rui Hou}, \bibinfo{person}{Hakan Inan}, \bibinfo{person}{Marcin Kardas}, \bibinfo{person}{Viktor Kerkez}, \bibinfo{person}{Madian Khabsa},
  \bibinfo{person}{Isabel Kloumann}, \bibinfo{person}{Artem Korenev}, \bibinfo{person}{Punit~Singh Koura}, \bibinfo{person}{Marie{-}Anne Lachaux}, \bibinfo{person}{Thibaut Lavril}, \bibinfo{person}{Jenya Lee}, \bibinfo{person}{Diana Liskovich}, \bibinfo{person}{Yinghai Lu}, \bibinfo{person}{Yuning Mao}, \bibinfo{person}{Xavier Martinet}, \bibinfo{person}{Todor Mihaylov}, \bibinfo{person}{Pushkar Mishra}, \bibinfo{person}{Igor Molybog}, \bibinfo{person}{Yixin Nie}, \bibinfo{person}{Andrew Poulton}, \bibinfo{person}{Jeremy Reizenstein}, \bibinfo{person}{Rashi Rungta}, \bibinfo{person}{Kalyan Saladi}, \bibinfo{person}{Alan Schelten}, \bibinfo{person}{Ruan Silva}, \bibinfo{person}{Eric~Michael Smith}, \bibinfo{person}{Ranjan Subramanian}, \bibinfo{person}{Xiaoqing~Ellen Tan}, \bibinfo{person}{Binh Tang}, \bibinfo{person}{Ross Taylor}, \bibinfo{person}{Adina Williams}, \bibinfo{person}{Jian~Xiang Kuan}, \bibinfo{person}{Puxin Xu}, \bibinfo{person}{Zheng Yan}, \bibinfo{person}{Iliyan Zarov}, \bibinfo{person}{Yuchen
  Zhang}, \bibinfo{person}{Angela Fan}, \bibinfo{person}{Melanie Kambadur}, \bibinfo{person}{Sharan Narang}, \bibinfo{person}{Aur{\'{e}}lien Rodriguez}, \bibinfo{person}{Robert Stojnic}, \bibinfo{person}{Sergey Edunov}, {and} \bibinfo{person}{Thomas Scialom}.} \bibinfo{year}{2023}\natexlab{}.
\newblock \showarticletitle{Llama 2: Open Foundation and Fine-Tuned Chat Models}.
\newblock \bibinfo{journal}{{\em CoRR\/}}  \bibinfo{volume}{abs/2307.09288} (\bibinfo{year}{2023}).
\newblock
\showDOI{%
\url{https://doi.org/10.48550/ARXIV.2307.09288}}
\showeprint{2307.09288}


\bibitem[\protect\citeauthoryear{Wang, Cai, Xie, Pan, Guan, Chu, Wang, Li, Huang, Cai, Hao, and Ding}{Wang et~al\mbox{.}}{2025}]%
        {wang2025wlb}
\bibfield{author}{\bibinfo{person}{Zheng Wang}, \bibinfo{person}{Anna Cai}, \bibinfo{person}{Xinfeng Xie}, \bibinfo{person}{Zaifeng Pan}, \bibinfo{person}{Yue Guan}, \bibinfo{person}{Weiwei Chu}, \bibinfo{person}{Jie Wang}, \bibinfo{person}{Shikai Li}, \bibinfo{person}{Jianyu Huang}, \bibinfo{person}{Chris Cai}, \bibinfo{person}{Yuchen Hao}, {and} \bibinfo{person}{Yufei Ding}.} \bibinfo{year}{2025}\natexlab{}.
\newblock \showarticletitle{WLB-LLM: Workload-Balanced 4D Parallelism for Large Language Model Training}.
\newblock \bibinfo{journal}{{\em arXiv preprint arXiv:2503.17924\/}} (\bibinfo{year}{2025}).
\newblock
\showURL{%
\url{https://doi.org/10.48550/arXiv.2503.17924}}


\bibitem[\protect\citeauthoryear{Yang, Li, Yang, Zhang, Hui, Zheng, Yu, Gao, Huang, Lv, Zheng, Liu, Zhou, Huang, Hu, Ge, Wei, Lin, Tang, Yang, Tu, Zhang, Yang, Yang, Zhou, Zhou, Lin, Dang, Bao, Yang, Yu, Deng, Li, Xue, Li, Zhang, Wang, Zhu, Men, Gao, Liu, Luo, Li, Tang, Yin, Ren, Wang, Zhang, Ren, Fan, Su, Zhang, Zhang, Wan, Liu, Wang, Cui, Zhang, Zhou, and Qiu}{Yang et~al\mbox{.}}{2025}]%
        {yang2025qwen3}
\bibfield{author}{\bibinfo{person}{An Yang}, \bibinfo{person}{Anfeng Li}, \bibinfo{person}{Baosong Yang}, \bibinfo{person}{Beichen Zhang}, \bibinfo{person}{Binyuan Hui}, \bibinfo{person}{Bo Zheng}, \bibinfo{person}{Bowen Yu}, \bibinfo{person}{Chang Gao}, \bibinfo{person}{Chengen Huang}, \bibinfo{person}{Chenxu Lv}, \bibinfo{person}{Chujie Zheng}, \bibinfo{person}{Dayiheng Liu}, \bibinfo{person}{Fan Zhou}, \bibinfo{person}{Fei Huang}, \bibinfo{person}{Feng Hu}, \bibinfo{person}{Hao Ge}, \bibinfo{person}{Haoran Wei}, \bibinfo{person}{Huan Lin}, \bibinfo{person}{Jialong Tang}, \bibinfo{person}{Jian Yang}, \bibinfo{person}{Jianhong Tu}, \bibinfo{person}{Jianwei Zhang}, \bibinfo{person}{Jianxin Yang}, \bibinfo{person}{Jiaxi Yang}, \bibinfo{person}{Jing Zhou}, \bibinfo{person}{Jingren Zhou}, \bibinfo{person}{Junyang Lin}, \bibinfo{person}{Kai Dang}, \bibinfo{person}{Keqin Bao}, \bibinfo{person}{Kexin Yang}, \bibinfo{person}{Le Yu}, \bibinfo{person}{Lianghao Deng}, \bibinfo{person}{Mei Li}, \bibinfo{person}{Mingfeng
  Xue}, \bibinfo{person}{Mingze Li}, \bibinfo{person}{Pei Zhang}, \bibinfo{person}{Peng Wang}, \bibinfo{person}{Qin Zhu}, \bibinfo{person}{Rui Men}, \bibinfo{person}{Ruize Gao}, \bibinfo{person}{Shixuan Liu}, \bibinfo{person}{Shuang Luo}, \bibinfo{person}{Tianhao Li}, \bibinfo{person}{Tianyi Tang}, \bibinfo{person}{Wenbiao Yin}, \bibinfo{person}{Xingzhang Ren}, \bibinfo{person}{Xinyu Wang}, \bibinfo{person}{Xinyu Zhang}, \bibinfo{person}{Xuancheng Ren}, \bibinfo{person}{Yang Fan}, \bibinfo{person}{Yang Su}, \bibinfo{person}{Yichang Zhang}, \bibinfo{person}{Yinger Zhang}, \bibinfo{person}{Yu Wan}, \bibinfo{person}{Yuqiong Liu}, \bibinfo{person}{Zekun Wang}, \bibinfo{person}{Zeyu Cui}, \bibinfo{person}{Zhenru Zhang}, \bibinfo{person}{Zhipeng Zhou}, {and} \bibinfo{person}{Zihan Qiu}.} \bibinfo{year}{2025}\natexlab{}.
\newblock \showarticletitle{Qwen3 technical report}.
\newblock \bibinfo{journal}{{\em arXiv preprint arXiv:2505.09388\/}} (\bibinfo{year}{2025}).
\newblock
\showURL{%
\url{https://doi.org/10.48550/arXiv.2505.09388}}


\bibitem[\protect\citeauthoryear{Zhai, He, Ma, Zong, Zhang, and Zhai}{Zhai et~al\mbox{.}}{2023}]%
        {zhai2023smartmoe}
\bibfield{author}{\bibinfo{person}{Mingshu Zhai}, \bibinfo{person}{Jiaao He}, \bibinfo{person}{Zixuan Ma}, \bibinfo{person}{Zan Zong}, \bibinfo{person}{Runqing Zhang}, {and} \bibinfo{person}{Jidong Zhai}.} \bibinfo{year}{2023}\natexlab{}.
\newblock \showarticletitle{$\{$SmartMoE$\}$: Efficiently training $\{$Sparsely-Activated$\}$ models through combining offline and online parallelization}. In \bibinfo{booktitle}{{\em 2023 USENIX Annual Technical Conference (USENIX ATC 23)}}. \bibinfo{pages}{961--975}.
\newblock


\bibitem[\protect\citeauthoryear{Zhao, Gu, Varma, Luo, Huang, Xu, Wright, Shojanazeri, Ott, Shleifer, Desmaison, Balioglu, Damania, Nguyen, Chauhan, Hao, Mathews, and Li}{Zhao et~al\mbox{.}}{2023}]%
        {DBLP:pytorch-fsdp}
\bibfield{author}{\bibinfo{person}{Yanli Zhao}, \bibinfo{person}{Andrew Gu}, \bibinfo{person}{Rohan Varma}, \bibinfo{person}{Liang Luo}, \bibinfo{person}{Chien{-}Chin Huang}, \bibinfo{person}{Min Xu}, \bibinfo{person}{Less Wright}, \bibinfo{person}{Hamid Shojanazeri}, \bibinfo{person}{Myle Ott}, \bibinfo{person}{Sam Shleifer}, \bibinfo{person}{Alban Desmaison}, \bibinfo{person}{Can Balioglu}, \bibinfo{person}{Pritam Damania}, \bibinfo{person}{Bernard Nguyen}, \bibinfo{person}{Geeta Chauhan}, \bibinfo{person}{Yuchen Hao}, \bibinfo{person}{Ajit Mathews}, {and} \bibinfo{person}{Shen Li}.} \bibinfo{year}{2023}\natexlab{}.
\newblock \showarticletitle{PyTorch {FSDP:} Experiences on Scaling Fully Sharded Data Parallel}.
\newblock \bibinfo{journal}{{\em Proc. {VLDB} Endow.\/}} \bibinfo{volume}{16}, \bibinfo{number}{12} (\bibinfo{year}{2023}), \bibinfo{pages}{3848--3860}.
\newblock
\showDOI{%
\url{https://doi.org/10.14778/3611540.3611569}}


\end{thebibliography}
